\documentclass{jpp}  

  \usepackage{comment}
  \usepackage{graphicx, color}
  \usepackage{mathtools}          
  \usepackage{amssymb,amsmath}
  \usepackage{bm,bbm}
  \usepackage{float}
  \usepackage[usenames,dvipsnames]{xcolor}

   \newcommand{\vct}[1]  {\ensuremath{\boldsymbol{#1}}}    

 \newcommand{\va} {\vct{a}}
 \newcommand{\vb} {\vct{b}}
 \newcommand{\vc} {\vct{c}}

 \newcommand{\vf} {\vct{f}}
 \newcommand{\vg} {\vct{g}}

 \newcommand{\vj} {\vct{j}}

 \def\vr{\vct{r}}               

 \newcommand{\vu} {\vct{u}}

 \newcommand{\vx} {\vct{x}}
 \newcommand{\vy} {\vct{y}}
 \newcommand{\vz} {\vct{z}}

 \newcommand{\vA} {\vct{A}}
 \newcommand{\vB} {\vct{B}}
 \newcommand{\vC} {\vct{C}}

 \newcommand{\vF} {\vct{F}}

 \newcommand{\vJ} {\vct{J}}

 \newcommand{\vS} {\vct{S}}

 \newcommand{\vX} {\vct{X}}
 \newcommand{\vY} {\vct{Y}}
 \newcommand{\vZ} {\vct{Z}}
 \newcommand{\vOmega}{\vct{\Omega}}
 \newcommand{\vSigma}{\vct{\Sigma}}

 \renewcommand{\d} {\mathrm{d}}

   \newcommand{\PD}    [2]    {\frac{\partial   {#1}} {\partial{#2}} }

 \newcommand{\avg}  [1]  { \left\langle #1 \right\rangle }
 
\newcommand{\squishlist}{
   \begin{list}{$\bullet$}
    { \setlength{\itemsep}{0pt}      \setlength{\parsep}{3pt}
      \setlength{\topsep}{3pt}       \setlength{\partopsep}{0pt}
      \setlength{\leftmargin}{1.5em} \setlength{\labelwidth}{1em}
      \setlength{\labelsep}{0.5em} } }

\newcommand{\squishlisttwo}{
   \begin{list}{$\bullet$}
    { \setlength{\itemsep}{0pt}    \setlength{\parsep}{0pt}
      \setlength{\topsep}{0pt}     \setlength{\partopsep}{0pt}
      \setlength{\leftmargin}{2em} \setlength{\labelwidth}{1.5em}
      \setlength{\labelsep}{0.5em} } }

\newcommand{\squishend}{
    \end{list}  }

   \renewcommand{\vec} [1] {\boldsymbol{#1}}    
  \newcommand{\vom}    {\vec{\omega}}

  \newcommand{\beq}{\begin{equation}}
  \newcommand{\eeq}{\end{equation}}
  \newcommand{\pa}  {\left(}
  \newcommand{\pc}  {\right)}
  \newcommand{\lan} {\langle}
  \newcommand{\ran} {\rangle}

  \newcommand{\eps}{\varepsilon}

   \newcommand{\ol}  [1] {\overline{#1}}     

   \newcommand{\oll} [1] {\overline{#1}^\ell}

  \newcommand{\ou}{\overline{\vu}^\ell}
  \newcommand{\ob}{\overline{\vb}^\ell}
  \newcommand{\oj}{\overline{\vj}^\ell}

  \newcommand{\ovA}{\overline{\bm A}}
  \newcommand{\ovB}{\overline{\bm B}}
  \newcommand{\ovC}{\overline{\bm C}}
  \newcommand{\ovS}{\overline{\bm S}}
  \newcommand{\ovJ}{\overline{\bm J}}
  \newcommand{\ovOm}{\overline{\bm \Omega}}
  \newcommand{\ovSS}{\overline{\bm \Sigma}}

  \newcommand{\Sbarij}{ \overline{S}^\ell_{ij}} 
  
   \newcommand{\tr}[1] 
      {\mathrm{Tr} \left\{ {#1} \right\}}

  \newcommand{\red}  [1]{\textcolor{red}{#1}}

  \newcommand{\remark}[3] {\textcolor{#1} { ... {#2}: #3}\newline}

    \newcommand{\damiano}[1]  {\remark{blue}{DC}{#1} }
    \newcommand{\sean}   [1]  {\remark{OliveGreen}{SO}{#1} }
    \newcommand{\moritz} [1]  {\remark{red}{ML}{#1} }

  \newcommand{\inhouse}[1] {\textcolor{OliveGreen}
                            {\textsf{\ldots WORKING COMMENTS: #1}\newline }}

\newcommand{\Sl}  {\overline{S}^\ell}

\newcommand{\Sb}  
   {\overline{\Sigma}}

\newcommand{\Sbl}  {\overline{\Sigma}^{\ell} }
\newcommand{\vSbl}  {\overline{\bm \Sigma}^{\ell} }
\newcommand{\Sul}  {\overline{S}^{\ell} }
\newcommand{\vSul}  {\overline{\bm S}^{\ell} }

\newcommand{\vSFb} {\overline{\bm \Sigma}^{\sqrt{\theta}} }

\newcommand{\vSFu} {\overline{\bm S}^{\sqrt{\theta}} }

\newcommand{\vSFFb}{\overline{\overline{\bm \Sigma}^{\sqrt{\theta}}}^\phi }

\newcommand{\vSFFu}{\overline{\overline{\bm S}^{\sqrt{\theta}}}^\phi }
\newcommand{\Ol}  {\overline{\Omega}^{\ell} }

\newcommand{\Ob}  {\overline{J}^{\ell} }
\newcommand{\vOb}  {\overline{\bm J}^{\ell} }

 \newcommand{\Obm} {\overline{J}^m}
 
 \newcommand{\Sbm} {\overline{\Sigma}^m}

\newcommand{\vOFb} {\overline{\bm J}^{\sqrt{\theta}} }

\newcommand{\vOFu} {\overline{\bm \Omega}^{\sqrt{\theta}} }

\newcommand{\vOFFb}{\overline{\overline{\bm J}^{\sqrt{\theta}}}^\phi }

\newcommand{\vOFFu}{\overline{\overline{\bm \Omega}^{\sqrt{\theta}}}^\phi }

\newcommand{\Al} {\overline{A}^\ell}
\newcommand{\Am} {\overline{A}^m}
\newcommand{\Bl} {\overline{B}^\ell}
\newcommand{\Bm} {\overline{B}^m}

\newcommand{\bfm} {\overline{b}^{m}}

\shorttitle{MHD energy flux decomposition}
\shortauthor{D. Capocci, P. Johnson, S. Oughton, L. Biferale, M. Linkmann}

\title{Energy Flux Decomposition in 
Magnetohydrodynamic Turbulence}

\author{Damiano Capocci\aff{1}
\corresp{\email{capocci@roma2.infn.it}},
Perry L. Johnson\aff{2},
Sean Oughton\aff{3}, \\
Luca Biferale\aff{1}
 \and Moritz Linkmann\aff{4}\corresp{\email{moritz.linkmann@ed.ac.uk}}}

\affiliation{\aff{1}Department of Physics and INFN, University of Rome Tor Vergata, Rome, Italy
\aff{2} Department of Mechanical and Aerospace Engineering, University of California, Irvine, USA
\aff{3} Department of Mathematics, University of Waikato, Hamilton, New Zealand
\aff{4} School of Mathematics and Maxwell Institute for Mathematical Sciences, \\ University of Edinburgh, Edinburgh, EH9 3FD, United Kingdom
}

\begin{document}

  \maketitle

\begin{abstract}
In hydrodynamic (HD) turbulence an exact decomposition of the energy flux
across scales has been derived that identifies the contributions
associated with vortex stretching and strain self-amplification (P.
Johnson, \emph{Phys. Rev. Lett.}, 124, 104501 (2020), \emph{J. Fluid
Mech.} 922, A3 (2021)) to the energy flux across scales.  
Here we extend this methodology to general
coupled advection-diffusion equations, in particular to homogeneous
magnetohydrodynamic (MHD) turbulence, and we show that several subfluxes are related to each other by kinematic constraints akin to the Betchov relation in HD. 
Applied to data from direct numerical simulations, this
decomposition allows for an identification of
physical processes 
and for the quantification of their respective contributions to 
the energy cascade, as well as a quantitative assessment of 
their multi-scale nature through a further decomposition into 
single- and multi-scale terms.
We find	that vortex stretching is strongly depleted in MHD compared to HD, and the 
kinetic energy is transferred from large to small scales almost exclusively by
the generation of regions of small-scale intense strain induced by the Lorentz force. 
In regions of large strain, current sheets are stretched by large-scale straining
motion into regions of magnetic shear. This magnetic shear in turn drives extensional
	flows at smaller scales. 
Magnetic energy is transferred from large to small scales, albeit with 
considerable backscatter, 
predominantly by the aformentioned current-sheet thinning in region of high strain while
the contribution from current-filament stretching -- the analogue to vortex stretching -- is negligible. 
Consequences of these results to subgrid-scale turbulence modelling are discussed. 
\end{abstract}

\section{Introduction}
    \label{sec:intro}

\bigskip
Turbulence in electrically conducting fluids and plasmas is of relevance to a variety of processes in geophysical and astrophysical situations, as well as in industry \citep{Weiss2014,Davidson2016} and for nuclear fusion under magnetic confinement. 
For example, the solar wind is turbulent \citep{BrunoCarbone13},
convection-driven turbulence occurs in planetary cores and in the outer layers of stars \citep{Jones2011}, 
turbulence on ion and electron scales affects plasma confinement in magnetic confinement fusion reactors \citep{Freidberg2007},
and the heat transfer in liquid metal cooling applications is dependent on the level of turbulence in the flow \citep{Davidson1999}.  
Even though these systems are very different in terms of features like the presence of strong background magnetic fields, 
the level of magnetic field fluctuations, 
temperature gradients, 
density fluctuations, 
domain geometry, 
or the level of collisionality, 
they nonetheless share fundamental nonlinear processes that define energy conversion and inter-scale energy transfers, at least on scales where the fluid approximation is applicable. Even in the simplest case of magnetohydrodynamic (MHD) turbulence, 
despite considerable theoretical process 
\citep[e.g.,][]{GoldreichSridhar95, Biskamp-turb, ZhouEA04, PetrosyanEA10, BrandenburgEA12, TobiasCattaneo13, Beresnyak19, Matt21-maxwell, Schekochihin-biased}, the physical nature of these processes remain opaque.  

%
Moreover, the typical parameter ranges in which MHD turbulence develops in
Nature are far from those attainable with direct numerical simulation (DNS)
\citep{plunian20131, MieschEA15, Schmidt15}.  As a consequence, the demand for
approximations and subgrid-scale (SGS) models for large-eddy simulations (LES) of 
MHD turbulence that are able
to capture the effects of unresolved small-scale fluctuations---that govern
important processes such as magnetic reconnection and plasma heating---is
increasing \citep{MieschEA15}. However, constructing such models is a challenge
due to small-scale anisotropy \citep{ShebalinEA83, OughtonEA94,
GoldreichSridhar97, TobiasCattaneo13}, strong intermittency in the magnetic
fluctuations as observed in numerical simulations \citep[e.g.,][]{MininniPouquet09-intermit,
SahooEA11, YoshimatsuEA11, RodriguezImazioEA13, MeyrandEA15} and in the solar
wind \citep[e.g.,][]{Veltri99, SalemEA09, WanEA12-kurt, MattEA15-philtran}, 
and whether magnetic-field fluctuations are 
maintained by the flow or by an external electromagnetic force \citep{AlexakisChibbaro22}.  For a summary of
the SGS modelling effort and its challenges, we refer to the
review articles by \citet{MieschEA15} and \citet{Schmidt15}.

Here, we focus on energy transfer across scales in
statistically stationary homogeneous 
MHD
turbulence in a saturated nonlinear dynamo
regime without a mean magnetic field, and with negligible levels or cross and magnetic helicity. 
The total energy cascade in this case is
direct \citep{AluieEyink10}, transferring energy from the large
to the small scales in a scale-local
  fashion. 
The aims of this paper are (i) to understand the physical
mechanisms that govern the MHD energy cascade and (ii) to quantify their
importance and provide guidance for SGS modelling \citep{Johnson2022}.  
In terms of
turbulence theory this corresponds to understanding physical properties of the
subgrid scale (SGS) stresses as a function of scale.  \cite{Eyink06-multiscale}
introduced a viable approach for this that involves expanding the SGS tensors
in terms of vector field gradients. Here we follow a filtering approach, generalising an
exact gradient-based decomposition of the hydrodynamic (HD) energy fluxes
\citep{Johnson20,Johnson21} to coupled advection-diffusion equations and hence
to the MHD equations. This methodology distinguishes between terms 
that are local in scale, corresponding to the first term in the gradient expansion 
and those which are truly multi-scale, providing a closed expression for the remainder of the series expansion
 \citep{Johnson20,Johnson21}. 
Expressing SGS stresses through vector-field gradients results in a decomposition of the energy fluxes 
in terms of different tensorial contractions between strain-rate, vorticity, current and magnetic strain, and as such 
facilitates the physical interpretation of such sub-fluxes. The provision of closed expressions allows for a quantification of the 
relative contribution of all terms to the energy cascade using data obtained by direct numerical simulation (DNS).
In homogeneous and isotropic HD turbulence, where only the inertial term is present, the decomposition identifies three processes that 
transfer kinetic energy across scales, vortex stretching, strain self-amplification and strain-vorticity alignment, 
and quantifies their relative contribution to the energy cascade \citep{Johnson20,Johnson21}. 
Similarly the direct cascade of kinetic helicity is carried by three different processes, vortex flattening, vortex twisting and
vortex entanglement \citep{CapocciEA23-Hk}.

In MHD, the total energy transfer can be
split into four subfluxes, Inertial, Maxwell, Dynamo, and Advection\footnote{We use these words capitalised to indicate that they refer to the SGS energy flux arising from the term with the lowercase version of the name in the momentum or induction equation.}, with the former 
two originating from the Reynolds and Maxwell stresses in the momentum equation and the latter two from 
stresses in the induction equation that result in the advection and bending/stretching of magnetic field lines by the flow. As Dynamo and Advection terms have a common physical origin, the electric field, often only their sum is considered in \emph{a-priori} analyses of DNS data \citep{Aluie17,OffermansEA18,AlexakisChibbaro22} and \emph{a-posteriori} in LES \citep{ZhouVahala91,Muller02,KessarEA16, VlaykovEA16,GreteEA16}. 
However, in the present work it will prove instructive to consider them separately.  
Here, we generalise and apply the aforementioned decomposition to
each of these four subfluxes.  
In doing so, we find that the average Inertial flux is strongly depleted in all components. That is, vortex-stretching and strain self-amplification are strongly suppressed at all length scales. 
Consequently, kinetic energy is transferred from large to small scales mostly by the Maxwell flux that encodes the effect of the Lorentz force on the flow, with the dominant contribution arising from  
small-scale strain amplification by current-sheet thinning. As the flow and magnetic field lines move together in the limit of negligible dissipation, 
a current-sheet thinning process will generate strong outflows in the direction that is not constrained by the current sheet. This creates regions of intense strain across smaller scales.    
We point out that this interpretation makes use of Alfv\'en's theorem that should hold in approximation in the inertial range, where diffusion is negligible. 
Magnetic energy is almost exclusively transferred downscale by current-sheet thinning by the straining motion of an incompressible flow.
Contributions from strain-induced current filament stretching, the formal analogue to vortex stretching, are negligible. Furthermore, several magnetic 
subflux terms can be related to one another through kinematic constraints akin to the Betchov relation in HD \citep{Betchov56}.  

The structure of the paper is as follows: we begin in sec.~\ref{sec:theory}
with an outline of how the generalised method can be applied to obtain MHD
energy subfluxes.  In sec.~\ref{sec:methods} we discuss the numerical details
and the associated datasets on which we performed the filtering analysis.  In
sec.~\ref{sec:analysis} we consider each subflux decomposition, showing results
for both mean terms and fluctuations.  
Ramifications of those results for SGS modelling are considered in sec.~\ref{sec:sgs_modelling}.
In sec.~\ref{sec:concl} we discuss our
main results and indicate future work directions.  
Several appendices flesh out some aspects of the derivations and analysis.


    \section{Theory}
    \label{sec:theory}

In this section we begin by sketching the derivation of the
coarse-grained energy equations for MHD
and giving the definitions of the scale-space energy fluxes
that appear in them.  Subsequently we show how each flux can be
decomposed in terms of physically distinct contributions, and discuss their physical interpretations.



        \subsection{Coarse-grained energy equations}
        \label{sec:coarse-graining}

Our starting point is incompressible three-dimensional (3D)
homogeneous MHD turbulence.
The primary dynamical variables are then the fluctuation
velocity $ \vu (\vx, t)$
and the fluctuation magnetic field $ \vb (\vx,t) $, where we
measure the latter in Alfv\'en speed units:
        $  \vb / \sqrt{4\pi \rho} \to \vb $,
with $\rho$ the uniform mass density.
We consider situations with no mean magnetic field since these are more likely to exhibit global isotropy.
The governing equations, with allowance for hyper-dissipation,
are
\begin{align}
    \partial_t u_i
        +  \partial_j \left( u_i u_j \right)
   & =  -  \partial_i \left( p + \frac{b^2}{2} \right)
        +  \partial_j \left( b_i b_j \right)
        +  \nu_\alpha (-1)^{\alpha +1} {\nabla}^{2\alpha} u_i
        + F_i ,
  \label{eq:mtm}
\\
    \partial_t b_i
        +  \partial_j \left (b_i u_j \right)
     & =   \hspace*{7.25em}
           \partial_j \left( u_i b_j \right)
        +  \mu_\alpha (-1)^{\alpha +1} \, {\nabla}^{2\alpha} b_i
  \label{eq:induction}
\\
    \nabla \cdot \vu   & = 0  ,
       \quad
    \nabla \cdot \vb = 0 .
  \label{eq:divzeros}
\end{align}
Here $p$ is the pressure,
 $ \vF $ is a (large-scale) velocity forcing,
 $ \nu_\alpha $ and $ \mu_\alpha $ are the hyper-viscosity and hyper-resistivity,
and
 $ \alpha $  denotes the power of the Laplacian operator employed in
 the hyper-dissipation.
Standard Laplacian dissipation corresponds to the case $\alpha =1 $.

The MHD variables,
and equations,
may be spatially coarse-grained using a suitable
filtering field,
         $ G^\ell (\vr ) $
        \citep{Germano92,Aluie17}.
The role of
         $ G^\ell (\vr ) $
is to strongly suppresses structure at scales less than the
filtering scale $\ell$.
For example, the filtered velocity field is
\begin{equation}
      \ol{\vu}^\ell(\vx)
         =
      \int \d^3 r  \,  G^\ell (\vr ) \, \vu ( \vx + \vr) .
  \label{eq:defn-ubar}
\end{equation}
This can be interpreted as a weighted average of $\vu$ centered on the
position $\vx$. 
The weighting function $ G^\ell$ decays very rapidly to
zero at distances greater than a few
        $ \ell$ from  $ \vx $
and
 satisfies some other weak restrictions, such as smoothness and having a volume integral of unity.
Filtering is a linear operation and commutes with differentiation, properties we will make considerable use of below.
From Section~\ref{sec:Gaussian} onwards we
will specialise to a Gaussian filter, but in this section a specific
choice of filter is not needed.

Coarse-graining of~\eqref{eq:mtm}--\eqref{eq:induction} introduces
four sub-gridscale 
  (SGS) 
stress tensors, $ \tau^\ell(\cdot, \cdot) $,
associated with the advective-type nonlinear terms
(those containing a $\partial_j$).
These each have the form
\begin{align}
      \tau^\ell (f_i, g_j)
       & =
        \ol{f_i g_j}^\ell
      - \ol{f_i}^\ell \, \ol{g_j}^\ell ,
 \label{eq:tau-fg}
\end{align}
where
        $\vf$ and $\vg$
are the solenoidal vectors appearing in a
        $ g_j \partial_j f_i $
advective-type term.  
We remark that with this notation the  {advecting}  field
is the \emph{second} argument in a
        $ \tau^\ell(\cdot,\cdot) $.

To obtain the equations governing the (pointwise) evolution of the coarse-grained
kinetic energy
        $ E^\ell_u (\vx,t)  = \frac{1}{2} \ou \cdot \ou $
and magnetic energy
        $ E^\ell_b (\vx,t)  = \frac{1}{2} \ob \cdot \ob $
one filters
        \eqref{eq:mtm}--\eqref{eq:induction}
and then multiplies by
        $\ou$ and $\ob$, respectively
        \citep[e.g.,][]{ZhouVahala91,KessarEA16,Aluie17,OffermansEA18,AlexakisChibbaro22}.
The result can be written as
\begin{align}
  \partial_t {E}^\ell_u
        + \nabla \cdot \vct{{\cal J}}_u
      & =
        - \Pi^{I,\ell} - \Pi^{M,\ell}
        - {\cal W}^\ell
        - {\cal D}_{u} 
        + \ou \cdot \ol{\vF}^\ell,
  \label{eq:Eu-ls}
 \\
  \partial_t {E}^\ell_b
        + \nabla \cdot \vct{{\cal J}}_b
      & =
        - \Pi^{A,\ell} - \Pi^{D,\ell}
        + {\cal W}^\ell
        - {\cal D}_{b} ,
  \label{eq:Eb-ls}
\end{align}
%
where the $\vct{{\cal J}}$ terms account for the 
spatial transport of energy
and
the $\Pi$ terms embody energy fluxes 
(i.e., transfer \emph{across} scale $\ell$).
Our sign convention for the definitions of the $\Pi$ (see below) means that $\Pi > 0 $ corresponds to forward transfer of energy, i.e., to scales smaller than $\ell$.
The ${\cal D}$ terms represent (hyper-)dissipative effects.
Also present is the
 \emph{resolved-scale conversion} (RSC) term,
   $ {\cal W}^\ell = \ol{b}_i \ol{b}_j \partial_j \ol{u}_i $,
here expressed as in
  \cite{Aluie17}.
This appears with opposite sign in each equation, and represents an \emph{exchange} between large-scale kinetic and magnetic energies.  An important point is that it is \emph{not} an energy flux term
since it does not involve energy transfer across scale $\ell$.
It supports several interpretations including
as 
(i) the rate of work done on the large-scale flow by the large-scale Lorentz force
and 
(ii) the energy gained by $\ob$ as it is distorted by $\ou$ or \emph{vice versa}. 
Detailed forms for the spatial transport currents, 
  $ \vct{{\cal J}}_u, \vct{{\cal J}}_b $, 
depend on the form employed for ${\cal W}^\ell$  
and are available elsewhere
        \citep[e.g.,][]{KessarEA16,Aluie17,OffermansEA18,AlexakisChibbaro22},
while the problem of Galilean invariance has been addressed in \cite{OffermansEA18}.

Our primary interest herein centers on
the pointwise energy flux (at scale $\ell$) terms, denoted by 
    $ \Pi^\ell (\vx, t)$,
together with their volume averages,
    $ \avg{ \Pi^\ell } $.
For any choice of filter, these can be expressed in terms of
contractions of filtered gradient tensors and SGS stress tensors: 
\begin{align}
  \Pi^{I,\ell} & =  - \PD{\ol{u}_i^\ell}{x_j} \, \tau^\ell(u_i,u_j) ,
 \label{eq:Pi-I} 
 \\
  \Pi^{M,\ell} & =  \;\; \PD{\ol{u}_i^\ell}{x_j} \, \tau^\ell(b_i,b_j) ,
 \label{eq:Pi-M} 
 \\
  \Pi^{A,\ell} & =  - \PD{\ol{b}_i^\ell}{x_j} \, \tau^\ell(b_i,u_j) ,
 \label{eq:Pi-A} 
 \\
  \Pi^{D,\ell} & =  \;\; \PD{\ol{b}_i^\ell}{x_j} \, \tau^\ell(u_i,b_j) ,
 \label{eq:Pi-D}
\end{align}

Like the
        $\tau^\ell ( \cdot, \cdot) $
these arise in connection with the four advection type
nonlinearities in
        \eqref{eq:mtm}--\eqref{eq:induction},
that we refer to as the Inertial, Maxwell (meaning from the Lorentz force), Advection, and Dynamo terms.
Note the capitalization.
 Taken together with eqns.~\eqref{eq:Eu-ls} and \eqref{eq:Eb-ls} 
 these definitions of the $\Pi$'s
 mean than the interpretation of the direction of an energy flux does not depend on which flux it is.
 This is why~\eqref{eq:Pi-M} and~\eqref{eq:Pi-D}
 lack a leading minus sign. 
Specifically, a positive value for 
any one of these fluxes
corresponds to transfer of energy from scales greater than $\ell$ to scales smaller than $\ell$.
Clearly,
        $ \Pi^{I,\ell} + \Pi^{M,\ell} $
is the net flux of $ E^\ell_u $,
and
        $ \Pi^{A,\ell} + \Pi^{D,\ell} $
that for        $ E^\ell_b $.
As is well known, 
   $ \Pi^{A,\ell} $ and  $ \Pi^{D,\ell} $ 
have a common origin and may be readily combined to obtain the magnetic energy flux associated with the curl of the induced electric field.


        \subsection{Gaussian filter}
        \label{sec:Gaussian}

Rather remarkably
the choice of a Gaussian filter enables
an analytic determination of the SGS stresses and fluxes in terms of
the gradients of the filtered variables
        \citep{Johnson20,Johnson21}.
In the remainder of the paper we therefore employ an
         isotropic
Gaussian filter
\begin{equation}
    G^\ell (\vr )
        =
    \frac{1}{(2\pi\ell^2)^{3/2}}
       \exp \left( -\frac{|\vr|^2}{2 \ell^2} \right) 
.       
 \label{eq:Gell}
\end{equation}

        \subsection{Exact expressions for the $\tau^\ell$ and $\Pi^\ell$}
        \label{sec:exact}



Employing the Fourier transform of the Gaussian filter \eqref{eq:Gell},
   \cite{Johnson20,Johnson21} 
showed that the filtered version of an arbitrary 
field $\vu(\vx,t)$
satisfies a diffusion equation,
\begin{align}
   \PD{ \ol{u}^\ell_j}{(\ell^2)}
  & =
   \frac{1}{2} \nabla^2 \ol{u}^\ell_j,
  \qquad 
    \ol{u}^\ell_j \Bigr\vert_{\ell=0}
   =
     u_j( \vx, t ),
 \label{eq:ubar-diffusion}
\end{align}
where $\ell^2$ is the time-like variable. 
It was further shown that the associated SGS stress tensor, 
  $\tau^\ell (u_i, u_j ) $,
obeys a forced version of this diffusion equation.
The forcing term is
    $ \ol{A}^\ell_{ik} \ol{A}^\ell_{jk} $
where 
    $ \ol{A}^\ell_{ik} = \partial_k \ol{u}^\ell_i $
is the gradient tensor for the filtered field.
An exact solution for
    $ \tau^\ell ( u_i, u_j) $
was obtained that depends on 
    $ \ol{A}^\phi_{ik} $
for all scales $\phi \le \ell $.
Substituting this into the equation for the SGS Inertial flux, 
    \eqref{eq:Pi-I},
produces an exact solution for this flux
   \citep{Johnson20,Johnson21} corresponding to the exact summation of the perturbation series proposed by \cite{Eyink06-multiscale}.

             

Happily, this approach is readily extended to MHD and may be used to calculate the elements contained in equations~\eqref{eq:Pi-I}--\eqref{eq:Pi-D}.
Below we outline how to achieve this for the particular case of the magnetic energy subflux~\eqref{eq:Pi-A} that originates with
the advection term in the induction equation 
(i.e., $\vu \cdot \nabla \vb $).  
More details, plus the general case of three distinct 
    solenoidal 
fields, are available in Appendix~\ref{app:general-case};
  see also Appendix~C of
  \cite{Johnson21} and
  \cite{CapocciEA23-Hk}.

We seek an exact solution for
    $ \tau^\ell( b_i, u_j) = \oll{b_i u_j} - \oll{b_i} \oll{u_j} $.
Clearly 
    $ \oll{ b_i u_j } $ 
will also satisfy~\eqref{eq:ubar-diffusion} since that equation holds for any (Gaussian) filtered field.
Together with the product rule expansion of 
   $ \partial ( \oll{b}_i \oll{u}_j ) / \partial {\ell^2} $
this yields
\begin{align}
   \left(
     \PD{}{\ell^2} - \frac{1}{2} \nabla^2
   \right)
   \tau^\ell (b_i, u_j)
  & =
     \oll{B}_{ik} 
     \oll{A}_{jk}
 ,
  \qquad
     \tau^{\ell=0} (b_i, u_j) = 0
 ,
 \label{eq:bu-diffusion}
\end{align}
where
    $ \oll{B}_{ik} = \partial_k \oll{b}_i $
is the gradient tensor for $ \oll{b}_i $.  
The solution can be written in the form
  \citep{Johnson21}
\begin{align}
  \tau^\ell (b_i, u_j) 
    & = 
  \ell^2 \; \oll{B}_{ik} \oll{A}_{jk} 
    \; + \;
    \int_0^{\ell^2} \d\theta 
       \left(
         \ol{ \ol{B}_{ik}^{\sqrt{\theta}} 
              \ol{A}_{jk}^{\sqrt{\theta}} \,}^\phi 
           -
         \ol{\ol{B}_{ik}^{\sqrt{\theta}}\,}^\phi \;
         \ol{\ol{A}_{jk}^{\sqrt{\theta}}\,}^\phi  
       \right),
 \label{eq:tau-ell-bu}
\end{align}
where  $\phi = \sqrt{\ell^2 - \theta} $.
The first term on the RHS is a `single-scale' piece as it contains only resolved scale terms 
\citep[cf.][]{ClarkEA79}.  
In terms of other work, it corresponds to the non-linear model employed in \cite{leonard1975}, \cite{BorueOrszag98}, and \cite{MeneveauKatz00},
and is the first-order term in the expansion of \cite{Eyink06-multiscale}. It is also the leading-order term of the power law expansion in the filter limit going to zero relative of any filter kernel with finite moments \citep[cf.\ sec 13.4.4 of][]{Pope}.

The second term, involving an integral over all scales smaller than $\ell$, is manifestly a multi-scale contribution.
This element of the exact solution was first presented in
    \cite{Johnson20}.
Note that the integrand in 
    \eqref{eq:tau-ell-bu}
can itself be written as an SGS stress, but one based on the field gradients rather than the fields themselves: 
   $ \tau^\phi \left( 
       \ol{B}_{ik}^{\sqrt{\theta}}, \ol{A}_{jk}^{\sqrt{\theta}} 
             \right) $.

Contracting with $\ol{B}_{ij}^\ell $ provides an exact expression 
for~\eqref{eq:Pi-A}:
\begin{align}
  \Pi^{A,\ell} 
     &= 
	- \ell^2 \, \ol{B}_{ij}^\ell \ol{B}_{ik}^\ell \ol{A}_{jk}^\ell 
    \; - \;
	\ol{B}_{ij}^\ell
      \int_0^{\ell^2} \d\theta 
       \left(
	 \ol{ \ol{B}_{ik}^{\sqrt{\theta}} 
	      \ol{A}_{jk}^{\sqrt{\theta}} \,}^\phi 
           -
	 \ol{\ol{B}_{ik}^{\sqrt{\theta}}\,}^\phi \;
	 \ol{\ol{A}_{jk}^{\sqrt{\theta}}\,}^\phi  
       \right)
 \\
    & =
    \Pi^{A,\ell}_s  +  \Pi^{A,\ell}_m 
 ,
  \label{eq:Pi-A-exact}  
\end{align} 
where the subscripts $s$ and $m$ denote the single- and multi-scale contributions, respectively.
It is evident that all the SGS energy fluxes,  
    \eqref{eq:Pi-I}--\eqref{eq:Pi-D},
and also other SGS fluxes (e.g., for helicities),
can be written strictly in terms of (multiscale) gradients of the velocity and magnetic vector fields.
Appendix~\ref{app:general-case} contains further details. 
When discussing the individual SGS energy fluxes in Section~\ref{sec:analysis} we will make regular reference to 
eq.~\eqref{eq:Pi-xyz-exact}, the generalised form of~\eqref{eq:Pi-A-exact}.

The tensor contractions present in~\eqref{eq:Pi-A-exact} 
can be expressed as the trace of the matrix products involved,
after appropriate use of the transpose operation 
  (superscript $t$).
For example,
    $ \oll{B}_{ij} \oll{B}_{ik} \oll{A}_{jk}
      = \tr{ \left(\oll{\bm B}\right)^t \, \oll{\bm B} \, \left(\oll{\bm A}\right)^t}
    $.


Further insight into the physics of the scale-space flux $\Pi^{A,\ell}$ may be extracted by expressing each gradient tensor as the sum of its index-symmetric and index-antisymmetric components
   \citep{Johnson20}.  
Let us write
    $ S_{ij} = ( A_{ij} + A_{ji} ) /2 $
and
    $ {\Omega}_{ij} = ( A_{ij} - A_{ji} ) / 2 $,
respectively the (velocity) rate-of-strain tensor and the rotation rate tensor, with
    $ {\Omega}_{ij} = -\epsilon_{ijk} {\omega}_k /2 $
in terms of the vorticity
    $ {\omega}_k = \epsilon_{ijk} \partial_i {u}_j $.
Similarly,
    $ B_{ij} = \Sigma_{ij} + {J}_{ij} $,
where the non-zero elements of 
    $ J_{ij} = ( B_{ij} - B_{ji} ) / 2 = -\epsilon_{ijk} j_k/2 $ 
are essentially the components of the electric current density 
    $ \vj = \nabla \times \vb $ and the magnetic strain-rate tensor $\Sigma_{ij} = (B_{ij} + B_{ji})/2$.
We will of course require the filtered versions of all of these quantities.

Ostensibly this decomposition gives eight single-scale and and eight multi-scale sub-fluxes; see eq.~\eqref{eq:XYZ_expanded}. 
However, properties of the trace of particular products of symmetric or antisymmetric matrices mean some of these may vanish, cancel, or be equivalent.
In the present case one obtains, in connection with the single-scale contributions,
\begin{align}
  \tr{ \left(\oll{\bm B}\right)^t \, \oll{\bm B} \, \left(\oll{\bm A}\right)^t}
    &=
  \tr{ 
     \left( \ovSS^\ell - \ovJ^\ell   \right)
     \left( \ovSS^\ell + \ovJ^\ell  \right)
     \left( \ovS^\ell - \ovOm^\ell \right)
    } 
  \label{eq:BBA-sym-anti-prelim}
 \\
     &=
   \tr{   
	  \ovSS^\ell   \ovSS^\ell   \ovS^\ell 
       -  \ovJ^\ell  \ovJ^\ell \, \ovS^\ell 
       + 2 \ovSS^\ell  \ovJ^\ell  \ovS^\ell
     },
  \label{eq:BBA-sym-anti}
\end{align}
so that there are only three distinct subfluxes;  see Appendix~\ref{app:general-case} for details.
We write the single-scale flux 
for the Advection term as
\begin{align}
  \Pi^{A,\ell}_s 
    & = 
        \Pi^{A, \ell}_{s, \Sigma \Sigma S}
    +   \Pi^{A, \ell}_{s, J J S}
    +  2\Pi^{A, \ell}_{s, \Sigma J S}
,
 \label{eq:Pi-A-ss}
\end{align}
where
 $ \Pi^{A,\ell}_{s, PQR} = -\ell^2 \tr{\left(\oll{\bm P}\right)^t \oll{ \bm Q} \left(\oll{ \bm R}\right)^t} $ and each of $ \bm P, \bm Q,\bm R$ are either symmetric or antisymmetric tensors.


The multi-scale 
flux contributions can also be so decomposed, with the details given in Appendix~\ref{app:general-case}.   The particular forms for 
    $ \Pi^{A,\ell}_m $
are discussed in sec.~\ref{sec:Pi-A}.




    \section{Methods and Data}
    \label{sec:methods}

To quantify the four energy fluxes
        $ \Pi^{Y,\ell}$
present in eqs.~\eqref{eq:Eu-ls}--\eqref{eq:Eb-ls},
and their decompositions, 
we employ outputs from numerical simulations
of the MHD equations~\eqref{eq:mtm}--\eqref{eq:divzeros}.
We consider both standard diffusive
($\alpha = 1$) 
and hyper-diffusive 
($\alpha = 5$) cases, always
with
        $ \nu_\alpha = \mu_\alpha $.
The fluctuation fields, $\vu$ and $\vb$, have zero means and there is
no background magnetic field (i.e., $B_0 = 0 $).
The equations are solved using fully dealiased Fourier pseudospectral
codes in a triply periodic $ (2\pi)^3 $ domain
        \citep{PattersonOrszag71,CanutoEA}.
The time advancement is via a second-order Runge--Kutta scheme
with dealiasing implemented using the two-thirds rule.

\begin{table}
  \begin{center}
\def~{\hphantom{0}}
   \begin{tabular}{ccccccccccccccccc}
        \hline
        \hline
		 id & $N$ &$\alpha$  & $E_u$ & $\nu_\alpha$ & $\eps_u$ & $\eps_b$ & $L_u$ & $\tau$  & $\mbox{Re}$ &  $\eta_\alpha/10^{-3}$  & $k_\text{max} \eta_{\alpha}^u$ & $k_\text{max} \eta_{\alpha}^b$ & $\Delta t / \tau$ & \# \\
        \hline
		 A1 &  2048 & 5 & 0.66  & $5 \times 10^{-26}$ & 0.33  & 0.43  & 0.51  & 0.81    & 9931 & 2.0 & 1.38 & 1.37 & 1.1  &  18  \\
       \hline
      H2 & 2048 & 4 & 3.84  & $5.7 \times 10^{-20}$ & 1.50  & - & 1.10  & 0.69  & 26000   & 2.3    & 1.57 & - & 1.0  & 6 \\

        \hline
      \hline
        \end{tabular}
        \caption{Simulation parameters and key observables, where
        $N$ is the number of collocation points in each coordinate,
        $\alpha$ is the power of $\nabla^2$ used in the hyper-diffusion,
        $E_u$ the (mean) total kinetic energy,
        $\nu_\alpha$ the kinematic hyperdiffusivity,
        $\eps_u$ and $\eps_b$ are the kinetic and magnetic energy dissipation rates,
        $L_u = (3 \pi/4 E_u) \int_0^{k_\text{max}} \d{k} E_u(k)/k$ the longitudinal integral scale,
        $\tau = L_u/\sqrt{2E_u/3}$ the large-scale eddy-turnover time, and
        $\mbox{Re}$ is the Reynolds number. 
	  Furthermore, $\eta_{\alpha} = (\nu_\alpha^3 / \varepsilon)^{1 / (6  \alpha -2)}$, $\eta^u_{\alpha} = (\nu_\alpha^3 / \varepsilon_u)^{1 / (6  \alpha -2)}$ and $\eta^b_{\alpha} = (\mu_\alpha^3 / \varepsilon_b)^{1 / (6  \alpha -2)}$, are the hyperdiffusive Kolmogorov scales calculated with respect to the total, viscous and Joule dissipation rates, respectively,
        $k_\text{max}$ the largest retained wavenumber component after de-aliasing, 
        $\Delta t$ the mean of the snapshots sampling intervals,
        and \# indicates the number of snapshots used in the averaging.
        The magnetic Prandtl number, 
        $Pm = \nu_\alpha / \mu_\alpha $, the ratio between the hyperviscosity and magnetic hyperdiffusivity, equals unity for A1.
	}
  \label{tab:datasets}
  \end{center}
\end{table}
As table~\ref{tab:datasets} indicates, we use up to $2048^3$
grid points.
The spatial resolution of the simulations is quantified by both the
grid spacing $\Delta{x} = 2\pi / N$
and the hyper-diffusive Kolmogorov scales
        $ \eta^u_{\alpha} = (\nu_\alpha^3 / \varepsilon_u)^{1/(6  \alpha -2)} $, and 
        $ \eta^b_{\alpha} = (\mu_\alpha^3 / \varepsilon_b)^{1/(6  \alpha -2)} $,  
where $\varepsilon_u$ and $\varepsilon_b$ are the mean kinetic and magnetic energy dissipation rates
        \citep{BorueOrszag95}.
For adequate resolution we require $ \eta^u_\alpha / \Delta{x} \gtrsim 1.3 $
and $ \eta^b_\alpha / \Delta{x} \gtrsim 1.3 $
  \citep[e.g.,][]{DonzisEA08, WanEA10-accuracy}.

The forcing $f_i$ applied to the system is
a drag-free Ornstein--Uhlenbeck process, active in the wavenumber band 
  $k \in [2.5,5.0] $ 
for the MHD simulations 
while the hydrodynamics dataset H2 is  forced in the band 
  $k \in [0.5,1.5]$. 
The snapshots, consisting of instantaneous velocity and magnetic fields, have been sampled about once per large-scale turnover time after the simulations reach statistically stationary states. 

\begin{figure}
	\begin{center}
         \includegraphics[width=.55\columnwidth]{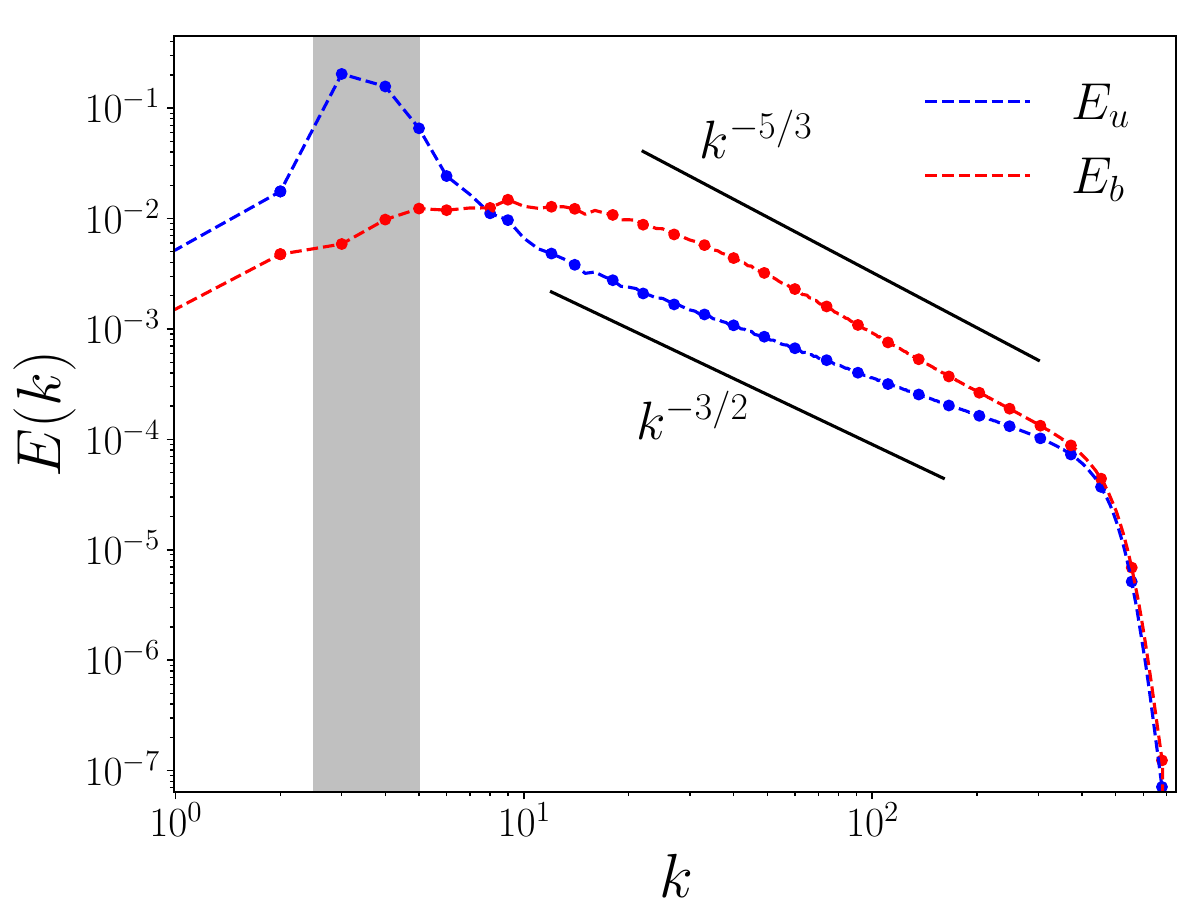} 
    \end{center}
	 \caption{ 
  Time-averaged omnidirectional spectra for the velocity and magnetic field (run A1). The gray region indicates the wavenumber band where the velocity field is forced: $k \in [2.5,5.0]$.
  }
\label{fig:kin_mag_spec}
\end{figure}

Figure~\ref{fig:kin_mag_spec} shows the time-averaged omnidirectional kinetic and magnetic spectra. The peak in the kinetic 
spectrum is due to the activity of forcing 
for $k \in [2.5,5.0]$,
while the magnetic spectrum is considerably lower over that interval since the induction equation is not forced.
In the (approximate) inertial range, both spectra have approximately powerlaw scaling, with  
     $ E_b(k) $ close to  $ k^{-5/3} $
   and $E_u(k) $ significantly shallower.
   As is typically seen in MHD simulations with no mean field, the Alfv\'en ratio, 
    $ E_u(k) / E_b(k) $, 
   is less than unity in the inertial range, i.e., magnetic energy predominates at these scales.
%
%
At high $k$ there is steep and roughly coincident decrease of both spectra. This is a consequence of two factors. 
First, the employment of hyperviscosity makes the dissipation range more concentrated in the small scales, leading to the sharp fall-off.
Second, unit Prandtl number ensures that the dissipation wavenumber-band is the same for the kinetic and magnetic spectra.

To calculate an effective Reynolds number for a hyperdissipative system we follow the approach described in \cite{BuzzicottiEA18-xfer}. 
There, the standard (Laplacian dissipation) integral-scale Reynolds number 
  $ Re = {U L_u} / {\nu} \propto (L_u /\eta_1)^{4/3}$
 \citep[e.g.,][]{BatchelorTHT,Pope}
is replaced with one based on the ratio between the integral scale $L_u$ and the effective dissipation range scale $I_d$. Specifically, we employ
\begin{equation}\label{eq:michele_reynolds}
   Re = C \left( \dfrac{L_u}{I_d} \right)^{4/3} ,
\end{equation}
where $I_d = \pi/\text{argmax}\left( k^2 {E_u}(k) \right)$ 
is the scale where the dissipation spectrum $k^2 E_u(k) $ 
has a maximum.
Here, $C$ is a fit parameter that has to be estimated by comparing eq.~\eqref{eq:michele_reynolds} with the common definition of the Reynolds number in a standard-viscosity run. 
Making use of this procedure we obtain
    $C=40$ and $Re=9931$ 
for run A1 (table~\ref{tab:datasets}).
Generalization of the Reynolds number for systems with hyper-dissipation has been discussed in \cite{SpyksmaEA12}.

\begin{figure}
	\begin{center}
         \includegraphics[width=1\columnwidth]{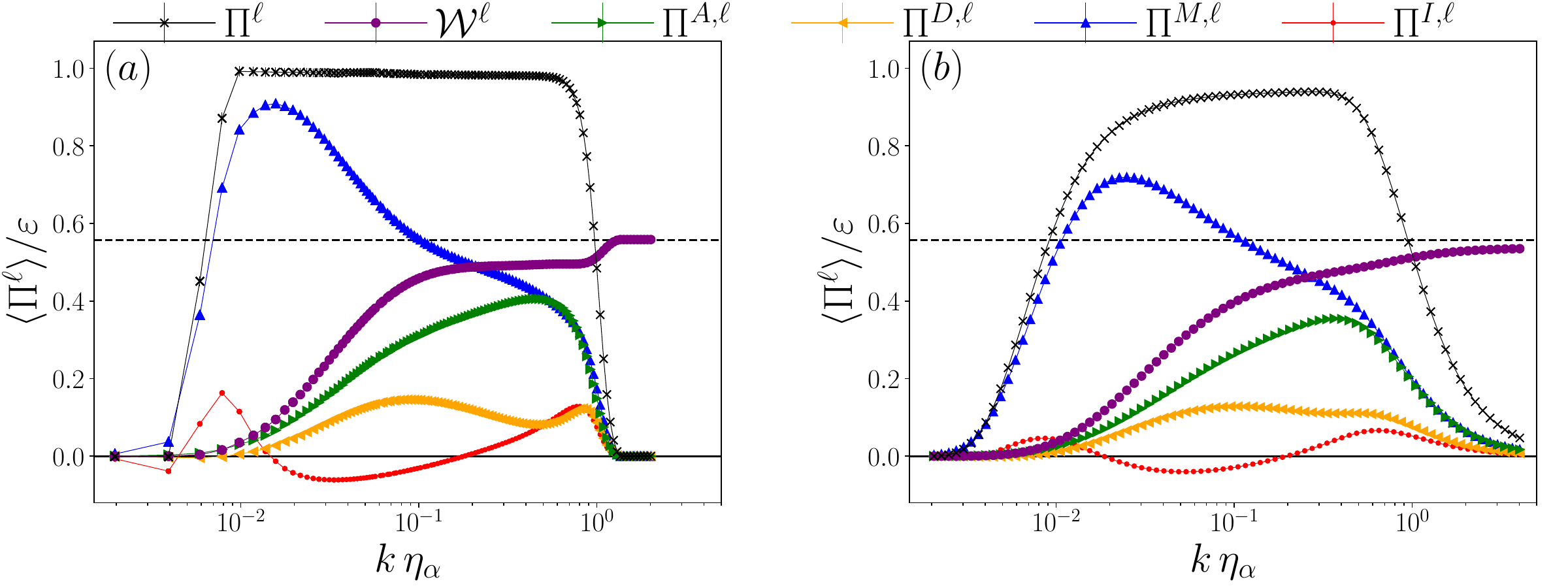} 
    \end{center}
    \caption{Terms contributing to the MHD filtered energy flux across scale $\ell$, along with the resolved-scale-conversion (kinetic to magnetic) term, as function of the adimensional parameter $k \eta_\alpha = \pi \eta_\alpha / \ell $:
    (a) Fourier filter, (b) Gaussian filter. 
    All terms are normalized by the mean total energy dissipation rate $\varepsilon = \varepsilon_u + \varepsilon_b$. 
     The dashed horizontal line indicates the normalized magnetic dissipation rate $\eps_b/\varepsilon$. 
     The errorbars, although not fully visible, indicate one standard error. 
     See eqns~\eqref{eq:Eu-ls}--\eqref{eq:Eb-ls}.
	}
  \label{fig:sharp}
\end{figure}
Figure~\ref{fig:sharp} displays the four MHD energy subfluxes, introduced in 
    eqs.~\eqref{eq:Eu-ls}--\eqref{eq:Eb-ls},
and their sum.  
Also shown is 
  $ {\cal W}^\ell $,
the resolved-scale-conversion of kinetic energy to magnetic energy 
 (recall this does not represent energy transfer across $\ell$).
The two panels present fluxes obtained through different filters, results shown 
in fig.~\ref{fig:sharp}(a) correspond to Galerkin truncation, and
those shown in fig.~\ref{fig:sharp}(a) 
to the Gaussian filter of eq.~\eqref{eq:Gell}.
The data shown in fig.~\ref{fig:sharp}(a) and (b) are qualitatively similar but with quantitative differences.
 Focusing on the similarities, 
 we see that the Inertial term $\lan \Pi^{I,\ell} \ran$ is relatively weak and is the only one to exhibit inverse transfer regions, in the intervals 
 $1.5 \times 10^{-2} \lesssim k \eta_\alpha \lesssim 2 \times 10^{-1}$ and 
 $4 \times 10^{-3} \lesssim k \eta_\alpha \lesssim 6 \times 10^{-3} $. 
 All the other subfluxes are associated with a direct cascade from the large to the small scales. The energy transfer from the momentum equation of eq.~\eqref{eq:Eu-ls} is almost entirely dominated by the Maxwell subflux 
    ($ - \Pi^{M,\ell}$)
 whose peak occurs in proximity to the forcing region. 
 In contrast, the advection term from the induction equation
    ($ \Pi^{A,\ell} $)
 is peaked at the small scales, close to the dissipative range. 
 The conversion term, ${\cal W}^\ell $, 
 is positive and
 increases monotonically with $k \eta_\alpha$.
 Particularly for the Fourier filter, it is roughly constant (i.e., scale independent) in the region 
    $ 0.2 \lesssim k \eta_\alpha \lesssim 0.9 $
 where kinetic and magnetic subfluxes are in equipartition \citep{BianAluie19}; 
 see fig.~\ref{fig:sharp_filter} in Appendix~\ref{app:subrange-pdf}. 
 Also evident is a sudden increase of 
   ${\cal W}^\ell $ 
 in the dissipative range ($k \eta_\alpha \gtrsim 1$), due to hyperdiffusion, where the conversion rate saturates to $\varepsilon_b $ as already pointed by \cite{BianAluie19}.

Turning to the differences between the two kinds of filtering, we observe that 
the bandwidth of the inertial range plateau is narrower for the Gaussian filter case, roughly
 $ k \eta_\alpha \in [0.04, 0.4] $
versus
 $ k \eta_\alpha \in [0.013, 0.7] $.
In general for the Gaussian filtering peaks are of lower amplitude and a little less localised. 
Linked to this is a more gradual roll-off of the fluxes at high $ k \eta_\alpha $
and a slower convergence of 
   $ {\cal W}^\ell $ 
to $ \varepsilon_b$ 
with increasing $ k$. 
These effects arise because Gaussian filtering at scale $\ell$ retains some effects from scales 
 $ \leq \ell $, 
unlike the situation for the sharp Galerkin truncation of the Fourier filter.
For
  $ \Pi^{D,\ell} $
there is also a qualitative difference, with the high-$k$ local minimum and maximum seen with the Fourier filter essentially absent when the Gaussian filter is used.

    \section{Analysis and Discussion}
    \label{sec:analysis}

   \subsection{Inertial flux and comparison with hydrodynamics}
     \label{sec:inertial}

The exact decomposition of the Inertial term 
    $\Pi^{I,\ell} $, eq.~\eqref{eq:Pi-I},
is
\begin{equation}
 \label{eq:inertial_dec}
    \Pi^{I,\ell} =  \Pi^{I,\ell}_{s, SSS}  +  \Pi^{I,\ell}_{m,SSS}  +   \Pi^{I,\ell}_{s, S \Omega \Omega}  +  \Pi^{I,\ell}_{m,S \Omega \Omega}  +  \Pi^{I,\ell}_{m, S \Omega S} 
 ,
\end{equation}
where $\Pi^{I,\ell}_{s, S \Omega S} $ is not included as it is identically zero.
This is the special case of eq.~\eqref{eq:Pi-xyz-exact} where all the fields are the velocity and naturally it coincides  with the original decomposition provided by \cite{Johnson20} for Navier--Stokes turbulence. It is thus of interest to investigate
whether there are differences between the HD and MHD instances of eq.~\eqref{eq:inertial_dec}, and how these might arise.


\begin{figure}
	\begin{center}
    \includegraphics[width=1.0\columnwidth]{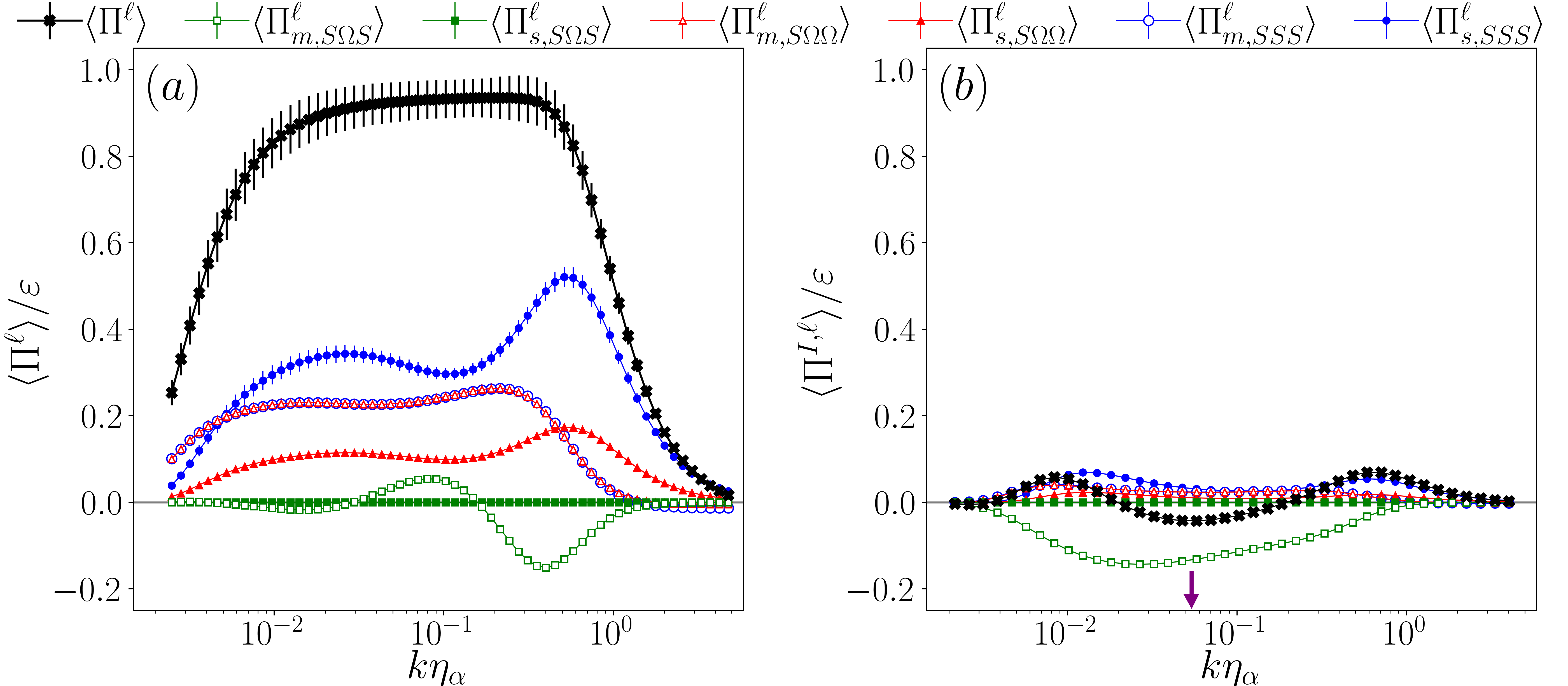} 
    \end{center}
	 \caption{Contributions to the Inertial energy flux $\langle \Pi^{I,\ell} \rangle$ for
    (a) the HD dataset H2 and (b) the MHD dataset A1,
     as a function of the (nondimensionalized) reciprocal scale $\ell$, i.e., 
      $ k \eta_\alpha = \pi \eta_\alpha / \ell $.    
    All fluxes are normalised by the mean total energy dissipation rate $\eps$ for the dataset. Filled symbols correspond to single-scale contributions while hollow markers indicate multi-scale contributions. Errorbars are for one standard error. 
    Both panels indicate that $ \langle\Pi^{I,\ell}_{m,SSS}\rangle \approx  \langle\Pi^{I,\ell}_{m,S\Omega\Omega}\rangle$.
    In (b) the purple arrow locates $k \eta_\alpha = 5.4 \times 10^{-2}$ 
    (equivalent to $k L_u \approx 14$),
    the value used for the
    p.d.f.s shown in figures~\ref{fig:pdf_inertial}, \ref{fig:pdf_Lorentz}, and~\ref{fig:pdf_advection}. 
    }
\label{fig:fluxes_hydro_vs_mhd}
\end{figure}
Figure~\ref{fig:fluxes_hydro_vs_mhd} compares these two cases.  
For the hydrodynamic case, panel (a), there is a relatively clear and extended plateau for the total flux (and some of the subfluxes) corresponding to an inertial range.
In contrast, the MHD case shown in panel (b) lacks such a plateau and the individual subfluxes are mostly much smaller than their hydrodynamic counterparts.
Indeed, only  
  $ \langle \Pi^{I,\ell}_{m,SS\Omega} \rangle$ 
has values comparable to its hydrodynamic counterpart, albeit with a different functional form, being negative for almost all $k$.
Intriguingly, this term is the only one that does not vanish point-wise in 2D turbulence, as \cite{Johnson21} has discussed.\footnote{The energy flux 
    decomposition for 2D MHD turbulence is considered in Appendix~\ref{app:mhd2d}.}
A depletion of the inertial flux in MHD turbulence has been observed by \cite{Alexakis13} and \cite{YangEA21-forcing} for configurations with large-scale electromagnetic forcing  
and by \cite{OffermansEA18} for a saturated dynamo at lower Reynolds number. 
We have verified that the \cite{Betchov56} relation,
  $  \langle{ \Pi^{I,\ell}_{s,SSS} } \rangle 
  = 3 \langle{ \Pi^{I,\ell}_{s,S\Omega \Omega} } \rangle$,
holds for both datasets, as it must.

Also of interest is that there appears to be an approximate \emph{multi}-scale analog of the Betchov relation, with
 $\langle \Pi^{I,\ell}_{m,SSS} \rangle \approx 
  \langle \Pi^{I,\ell}_{m,S\Omega \Omega } \rangle $.
For hydrodynamics, this was already noted in \cite{Johnson21} and justified by \cite{YangEA23-strain}.  
Evidently this approximate degeneracy also holds in this MHD situation, although the smallness of the terms makes this difficult to appreciate from Figure~\ref{fig:fluxes_hydro_vs_mhd}(b).

\begin{figure}
\begin{center}
         \includegraphics[width=.48\columnwidth]{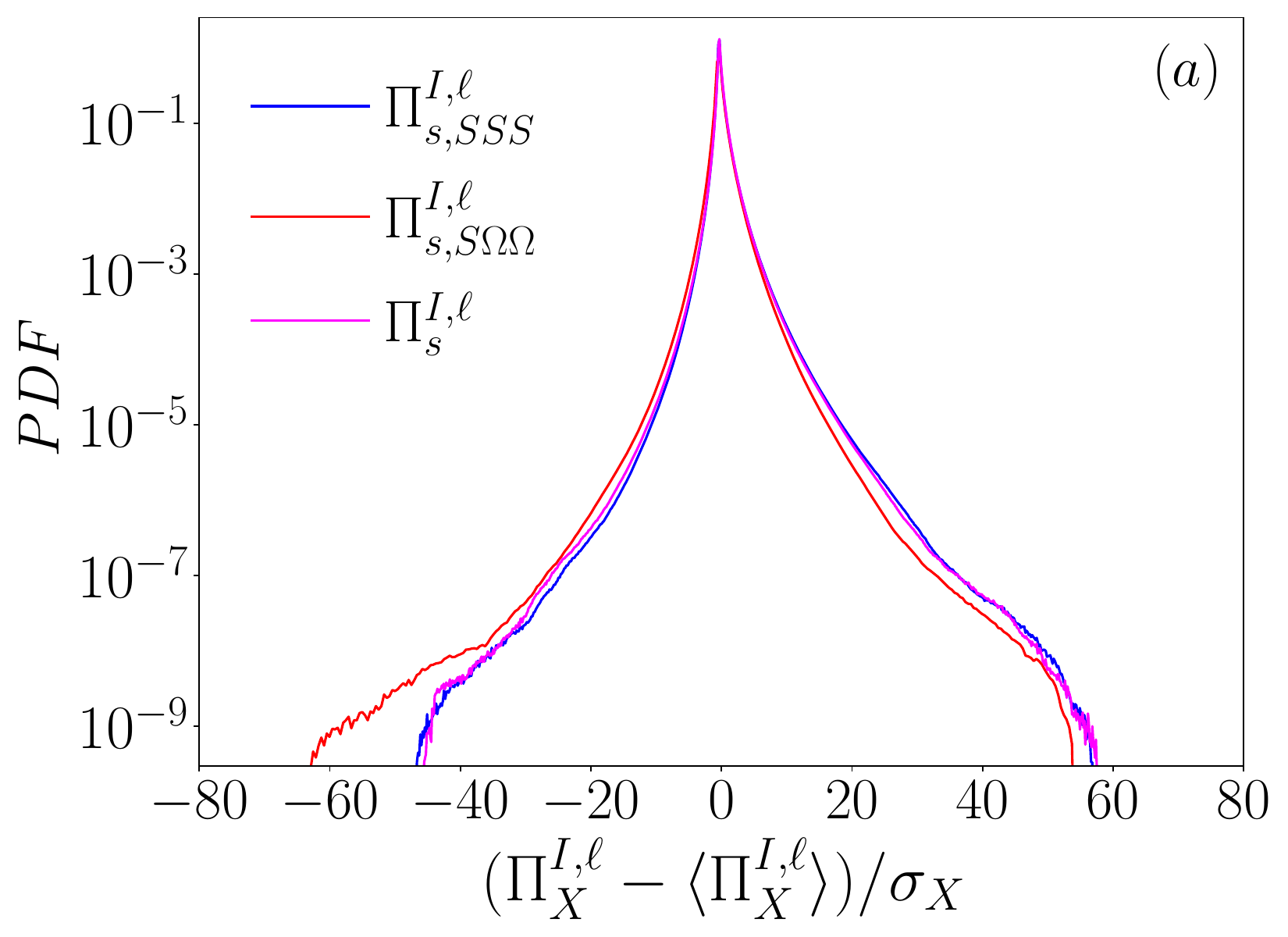} 
         \includegraphics[width=.48\columnwidth]{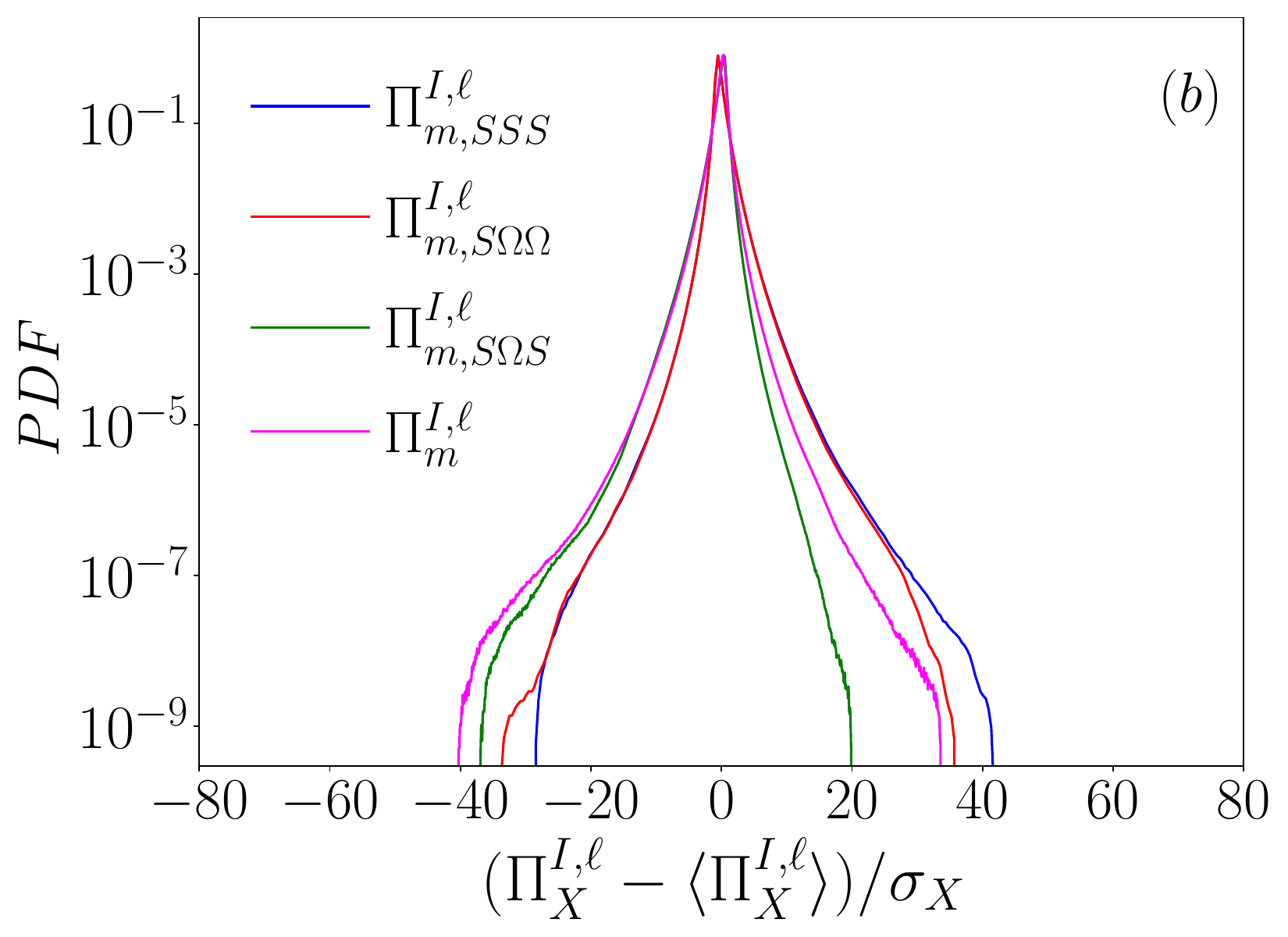} 
    \end{center}
	 \caption{Standardised p.d.f.s of the MHD Inertial subfluxes $\Pi^{I,\ell}_{X}$ at 
   $ k \eta_\alpha = 5.4 \times 10^{-2}$ 
   (equivalent to $ k L_u = 14$) 
   from dataset A1, where $X$ identifies the specific subflux. (a) single-scale fluxes; (b) multi-scale fluxes. 
	}
\label{fig:pdf_inertial}
\end{figure}

Having discussed mean fluxes, we now examine some statistical properties of their pointwise contributions.
Figure~\ref{fig:pdf_inertial} presents standardised probability density functions (p.d.f.s) of the MHD Inertial subfluxes at 
 $ k \eta_\alpha = 5.4 \times 10^{-2} $.
It is evident that the distributions are strongly non-Gaussian and exhibit very wide tails, with fluctuations at tens of standard deviations.  
(As we shall see, this is a common characteristic for all the MHD energy fluxes and subfluxes.)
The  variance, skewness, and kurtosis 
for each p.d.f.\
are reported in
    table~\ref{tab:table_moments}. 
Although the p.d.f.\ of 
 $  \Pi_{s,S \Omega \Omega}^{I,\ell} $, 
that corresponds to the scale-local vortex stretching, has more asymmetrical tails than those of $\Pi_{s,SSS}^{I,\ell}$, 
the skewness for the latter is nonetheless larger. 
The term  $\Pi_{s,S\Omega S}^{I,\ell}$ does not show up in fig.~\ref{fig:pdf_inertial} because it is identically zero due to the symmetries of the tensors involved in the corresponding trace of eq.~\eqref{eq:Pi-xyz-exact}. On the contrary the right panel shows that $\Pi_{m,S \Omega S}^{I,\ell}$ is the most negatively skewed p.d.f.\ among the Inertial ones.

In connection with
the approximate degeneracy between 
 $ \langle \Pi^{I,\ell}_{m,SSS} \rangle $
and $ \langle \Pi^{I,\ell}_{m,S \Omega \Omega} \rangle $ discussed above,  
  fig.~\ref{fig:pdf_inertial}(b)
reveals that their p.d.f.s coincide up to events with a standardised
probability density of $ \sim 10^{-7}$, potentially indicating that the
approximate identity is true not just on average but also for higher-order moments,
although further analysis is required.

   \subsection{Maxwell flux}
     \label{sec:Pi-M}

\begin{figure}
	\begin{center}
         \includegraphics[width=\columnwidth]{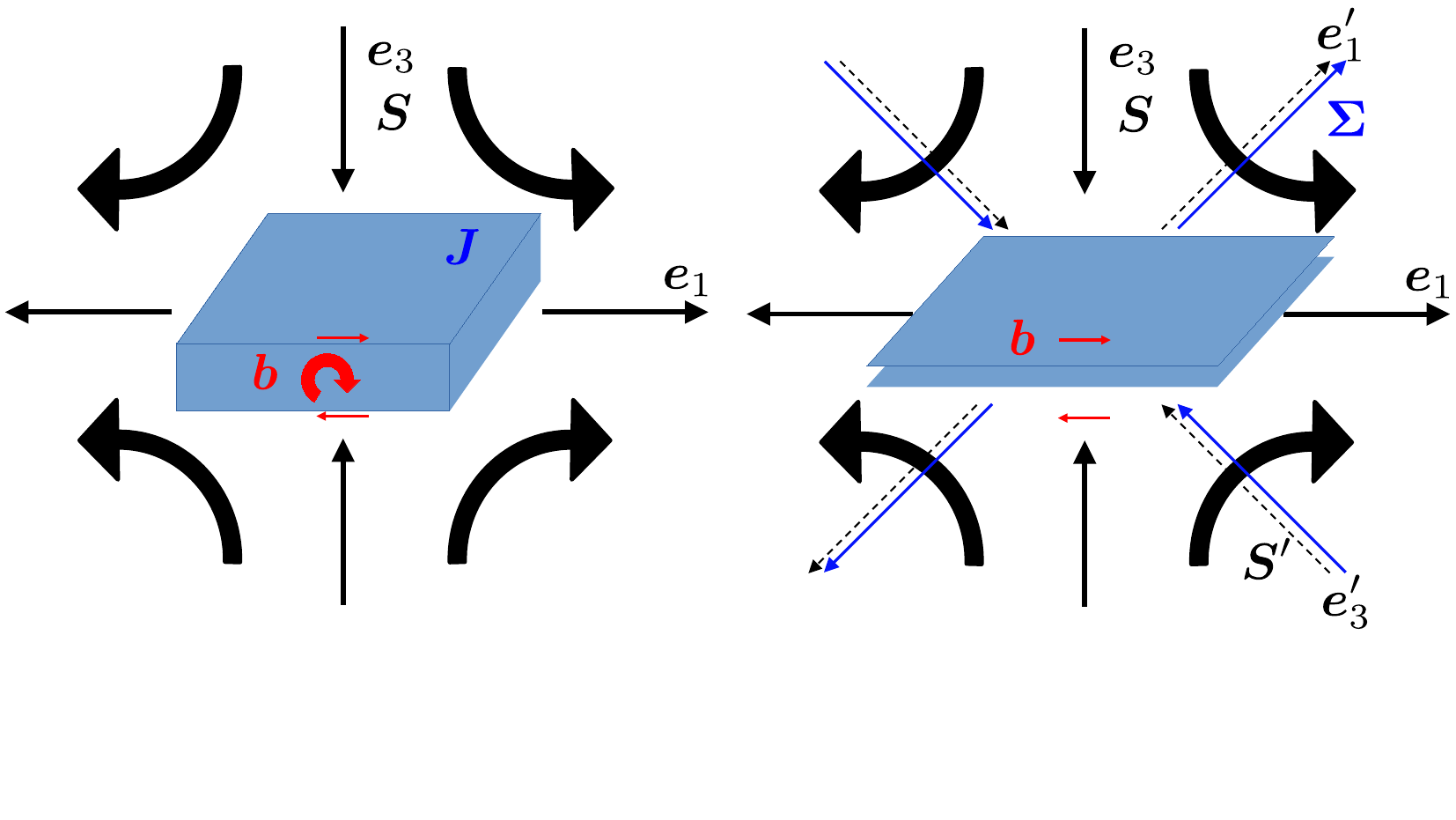} 
    \end{center}
	\vspace{-5em}
	 \caption{
		 Two-dimensional sketch of current-sheet thinning and
		 strain rate amplification by the latter. A
		 current sheet, $\bm{J}$, is stretched by large-scale strain,
		 $\bm{S}$ (left) into a magnetic shear layer (right, red
		 arrows).  This induces a stretching of the magnetic flux
		 tubes in the sheet. By conservation of magnetic flux, the
		 magnetic field strength at the thereby generated smaller
		 scales increases.  That is, magnetic energy is transferred
		 from large to small scales.  
   The resulting magnetic shear layer has an associated  magnetic strain rate field,
		 $\bm{\Sigma}$, whose principal axes (solid blue arrows) are at $45^\circ$ to those of the 
	large-scale strain rate tensor (straight black arrows). 
   As the magnetic shear will align with the extensional direction of 
		 the (velocity) strain rate tensor, this causes the fluid to be accelerated along these extensional
		 directions and slowed down in the compressional directions,
		 thereby generating a stronger rate-of-strain field across smaller
		 scales, $\bm{S}^\prime$, indicated by the dashed arrows. 
   The principal axes of the large-scale strain rate tensor are denoted by $\bm{e}_1$ in one extensional direction and $\bm{e}_3$ in one compressional direction and analogously for the small-scale strain rate.
	}
\label{fig:sketch-current-thinning}
\end{figure}

The decomposition of the energy flux associated with the Lorentz force, which we refer to as the Maxwell term, 
 $ \Pi^{M,\ell} $ of  eq.~\eqref{eq:Pi-M}, 
contains an extra (single-scale) term with respect to the Inertial flux.  Specifically, from eq.~\eqref{eq:Pi-xyz-exact} we obtain
\begin{equation}
\label{eq:dec_maxwell}
    \Pi^{M,\ell} =  \Pi^{M,\ell}_{s,S  \Sigma  \Sigma}  +  \Pi^{M,\ell}_{m,S  \Sigma  \Sigma}  +   \Pi^{M,\ell}_{s, S J J}  +  \Pi^{M,\ell}_{m, S J J}  +  \Pi^{M,\ell}_{s, S J  \Sigma}  + \Pi^{M,\ell}_{m, S J  \Sigma}
 .
\end{equation}
Terms of type $S  \Sigma  \Sigma$ can be associated with strain rate amplification by magnetic shear, while
terms of type $S J J$ correspond to current-filament stretching that is analogous to vortex stretching in HD. 
The last two terms are of type $S J  \Sigma$ and describe the back-reaction of the magnetic field on the flow, 
or more specifically how the velocity strain rate is modified (typically amplified) in connection with 
a current-sheet thinning process. 
As we shall see, this is by far the dominant process. 
It proceeds as follows (fig.~\ref{fig:sketch-current-thinning}). 
First, a current sheet is stretched by large-scale straining motions into a magnetic shear layer, in a process similar to vortex thinning in HD \citep{Kraichnan76,Chen06inverse,Johnson21}. 
This results in a stretching of the magnetic flux tubes in the sheet. By conservation of magnetic flux, the magnetic field strength at the thereby generated smaller scales must \emph{increase}. 
That is, magnetic energy is transferred from large to small scales. 
 (We will revisit this process in section \ref{sec:Pi-A} in the context of the inter-scale transfer of magnetic energy.) 
The magnetic rate-of-strain field associated with the resulting magnetic shear layer now accelerates fluid along its extensional directions and slows it down in the compressional directions, thereby 
generating a stronger rate-of-strain field across smaller scales. 

It is instructive to consider the process in two dimensions, in analogy to the vortex thinning of 2D HD \citep{Johnson21}. In the reference frame of the rate-of-strain tensor at scale $\ell$, the associated terms are
\begin{align}
\label{eq:2d-current-sheet-thinning-s}
\Pi^{M,\ell}_{s,SJ\Sigma}(\bm{x}) 
& = 2  
\lambda_S^\ell(\bm{x})\lambda^{\ell}_\Sigma (\bm{x}) \overline{j}^\ell (\bm{x}) \sin{2\psi(\bm{x})} 
\ , \\
\label{eq:2d-current-sheet-thinning-m}
\Pi^{M,\ell}_{m,SJ\Sigma}(\bm{x}) 
& = 
2 \int_0^{\ell^2} \d\theta \int_{\mathbb{R}^3} G^\phi(\bm{r}) \lambda_S^\ell(\bm{x})\lambda^{\sqrt{\theta}}_\Sigma (\bm{x} + \bm{r}) \overline{j}^{\sqrt{\theta}} (\bm{x} + \bm{r}) \sin{2\psi(\bm{x} + \bm{r})} 
\ , 
\end{align}
where $\pm\lambda_S$ and $\pm\lambda_\Sigma$ are the eigenvalues of the velocity and magnetic rate-of-strain tensors, respectively, $\psi$ the angle between the respective eigenvectors, and $j$ the out-of-plane component of the current density. As can be seen from these formulae, a maximum energy transfer occurs when the principal axes of the magnetic rate-of strain tensor have a $\pm45^\circ$ angle to those of the velocity rate-of-strain tensor.   
Depending on the sign of the out-of-plane current density and $\sin(2\psi)$, eqs.~\eqref{eq:2d-current-sheet-thinning-s} and \eqref{eq:2d-current-sheet-thinning-m} result in a direct or an inverse cascade. 
Figure~\ref{fig:sketch-current-thinning} presents a schematic depiction of the process. 
A direct cascade occurs if the angle between the principal axes of velocity and magnetic strain-rate tensors is in the same rotational direction as the out-of-plane current. A similar, albeit less straightforward,  assessment is possible in three dimensions, where
\beq
\Pi^{M,\ell}_{s,SJ\Sigma}(\bm{x}) = 2 \ell^2 \overline{S}_{ij}^\ell \overline{J}^\ell_{ik}\overline{\Sigma}^\ell_{kj} 
= 2 \ell^2 \sum_{i=1}^3  \sum_{j=1}^3 \lambda^\ell_i \mu_j^\ell\cos^2{\psi_{ij}} ,     
\eeq
and similarly for the multi-scale term. Here, $\mu^\ell_j$ is the $j$-th eigenvalue of the
symmetric part of the product matrix 
$\overline{J}^\ell_{ik} \overline{\Sigma}^\ell_{kj} $
 (a contribution to the Maxwell SGS stress tensor)
and $\psi_{ij}$ the angle between the $i$-th eigenvector of $\ovS^\ell$ and the $j$-th eigenvector of the symmetric part of 
$\overline{J}^\ell_{ik} \overline{\Sigma}^\ell_{kj} $. 
For a forward cascade of kinetic energy,  $\langle \Pi^{M,\ell} \rangle >0$, which implies that $\overline{S}^\ell_{ij} \overline{J}^\ell_{ik}\overline{\Sigma}^\ell_{kj}$
should be preferentially positive. Due to the presence of the cosine squared factor, this implies that the principal axes of the rate-of-strain tensor and those of the subscale stress must preferentially align, resulting in a stretching of the magnetic flux tubes along the extensional directions of the strain-rate tensor, as discussed above.

\begin{figure}
	\begin{center}
         \includegraphics[width=.65\columnwidth]{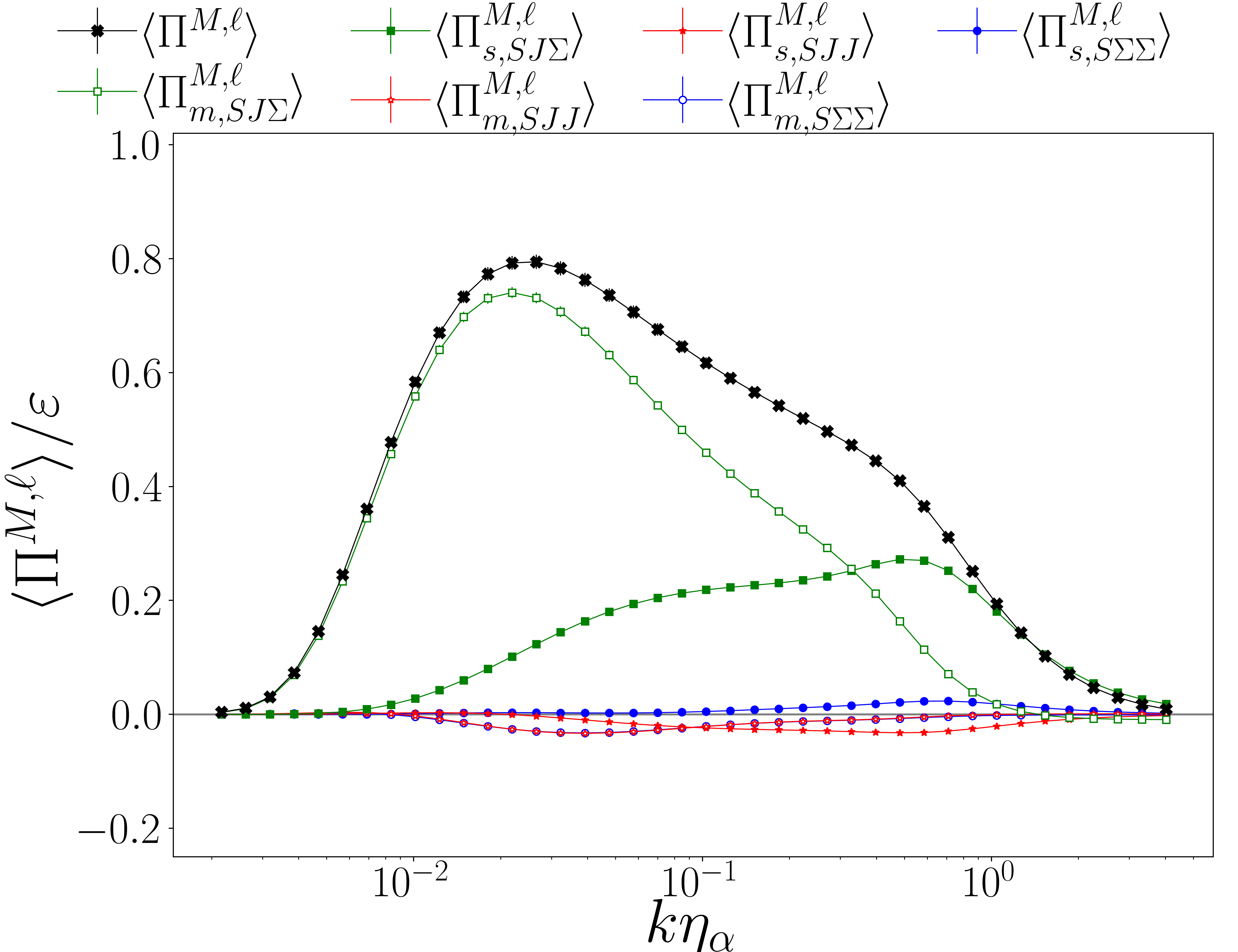} 
    \end{center}
	 \caption{Contributions to the 
  Maxwell energy flux $  \langle \Pi^{M,\ell} \rangle$.  Data is from dataset A1 and normalised by the mean total energy dissipation rate $\eps$. 
       Errorbars indicate one standard error. 
    }
\label{fig:fluxes_lorentz}
\end{figure}

The subfluxes on the RHS of eq.~\eqref{eq:dec_maxwell} and the total Maxwell flux are shown in fig.~\ref{fig:fluxes_lorentz},
 as a function of (nondimensionalised) reciprocal $\ell$.
We see immediately that 
the net energy transfer proceeds from large scales to small scales with the total Maxwell flux 
   $ \langle \Pi^{M,\ell} \rangle $ 
being the dominant energy subflux for MHD, carrying approximately $ 80\% $ of the total energy dissipation rate at its peak.
At large scales, the major contribution is from the multi-scale term
 $ \langle \Pi^{M,\ell}_{m, S J \Sigma} \rangle $,
 switching to its single-scale partner,
$ \langle \Pi^{M,\ell}_{s, S J \Sigma} \rangle $,
as the dissipation scale is approached. 
All remaining terms in eq.~\eqref{eq:dec_maxwell}, are negligible.
Summarising the mean Maxwell flux behaviour, we may say that the net kinetic energy transfer in MHD proceeds by the back-reaction of the magnetic field on the flow
during the aforementioned current-sheet thinning process, while the contribution from current filament stretching and strain-amplification by magnetic shear 
are negligible.

\begin{figure}
	\begin{center}
         \includegraphics[width=.48\columnwidth]{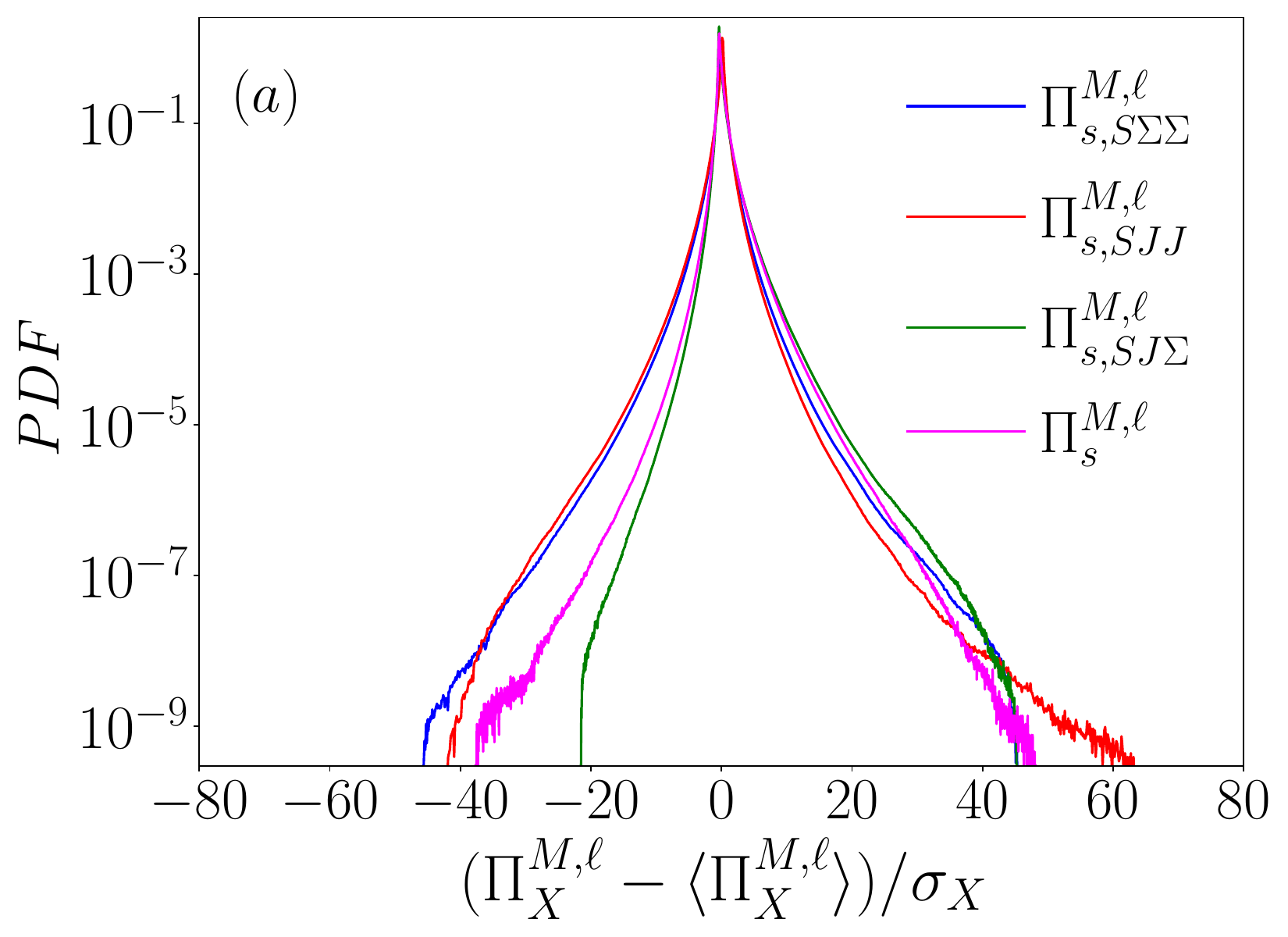} 
         \includegraphics[width=.48\columnwidth]{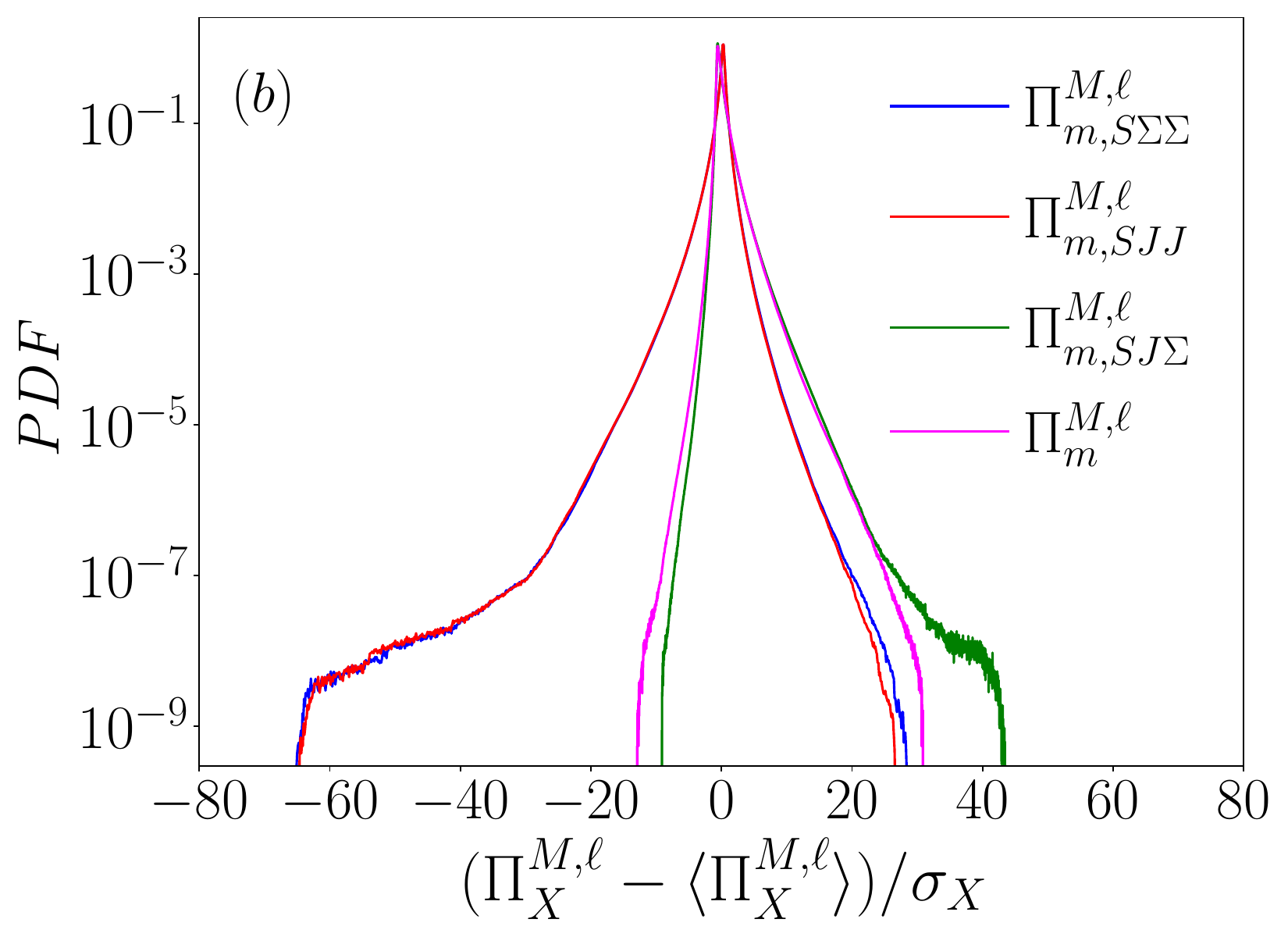} 
    \end{center}
	 \caption{ Standardised p.d.f.s for Maxwell subfluxes $\Pi^{M,\ell}_{X}$, at $k \eta_\alpha = 5.4 \times 10^{-2}$, from dataset A1, where $X$ represents the subflux identifier.
  (a) single-scale fluxes;
  (b) multi-scale fluxes.
  Note that the p.d.f.s for $\Pi^{M,\ell}_{m,S \Sigma \Sigma }$ and $\Pi^{M,\ell}_{m,S J J}$  are approximately coincident. 
  }
\label{fig:pdf_Lorentz}
\end{figure}

The p.d.f.s for the Maxwell energy fluxes are shown in fig.~\ref{fig:pdf_Lorentz}.
The predominance of
$ \langle \Pi^{M,\ell}_{m,S J \Sigma} \rangle $ 
in the direct cascade can be also appreciated by examining its p.d.f.,
which is the most positively-skewed among the multi-scale terms of
fig.~\ref{fig:pdf_Lorentz}; 
see also table~\ref{tab:table_moments} for p.d.f.\ moments. 
As can be seen from the data shown in fig.~\ref{fig:pdf_Lorentz}, while all terms 
of type $S  \Sigma  \Sigma$ and $S J J$ nearly vanish on average, fluctuations around 
60 standard deviations are not uncommon, and their multi-scale contributions
show considerable backscatter.

In Appendix~\ref{app:extended-betchov} we show that the averages of the subfluxes $\Pi^{M,\ell}_{m,S\Sigma \Sigma} $ and $ \Pi^{M,\ell}_{m,S J J} $ can be connected 
via an exact Betchov-like relation that holds for all homogeneous flows
\begin{equation}
  \langle \Pi^{M,\ell}_{m,S\Sigma \Sigma} \rangle 
  = 
  \langle \Pi^{M,\ell}_{m,S J J }\rangle 
  + 
  2\langle\Pi_{m,\Omega \Sigma J}^{M,\ell} \rangle   
  .
  \label{eq:extended_betchov}
\end{equation}
Note that the final term,
 $ \langle\Pi_{m,\Omega \Sigma J}^{M,\ell} \rangle $,
does not appear in the Maxwell flux, 
  eq.~\eqref{eq:dec_maxwell}, 
and our numerical results indicate 
 $ \langle \Pi^{M,\ell}_{m,S\Sigma \Sigma} \rangle 
  \approx 
  \langle \Pi^{M,\ell}_{m,S J J } \rangle $, see fig.~\ref{fig:fluxes_lorentz}. 
Physically, we may
interpret this approximate identity 
as indicating that the net ``strain-production'' by magnetic shear is almost equal to the net strain production by 
current-filament stretching. However we stress 
again that these contributions to the interscale kinetic energy transfer are negligible.  
Further discussion on terms associated with strain production and current-filament stretching is provided in Appendix ~\ref{app:extended-betchov}.

   \subsection{Advection and Dynamo fluxes}
     \label{sec:Pi-A} 

In this section we focus on the decomposition of both the Advection term, $\Pi^{A,\ell}$, 
and the Dynamo term, $\Pi^{D,\ell}$, 
as defined in
  eqs.~\eqref{eq:Pi-A}--\eqref{eq:Pi-D}. 
As is well known 
these two SGS fluxes share the same physical origin, 
namely the induced (fluctuation) electric field,
and this is associated with certain symmetries and equivalences between the Advection and Dynamo subfluxes.

From the application of eq.~\eqref{eq:Pi-xyz-exact}, we find:
\begin{align}
   \Pi^{A,\ell} =\ &  
        \Pi^{A,\ell}_{s, \Sigma \Sigma S} 
   +    \Pi^{A,\ell}_{m, \Sigma \Sigma S}
   +    \Pi^{A,\ell}_{m, \Sigma \Sigma \Omega}
   +    \Pi^{A,\ell}_{s, \Sigma J S}
   +    \Pi^{A,\ell}_{m, \Sigma J S}
   +    \Pi^{A,\ell}_{s, \Sigma J \Omega}
  \\
   +&    \Pi^{A,\ell}_{m, \Sigma J \Omega}
   +   \Pi^{A,\ell}_{s, J \Sigma S}
   +    \Pi^{A,\ell}_{m, J \Sigma S}
   +    \Pi^{A,\ell}_{s, J \Sigma \Omega}
   +    \Pi^{A,\ell}_{m, J \Sigma \Omega}
   +    \Pi^{A,\ell}_{s, J J S}
   +    \Pi^{A,\ell}_{m, J J S}
   +    \Pi^{A,\ell}_{m, J J\Omega} 
 ,
  \label{eq:adv_total} 
  \notag \\
 & \notag \\
    \Pi^{D,\ell} =\ &  
    \Pi^{D,\ell}_{s,\Sigma S \Sigma} 
+   \Pi^{D,\ell}_{m,\Sigma S \Sigma}   
+   \Pi^{D,\ell}_{s,\Sigma S J} 
+   \Pi^{D,\ell}_{m,\Sigma S J} 
+   \Pi^{D,\ell}_{m,\Sigma \Omega \Sigma}
+   \Pi^{D,\ell}_{s,\Sigma \Omega J}         
+   \Pi^{D,\ell}_{m,\Sigma \Omega J}\\
+&   \Pi^{D,\ell}_{s,J S \Sigma}
+    \Pi^{D,\ell}_{m,J S \Sigma}
+  \Pi^{D,\ell}_{s, J S J} 
+   \Pi^{D,\ell}_{m, J S J}
+   \Pi^{D,\ell}_{s, J \Omega \Sigma}        
+   \Pi^{D,\ell}_{m, J \Omega \Sigma}
+   \Pi^{D,\ell}_{m,J \Omega J} 
\notag
,
 \label{eq:dyn_total}
\end{align}
where we do not list terms that vanish identically, see Appendices \ref{app:general-case} and \ref{app:defintions}.

\begin{figure}
  \begin{center}
    \includegraphics[width=.98\columnwidth]{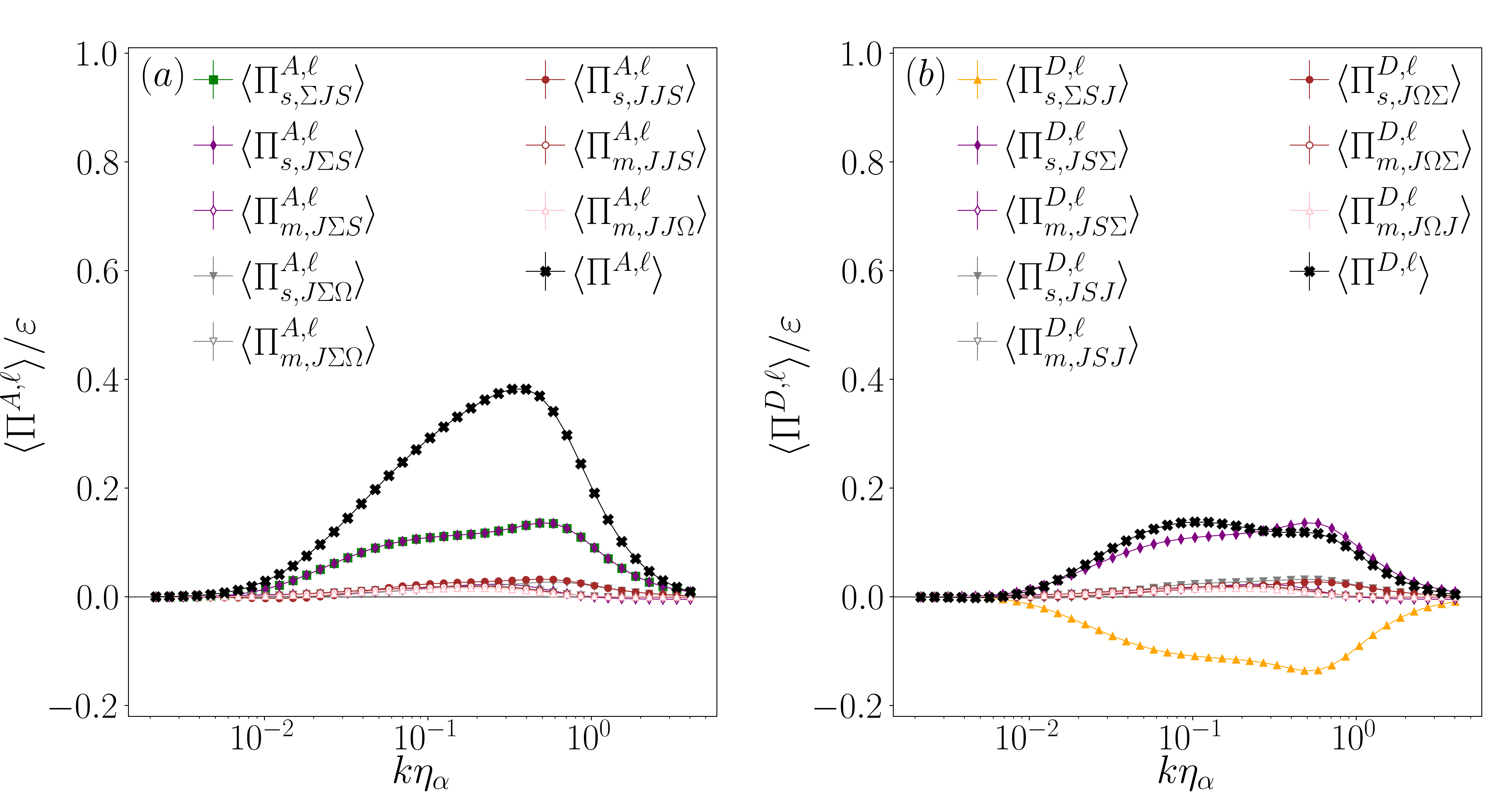}
  \end{center}
	 \caption{Decomposed fluxes for the 
  (a) Advection term $\langle \Pi^{A,\ell} \rangle$
  and (b) Dynamo term $  \langle \Pi^{D,\ell} \rangle$. 
  Fluxes are normalised by the mean total energy dissipation rate $\eps$ for dataset A1. In panel (a) $ \langle \Pi^{A,\ell}_{s,\Sigma J S} \rangle $ and $\langle  \Pi^{A,\ell}_{s,J \Sigma S} \rangle $ perfectly coincide. The errorbars indicate one standard error. 
  }
\label{fig:fluxes_advection_dynamo}
\end{figure}

Figure~\ref{fig:fluxes_advection_dynamo}
displays most of the Advection and Dynamo (sub)fluxes in separate panels.
For clarity, only the subfluxes relevant to the discussion below and to the net energy flux are shown.
The cyclic property of the trace can be used to show that some terms vanish identically and that each Advection subflux (both single-scale and multi-scale)
is equal to (plus or minus) a partner Dynamo subflux; see the subfluxes expressions in Appendix~\ref{app:defintions}. 
For this reason in
fig.~\ref{fig:fluxes_advection_dynamo} 
the following subflux pairs are not displayed since they are opposite in sign, 
$\langle \Pi^{A,\ell}_{s,\Sigma \Sigma S} \rangle$ 
and $\langle \Pi^{D,\ell}_{s,\Sigma S \Sigma} \rangle$ as well as 
$\langle \Pi^{A,\ell}_{s, \Sigma J \Omega} \rangle$ 
and $\langle \Pi^{D,\ell}_{s, \Sigma \Omega J} \rangle$
together with their multi-scale counterparts.
Hence, these cancel pairwise and make no contribution to the net magnetic energy flux
 $ \Pi^{A, \ell} + \Pi^{D,\ell} $; 
    see Appendix~\ref{app:defintions}. 
In contrast, 
$\langle \Pi^{A,\ell}_{s,J J S} \rangle$ 
and $\langle \Pi^{D,\ell}_{s,J S J} \rangle,$ 
and the related multi-scale terms, 
are equal 
and thus do contribute to the net flux.
%
Note, too, that the two Betchov relations
eqs.~\eqref{eq:multisc_betchov_local}--\eqref{eq:second_betchov}
can provide another source of symmetry, or approximate symmetry.


The physical interpretation of the respective terms is very similar to
what has been discussed for the Maxwell flux, except that now the effect of the flow on the magnetic field must be considered.
Recall from sec~\ref{sec:Pi-M}
that the Maxwell flux terms of type 
  $ S J \Sigma $ 
correspond to the stretching and, by incompressibility, thinning of current sheets into magnetic shear layers.
And that the back-reaction on the flow induced by this process is responsible for the bulk of the kinetic energy transfer to smaller
scales.
As we shall see, and as expected from the
discussion of current-sheet thinning in section~\ref{sec:Pi-M}, this
process also transfers most magnetic energy from large to small scales.  
However, in contrast to the Maxwell flux situation
(dominated by a multiscale term),
here it is two of the 
Advection
single-scale terms,
   $ \Sigma J S $ and
   $ J \Sigma S $,
that carry most of the magnetic energy flux.

 
Focusing  on the Advection term, from 
  fig.~\ref{fig:fluxes_advection_dynamo}(a) 
we observe that
the net flux is everywhere positive and
peaked at smaller scales, roughly at the end of the inertial range. 
Recall that the Maxwell flux is peaked at larger scales
  (fig.~\ref{fig:fluxes_lorentz}).
Due to the cyclic property of the trace the terms 
    $\Pi^{A,\ell}_{s, \Sigma J S} $ 
and $\Pi^{A,\ell}_{s, J \Sigma S} $ 
are equal while, because of the symmetry of the tensors involved, the subfluxes 
    $\Pi^{A,\ell}_{s,JJ\Omega}$ 
and 
    $\Pi^{A,\ell}_{s, \Sigma \Sigma \Omega}$ 
together with 
    $\Pi^{D,\ell}_{s,J\Omega J}$ 
and 
    $\Pi^{D,\ell}_{s, \Sigma \Omega \Sigma}$ 
are identically zero.
 It is also apparent from fig~\ref{fig:fluxes_advection_dynamo}(a) that the remaining Advection terms make negligible contributions to the net flux. 
 We note that
$ \Sigma \Sigma S $ type terms correspond to magnetic shear
amplification due to straining motions and those of
type $J J S$ to current filament stretching.
Terms of type $\Sigma J \Omega$ encode a
correlation between current and vorticity.
Thus, the net Advection term is primarily due to single-scale contributions, being approximately equal to 
$ 2\Pi^{A,\ell}_{s,\Sigma J S} $, and it carries about $40\%$ of the total energy flux at its peak.

For the Dynamo term, $\Pi^{D,\ell}$, we observe that the net flux is almost
flat in the inertial range, substantially positive definite for these scales
(fig.~\ref{fig:fluxes_advection_dynamo}(b)) and responsible for about $15\%$ of the total energy flux. 
All subfluxes except $
\Pi^{D,\ell}_{s,\Sigma S J} $ shown in yellow and $\Pi^{D,\ell}_{s, J S \Sigma
}$ indicated by the filled purple symbols are negligible. However, using the
cyclic property once again, one can show that $ \Pi^{D,\ell}_{s,\Sigma S J} =
-\Pi^{D,\ell}_{s, J S \Sigma }$ hence these two terms cancel out and do not
contribute to the net flux.  
In contrast to the Advection 
term, the major contributions to the net Dynamo term are from subleading single and multiscale subfluxes of various types, 
with each contributing only 1--2\% to the total energy flux adding up to a total of around $15\%$ of the total energy flux. 
In summary, single-scale current-sheet thinning is the dominant process transferring  magnetic energy across scales. 
It solely originates from the advective term in the induction equation.

\begin{figure}
 \begin{center}
     \includegraphics[width=.48\columnwidth]{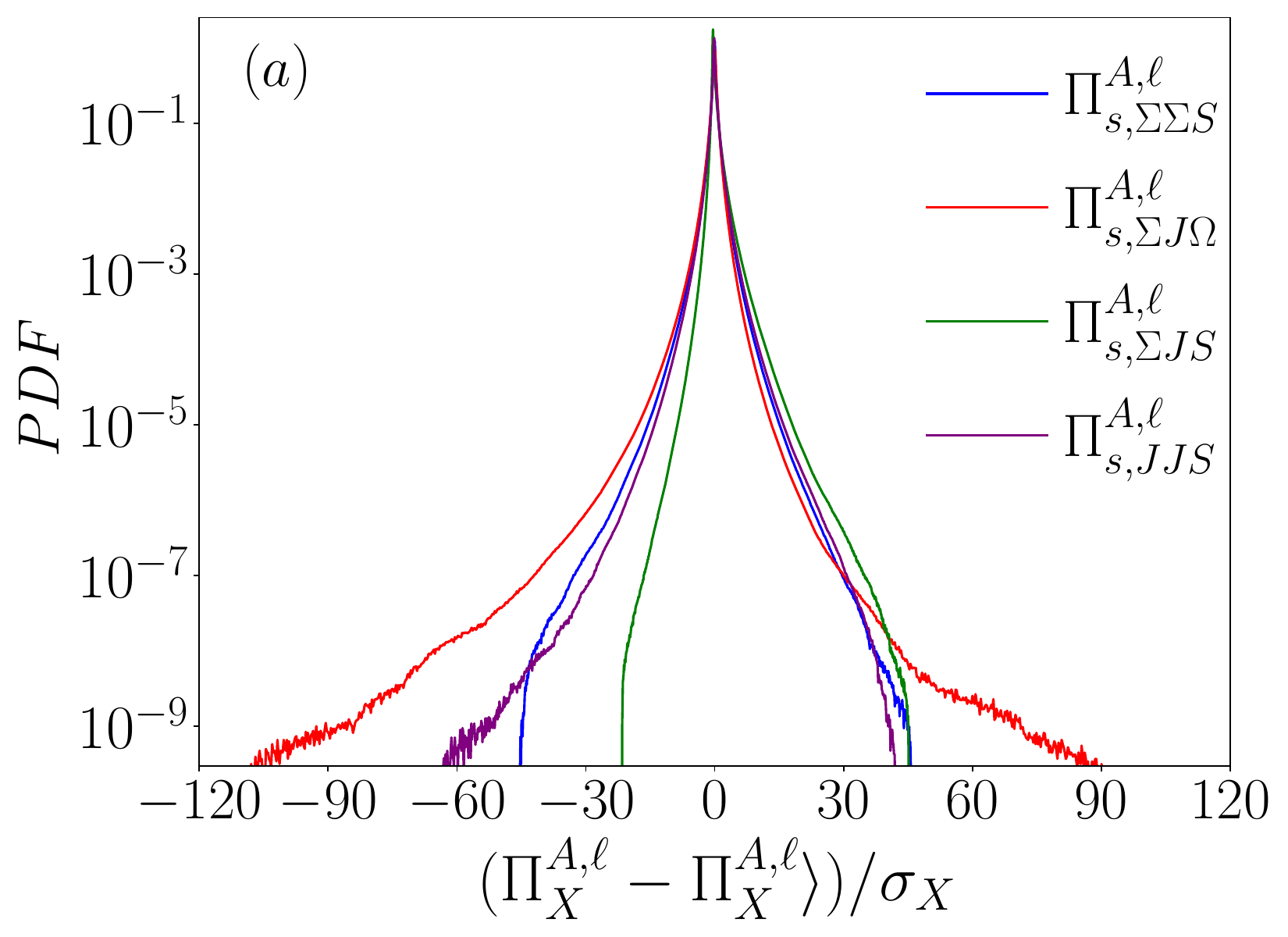} 
     \includegraphics[width=.48\columnwidth]{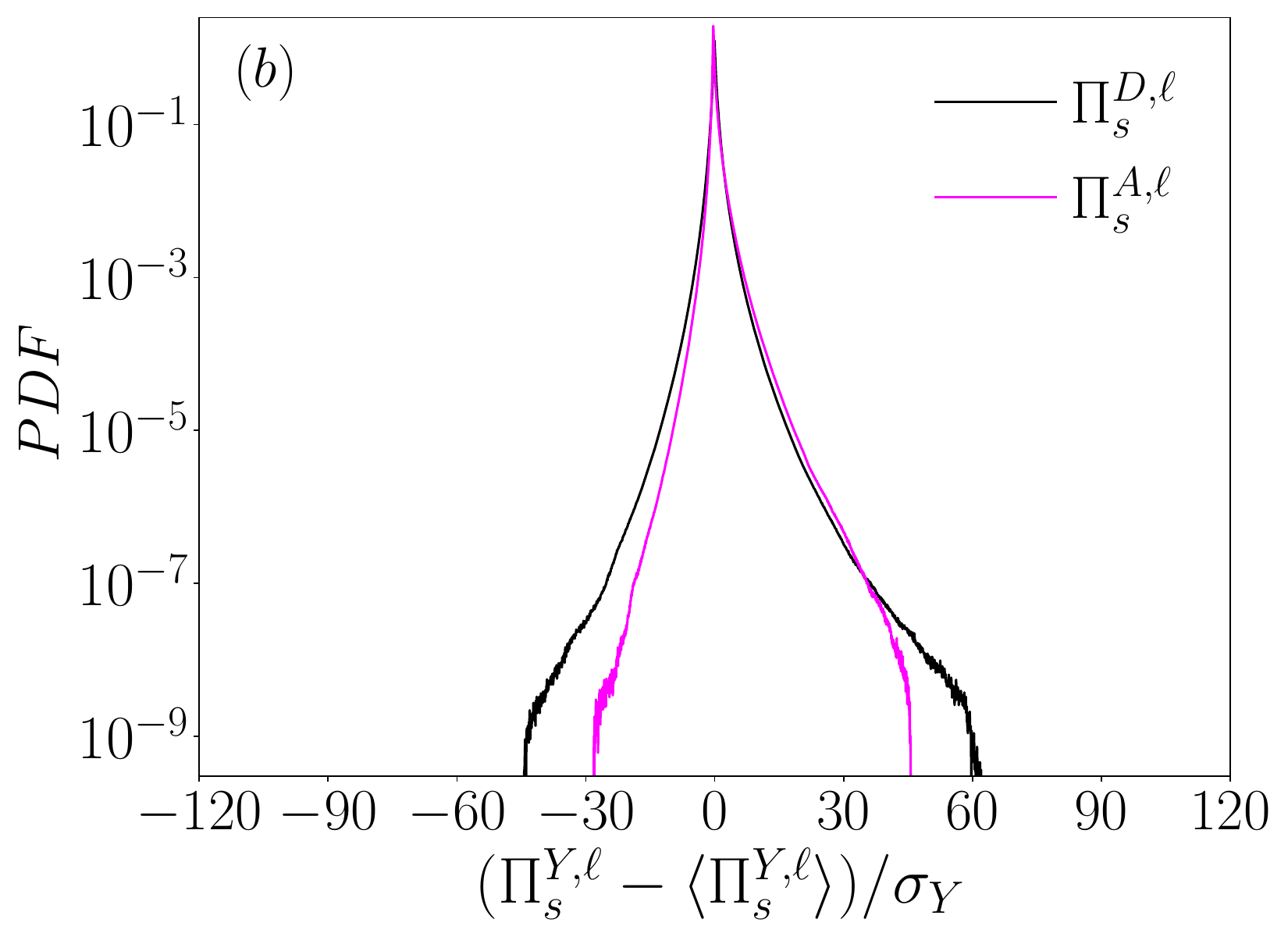} 
     \includegraphics[width=.48\columnwidth]{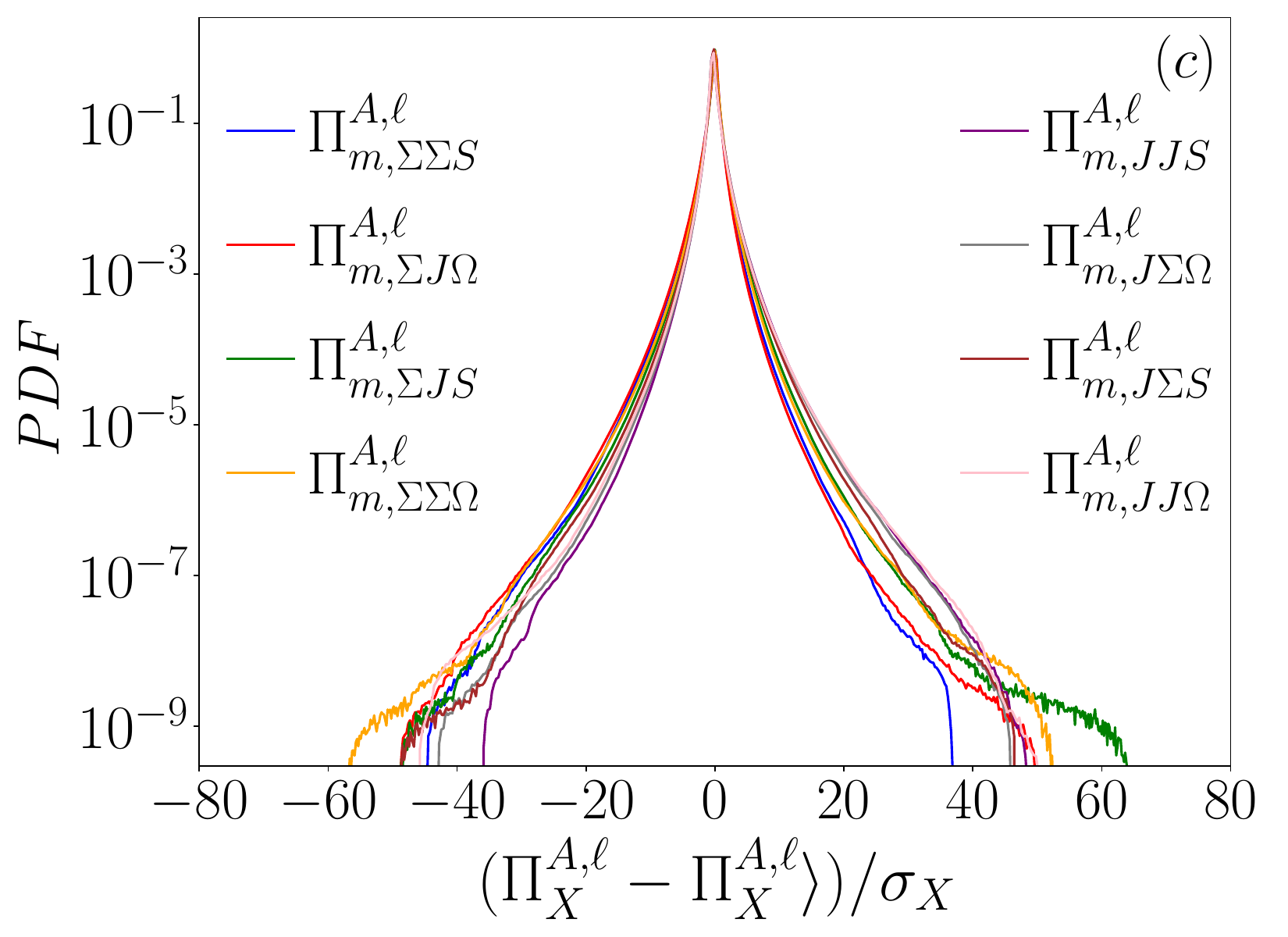} 
     \includegraphics[width=.48\columnwidth]{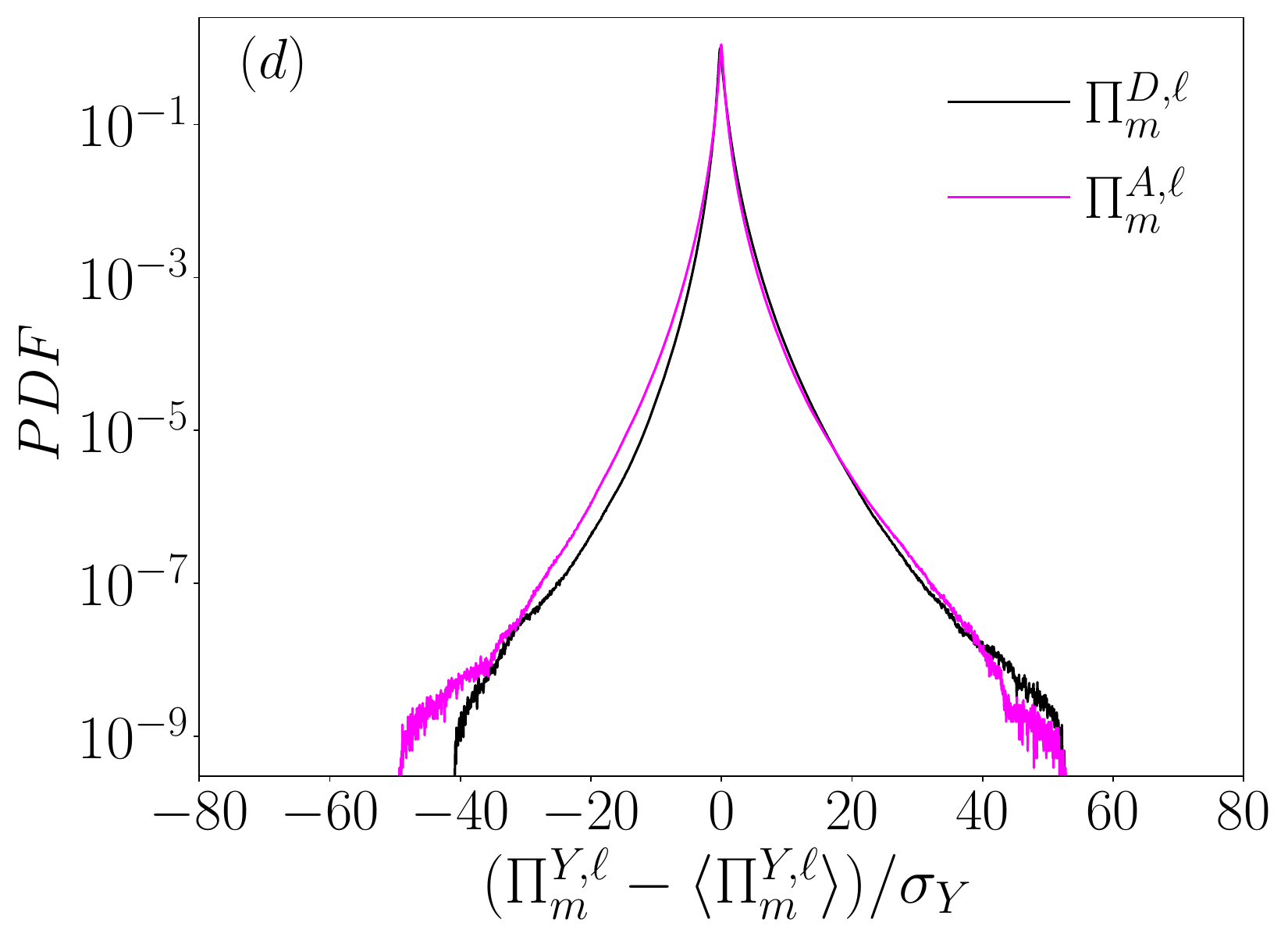} 
 \end{center}
 \caption{Standardised p.d.f.s for dataset A1 at
  $k \eta_\alpha = 5.4 \times 10^{-2} $.
(a) single-scale Advection subfluxes $\Pi^{A,\ell}_{X}$,  
where $X$ represents the subflux identifier, and 
(b) net single-scale fluxes for the Dynamo ($ \Pi^{D,\ell}_{s}$)
and Advection terms ($\Pi^{A,\ell}_{s}$).
In the $x$-axis titles for panels (b) and (d), 
    $Y =A$  or $D$, as appropriate. 
Note that
$\Pi^{A,\ell}_{s, \Sigma J  S} = \Pi^{A,\ell}_{s, J S \Sigma}$ 
  (see  fig.~\ref{fig:fluxes_advection_dynamo}).
(c)--(d): as for panels (a)--(b) except for the \emph{multi}-scale subfluxes of 
   $\Pi^{A,\ell}_{} $ and  $ \Pi^{D,\ell}_{}$. 
   Note that panels (c)-(d) have the same x-axis range as figures \ref{fig:pdf_inertial}--\ref{fig:pdf_Lorentz}.}
\label{fig:pdf_advection}
\end{figure}

Consider now the p.d.f.s.
As a consequence of the symmetries between Dynamo and Advection subfluxes, only the Advection p.d.f.s are shown in panels 
(a) and (c) of figure~\ref{fig:pdf_advection}, respectively for the single-scale terms and the multi-scale terms.  
It is striking that 
the p.d.f.\ of $\Pi^{A,\ell}_{s,\Sigma J \Omega}$ is by far the 
most strongly fluctuating 
with huge fluctuations of more than 100 standard deviations. 
The other Advection subflux p.d.f.s, both single-scale and multi-scale, span a range comparable with those associated with the p.d.f.s for the Inertial and Maxwell fluxes. 
Moreover, all the multi-scale fluxes (Advection and Dynamo) have quite similar p.d.f.s, and thus so do the net multi-scale flux p.d.f.s 
 (fig.~\ref{fig:pdf_advection}(d)).
In the case of the net single-scale p.d.f.s, 
the Advection--Dynamo agreement is still good in the cores of the distributions, but there is a significant difference at larger negative fluctuations
 (fig.~\ref{fig:pdf_advection}(b)). 



    \subsection{Total energy flux, $\Pi^{\ell}$}
    \label{sec:total-flux}

In the preceding subsections we analysed the four contributions to the incompressible MHD energy flux, finding that each of them may be reasonably well approximated using just some of the subfluxes. These approximation may now be assembled to give an approximate form for the  
mean total MHD energy flux.  Specifically we suggest that
\begin{align}
 \lan \Pi^{\ell} \ran 
  & =
   \lan \Pi^{I, \ell} \ran + \lan \Pi^{M, \ell} \ran +  \lan \Pi^{A, \ell} \ran  + \lan  \Pi^{D, \ell} \ran
 \approx 
  - 2 \langle \Pi^{M,\ell}_{s,SJ\Sigma} \rangle 
  -   \langle \Pi^{M,\ell}_{m,SJ\Sigma} \rangle 
 \\
 & = 4 \ell^2 \left\lan \tr{ (\ovS^\ell)^t \ovJ^\ell (\ovSS^\ell)^t} \right\ran\\  
 &+ 2\left\lan \int_0^{\ell^2}  \d\theta \,
       \tr{\big( \overline{\vS}^\ell \big)^t \Big(
       \ol{ \overline{\vJ}^{\sqrt{\theta}} 
              \big(\overline{\vSigma}^{\sqrt{\theta}} \big)^{t} \,}^\phi 
           -
            \ol{\overline{\vJ}^{\sqrt{\theta}}\,}^\phi\;
             \ol{ \big( \overline{\vSigma}^{\sqrt{\theta}} \big)^{\! t} \,}^\phi  
       \Big)} \right\ran
  \label{eq:Pi-ell-approx}
\end{align}
is a suitable expression.  Note that, after using cyclic properties of the trace, this is expressed only in terms of two of the Maxwell subfluxes, one single-scale and one multi-scale.  These are discussed in sec.~\ref{sec:Pi-M} and 
we remind the reader that all subflux definitions can be found in Appendix~\ref{app:defintions}. Interestingly, the terms on the RHS of eq.~\eqref{eq:Pi-ell-approx} are part of a group of terms that remain non-zero in 2D MHD; see Appendix \ref{app:mhd2d}.
We intend to explore this intriguing feature in future work.

Equation \eqref{eq:Pi-ell-approx} demonstrates that the mean total energy flux in MHD turbulence is largely given by 
the stretching and thinning 
of current-sheets into magnetic shear layers by large-scale strain,
resulting in a transfer of magnetic energy from large to small scales.
In addition, there is
a back-reaction of this process on the flow, whereby the ensuing magnetic strain-rate field accelerates fluid along its extensional directions and slows it down in the compressional
directions, thereby generating a stronger strain-rate field across smaller scales, as shown schematically in fig.~\ref{fig:sketch-current-thinning}.

    \subsection{Current-sheet thinning and magnetic reconnection}
    \label{sec:reconnection}
As a means of inter-scale energy transfer,
the current-sheet thinning process
can only proceed as described in an inertial range, i.e., at scales where Joule and viscous dissipation are negligible. As the dissipative scales are approached Alfv\'en's theorem ceases to hold and magnetic reconnection can occur, with associated changes to the topology of the magnetic field.  Interestingly, models of MHD reconnection  are geometrically very similar to the inertial-range energy transfer process described here. For example in the Sweet--Parker model a current sheet is thinned by a flow that pushes magnetic field lines closer and closer together. Eventually, when the distance between the field lines approaches scales where Joule dissipation becomes important, the topological conservation of the magnetic field is broken and magnetic field-lines reconnect. That is, the continuation of the current-sheet thinning process described herein, 
and shown conceptually in fig.~\ref{fig:sketch-current-thinning}, 
to smaller and smaller scales can lead naturally to magnetic reconnection. 


\begin{table}
  \begin{center}
\def~{\hphantom{0}}
   \begin{tabular}{l cccccccccc}
        \hline
        \hline
		 & $ \Pi^{I,\ell}_{s} $ & $ \Pi^{I,\ell}_{s,SSS} $  & $ \Pi^{I,\ell}_{s,S\Omega \Omega} $ &
        &
        $\Pi^{I,\ell}_{m} $ & $ \Pi^{I,\ell}_{m,SSS} $ &  $ \Pi^{I,\ell}_{m,S\Omega S} $ & $ \Pi^{I,\ell}_{m,S\Omega \Omega} $ \\
        \hline
		 $(\sigma_X^I)^2$ & 1.107  & 0.652  & 0.114 &   & 2.522  & 0.107  & 0.303  & 0.109 \\
          $S^I_X$ & 2.958 & 3.234 & 1.483 &  & -1.558 & 1.424 & -2.155 & 1.301\\ 
         $K^I_X$ & 39.18  & 41.48 & 29.68 &  & 17.15 & 18.96 & 15.42 & 17.50 \\
        \hline
        \hline
		 & $ \Pi^{M,\ell}_{s} $ &   $ \Pi^{M,\ell}_{s,S\Sigma \Sigma} $  & $ \Pi^{M,\ell}_{s,S J J} $& $ \Pi^{M,\ell}_{s,S J \Sigma } $ & $ \Pi^{M,\ell}_{m} $ & $ \Pi^{M,\ell}_{m,S \Sigma \Sigma} $ &  $ \Pi^{M,\ell}_{m,S J \Sigma} $ & $ \Pi^{M,\ell}_{m,S J J} $ \\
        \hline
		 $(\sigma_X^M)^2$ & 0.112 & 0.014  & 0.0113  & 0.119 & 0.353 & 0.014 & 0.467   & 0.014 \\
          $S^M_X$ & 3.207 & 0.326 & -0.827 & 4.030 & 3.105 & -2.176 & 3.417 & -2.289 \\
         \;$K^M_X$ & 32.22  & 32.71 & 31.09 & 39.88 & 21.34 & 27.96 & 24.60 & 28.46 \\
        \hline
        \hline
		 & $ \Pi^{A,\ell}_{s} $ &   $ \Pi^{A,\ell}_{s,\Sigma \Sigma S} $  & $ \Pi^{A,\ell}_{s, \Sigma J \Omega} $ & $\Pi^{A,\ell}_{s,\Sigma J S}$ & $\Pi^{A,\ell}_{s, J \Sigma S}$ & $\Pi^{A,\ell}_{s,J J S}$ \\
        \hline
		 $(\sigma_X^A)^2$ &  0.206 & 0.014   & 0.004  & 0.030  & 0.030  & 0.011   \\
          $S^A_X$ & 3.801 & -0.333  & -2.349  & 3.843  & 3.843  & 0.82  \\
         $K^A_X$ & 42.16  & 32.71 & 53.12 & 38.90 & 39.90 & 31.08   \\
        \hline
      \hline
		 & $ \Pi^{A,\ell}_{m} $ &   $ \Pi^{A,\ell}_{m,\Sigma S \Sigma} $  &  $ \Pi^{A,\ell}_{m,\Sigma \Sigma \Omega} $  & $ \Pi^{A,\ell}_{m,\Sigma J \Omega} $ & $\Pi^{A,\ell}_{m,\Sigma J S}$ & $\Pi^{A,\ell}_{m, J \Sigma S}$ & $\Pi^{A,\ell}_{m, J J S}$ & $\Pi^{A,\ell}_{m, J \Sigma \Omega}$  & $\Pi^{A,\ell}_{m, J J \Omega}$ \\
        \hline
		 $(\sigma_X^A)^2$ &  0.084   & 0.003 & 0.006  & 0.002  & 0.007   & 0.003   &  0.002 & 0.002 & 0.001    \\
          $S^A_X$ & 0.334 & -1.616 & -0.743   & -2.011   & -0.283   & 0.534  & 1.34 & 1.13 & 1.02 \\
         $K^A_X$ &  28.99 & 24.04 & 25.74 & 25.92 & 24.45 & 26.22 & 27.48 & 27.20  & 29.75   \\
        \hline
        \hline
        & $ \Pi^{D,\ell}_{s} $ & $ \Pi^{D,\ell}_{m} $ &      &   &   &   &   \\
             \hline
         $(\sigma_X^D)^2$ &  0.039   & 0.055  &     &    &      &     \\
        $S^D_X$ & -1.918   &  1.804  &    &    &    &      \\
        $K^A_X$ &  35.61 & 25.41 & & &   &  \\ 
     \hline
     \hline
  \end{tabular}
  \caption{Values of variance $(\sigma_X^Y)^2$, 
  skewness 
  $S^Y_X = {\left \langle \left (\Pi^{Y,\ell}_{X}  - \langle \Pi^{Y,\ell}_{X}  \rangle \right )^3 \right \rangle}/{(\sigma_X^Y)^3} $,
    and kurtosis 
  $K^Y_X = {\left \langle \left (\Pi^{Y,\ell}_{X}  - \langle \Pi^{Y,\ell}_{X}  \rangle \right )^4 \right \rangle}/{(\sigma_X^Y)^4}$ 
    for the subflux p.d.f.s 
    shown in figs.~\ref{fig:pdf_inertial}--\ref{fig:pdf_advection}, 
  where $X$ indicates the subflux identifier and $Y$ denotes the term identifier.
   }

 \label{tab:table_moments}
\end{center}
\end{table}

 \section{Consequences for MHD subgrid-scale modelling}
\label{sec:sgs_modelling}

MHD LES modelling usually proceeds through variations on the Clark and Smagorinsky models. This is the case for incompressible MHD \citep{ZhouVahala91,MuellerCarati02,KessarEA16}.
Moreover, adaptations have been made  
for the compressible case \citep{Chernyshov2010,VlaykovEA16,GreteEA16}, 
for channel flow at high magnetic Reynolds number \citep{Hamba10,JadhavChandy21-sgs,Jadhav2023},
and
for extensions of MHD taking various levels of microphysics terms into account, 
such as Hall MHD \citep{Miura2016} or Braginskii-extended (two-fluid) MHD \citep{Miura2017}, with the latter specifically focussed on the ballooning instability in stellarator devices.  
Most such approaches use the same SGS model for the inertial and Maxwell stresses based on velocity-field gradients with eddy viscosities involving either only the strain rate tensor or a weighted sum of the squared strain rate and the squared current, and similarly structured SGS models for the 
induction equation, with either heuristics or trial-and-error approaches to find suitable values for model constants. In what follows, we briefly discuss how the present results
can be used to construct suitably structured SGS models for each SGS-stress in the MHD equations. 


In terms of SGS modelling, our simulation-based results suggest that the Inertial terms can be neglected and a dissipative model 
for the Maxwell stresses should suffice to capture the (leading-order) mean effects. 
In terms of fluctuations, we find that the observed mean Inertial flux depletion is caused by considerable backscatter in all Inertial subfluxes. This could suggest that a more sophisticated model would be required for the Inertial term. 
However, in 3D HD turbulence, backscatter-free
SGS models such as the standard static Smagorinsky closure perform well in capturing high-order
statistics, that is anomalous exponents and multifractal predictions for the
correlation between velocity-field increments and SGS stresses
    \citep{LinkmannEA18}. 
Furthermore,  as the SGS stresses enter the filtered
Navier--Stokes equation only through their divergence, this results in a degree of (gauge) freedom to determine
 model
stress tensors that produce much less backscatter than
those constructed using the standard definition \citep{VelaMartin2022}.  
In fact, backscatter can be traced back to spatial fluxes disconnected from the (scale-space) energy transfer and as such does not require modelling \citep{VelaMartin2022}.

For the magnetic energy transfers, we observe the net transfers are from large to small scales, suggesting again that dissipative models should suffice. 
Indeed, as  multi-scale terms in our flux decomposition are negligible, a
Clark-type model involving only the coupling between current and strain-rate tensors  may work well for a nonlinear saturated non-helical (small-scale) dynamo.
However, additional stabilising terms may be required
as the Clark (or gradient) model is know to result in numerically unstable LES of a mixing layer in HD 
    \citep{Vreman1996b,Vreman1997} 
and for MHD 
    \citep{Muller02,KessarEA16}, 
as discussed in further detail below. 
Similar to the Inertial term results, we observe considerable backscatter in the 
magnetic energy fluxes, and it remains to be seen if the aforementioned results by \citet{LinkmannEA18} 
on the effect of SGS closures on high-order statistics carry over from HD to MHD.

For a nonhelical saturated dynamo, as is the case here,
\citet{Muller02} and \citet{KessarEA16} carried out 
    MHD LES 
with Clark-type models constructed from the full velocity and magnetic-field gradients for the sum of
the Reynolds and Maxwell SGS stress in the momentum equation and for the
magnetic stresses, resulting in unstable simulations as in the HD case. Using
\emph{a-priori} analyses of DNS data, \citet{KessarEA16} trace the instabilities
back to the Clark terms 
 (which correspond to what we have herein called single-scale flux contributions)
transferring an insufficient amount of kinetic and magnetic
energy to small scales, and to a production of backscatter.  According to our
analysis, for the momentum equation the former effect is due to  the Maxwell
stress having a significant multi-scale component which is not captured in the
Clark model. Furthermore, a Clark model based on the full gradients will
introduce effects that are not present in the full MHD dynamics especially
concerning the Maxwell and magnetic stresses, as all combinations of vorticity,
strain, current, and magnetic strain are included in the model and equally weighted. 
Our analysis, however, 
shows that only terms stemming from the coupling between current and strain-rate tensors are significant, 
with all remaining contributions to the net Maxwell and
magnetic fluxes being negligible. 
Thus, retaining these contributions in a SGS model may substantially (and inappropriately) affect the small-scale structure of the flow and the magnetic field. 


A further challenge for LES modelling of MHD turbulence is to accurately capture the 
    transfer 
of magnetic to kinetic energy and \emph{vice versa}. 
As discussed by
\citet{OffermansEA18}, the resolved-scale conversion term must either be fully accounted for in LES, resulting in the need to resolve all scales where this term is active, or in the present case of a saturated dynamo,
a model including  an extra term accounting for 
the under-resolved dynamo effect must be provided.

In this paper we have only considered the no mean magnetic field situation.
The presence of a strong background magnetic field is likely
to require an SGS modelling approach that differs from those just discussed, as the ensuing anisotropy and
two-dimensionalisation of magnetic and velocity-field fluctuations may result
in partial inverse fluxes. We will report results on configurations with strong
background magnetic fields in due course.  Similarly, the large-scale (helical)
dynamo requires an investigation in its own right, and the magnetic-field
growth at large scales is likely to require a different type of SGS modelling
approach, as \citet{KessarEA16} 
report that the Clark and even a standard static
Smagorinsky model result in unstable LES. 

The method discussed here can be readily extended to Hall- and two-fluid
MHD and other fluid models for plasma turbulence. For instance, the applicability of the
Smagorinsky closure to Hall MHD has been assessed by an \emph{a-priori} analysis
using sharp filtering \citep{Miura2012} prior to the deployment of said
closure \citep{Miura2016}. An \emph{a-priori} analysis and
decomposition of the Hall flux in analogy to the results presented here could
lead  to  a better understanding of the physics of the interscale magnetic enery transfer induced by 
the Hall effect and thus to a refinement of Hall-MHD SGS models. 
Finally, we point out that a new LES method, so-called \emph{physics-inspired coarsening} (PIC), has been 
devised recently \citep{Johnson2022}.
In the homogeneous case 
this approach reduces to Gaussian filtering and the 
representation of SGS-stresses in terms of field gradients \citep{Johnson20,Johnson21} as generalised herein. 
In PIC, the velocity field advanced in LES is formally obtained by artificial viscous smoothing, with the 
required pseudo-diffusion being introduced through an auxiliary Stokes equation. This approach may be generalisable 
to MHD and more complex fluid models applicable to plasma turbulence.

\section{Conclusions}
      \label{sec:concl}


Generalising a method introduced by \citet{Johnson20},  
we have presented  a
general analytical method for obtaining exact forms for inter-scale fluxes in advection--diffusion
equations through products of vector-field gradients, and applied it to kinetic
and magnetic energy fluxes in homogeneous MHD turbulence. 
The aim was to provide
expressions for subfluxes that are physically interpretable in terms of the action of the magnetic field on the flow and \emph{vice versa}. 
A quantification thereof is of interest for the fundamental
understanding of cascade processes in MHD turbulence, and also provides guidance
as to what physics needs to be captured in subgrid-scale models and how such models
should be constructed so that they preserve, at least approximately,
empirical features
of the mean energy fluxes
and their fluctuations.  

In MHD, scale-space energy fluxes are defined as the contraction of velocity- or
magnetic-field gradients with the appropriate subgrid-scale stresses 
Rewriting these in terms of symmetric and
antisymmetric components of field-gradients tensors yields terms with clear physical meanings. For example, strain and vorticity in case of the velocity field, and
current and magnetic strain/shear for the magnetic field. Expressing the MHD SGS
stresses in terms of vorticity, rate-of-strain, current, and magnetic rate-of-strain results in
an exact decomposition of magnetic and kinetic energy fluxes in terms of
interactions between the symmetric and antisymmetric components of velocity-
and magnetic-field gradients. 

The kinetic energy flux comprises two terms, the Inertial flux (as in hydrodynamics) and a flux term associated with the action of the Lorentz force on the flow. 
The former is decomposed into terms associated with vortex
stretching, strain self-amplification, and strain-vorticity alignment
\citep{Johnson20,Johnson21}. 
A term-by-term comparison between the Inertial
fluxes in HD and MHD turbulence shows that all Inertial subfluxes are depleted and indeed almost negligible in MHD turbulence.  That is, the physics of the kinetic
energy cascade is very different in statistically steady MHD turbulence as
compared to HD turbulence, as vortex stretching and strain
self-amplification have on average very little effect. 
In MHD turbulence, almost all
kinetic energy is transferred downscale by a current-sheet thinning process: in regions of large strain, current sheets are stretched
by large-scale straining motion into regions of magnetic shear. This magnetic shear in
turn drives extensional flows at smaller scales.  
The magnetic energy, is mainly transferred from large to small scales, albeit with considerable backscatter, by the aforementioned
current-sheet thinning in regions of high strain, while the contribution from current-
filament stretching --- the analogue to vortex stretching --- is negligible.

Finally, we note that the method can be further expanded in various directions, to include temperature fluctuations, for instance. 
An extension or application to compressible flows would be of interest especially for astrophysical plasmas.
As flux terms associated with any advective nonlinearity can be analysed by this method, a decomposition of Hall MHD and of fluid models of ion- or
electron-temperature-gradient turbulence 
\citep{Ivanov2020, Ivanov2022, Adkins2022} 
may be of interest for the magnetic confinement fusion community.  


\appendix
   \section{General formulation of advective-type SGS flux terms}
   \label{app:general-case}

In sec.~\ref{sec:exact} we derived, 
following \cite{Johnson20,Johnson21},
the form for the scale-filtered magnetic energy flux associated with the $  \vu \cdot \nabla \vb$ term of the induction equation.  
Here we outline how this approach is generally applicable for flux terms involving three distinct fields, connected with an (unfiltered) term of the form
    $ \vx \cdot (\vz \cdot \nabla) \vy $,
where 
    $\vx$, $\vy$, and $\vz$ 
are solenoidal, but otherwise arbitrary, 3D vector fields
 (they are not coordinate vectors). 
For appropriate mappings of $\vx$, $\vy$, $\vz$ 
to $\vu$ and $\vb$ 
this will yield any of the desired MHD SGS energy fluxes,   
   equations~\eqref{eq:Pi-I}--\eqref{eq:Pi-D}.
Moreover, the SGS fluxes associated with helically decomposed hydrodynamics and MHD \citep{Waleffe93, LessinnesEA11,LinkmannEA15, alexakis2017, AlexakisBiferale18, YangEA21-forcing}
and the kinetic, magnetic, and cross helicities may be obtained using similar special cases, see 
    \cite{CapocciEA23-Hk} for a decomposition of the kinetic helicity flux in Navier-Stokes turbulence and appendix~\ref{app:general-specialcases}.

The SGS stresses at scale $\ell$ associated with 
    $ \vz \cdot \nabla \vy $
are
\beq
  \tau^\ell(y_i, z_j)
     =
         \ol{y_i z_j}^\ell 
       - \ol{y_i}^\ell \ol{z_j}^\ell ,
\eeq
where the choice of the filter kernel is for now arbitrary. 
Clearly, 
  $ \left( \tau^\ell (y_i, z_j) \right)^t = \tau^\ell (z_i, y_j)$ 
where $(\cdot)^t$ denotes matrix transpose.
Note that the advect\emph{ing} field is $\vz$.
Contracting the SGS stress against the gradient tensor of a third arbitrary field, $\vx$, yields the general 
    SGS \emph{flux} term 
\beq
\label{eq:general_flux}
  \Pi_{xyz}^\ell
      =
   - \left( \p_j \ol{x}^\ell_i  \right)
      \tau^\ell(y_i, z_j) 
 .     
\eeq 
Here we have included a leading minus sign in eq.~\eqref{eq:general_flux}. 
However, if one wishes to have $ \Pi_{xyz}^\ell > 0 $ always correspond to forward transfer---as we have elected to do herein---this may not be correct.  It depends on the sign the $ \vz \cdot \nabla \vy $ term has when it is written on the RHS of the underlying advection-diffusion equation.  
For the MHD momentum equation, for example, the Lorentz force term has $ \vy = \vz = \vb $ and the minus sign for the associated kinetic energy flux (with $\vx = \vu$) should be absent.  See sections~\ref{sec:coarse-graining} and \ref{sec:Pi-M}.
When it is appropriate to do so, the minus sign and its propagation into other equations in this Appendix is easily removed.

In the special case that $\tau^\ell(y_i, z_j)$ is index-symmetric, only the index-symmetric part of 
  $ \partial_j \ol{x}^\ell _i $ 
contributes.  
This is the situation for the kinetic energy flux in HD 
  \citep{Germano92}
and by analogy in MHD, see e.~g. \citep{ZhouVahala91,KessarEA16,Aluie17,OffermansEA18,AlexakisChibbaro22}.
In general, however, the index-antisymmetric part of 
  $ \partial_j \ol{x}^\ell _i $ 
is also needed.


As shown in \cite{Johnson21}, eq.~\eqref{eq:general_flux}
may also be expressed entirely in terms of (products of) the gradient tensors for $\vx, \vy, \vz$, 
and integrals over them.
Denoting the respective gradient tensors as 
  $ \p_j x_i = X_{ij} $, 
  $ \p_j y_i = Y_{ij} $, and 
  $ \p_j z_i = Z_{ij}$, 
we have
\begin{align}
  \Pi^\ell_{xyz}
     & = 
    -\ell^2 \, \ol{X}_{ij}^\ell \,\ol{Y}_{ik}^\ell \, \ol{Z}_{jk}^\ell 
    \; - \;
      \ol{X}_{ij}^\ell 
      \int_0^{\ell^2} \d\theta 
       \left(
         \ol{ \ol{Y}_{ik}^{\sqrt{\theta}} \,\,
              \ol{Z}_{jk}^{\sqrt{\theta}} \,}^\phi 
           -
         \ol{\ol{Y}_{ik}^{\sqrt{\theta}}\,}^\phi \;
         \ol{\ol{Z}_{jk}^{\sqrt{\theta}}\,}^\phi  
       \right)     
  \label{eq:Pi-xyz-exact}
 \\ 
  & = 
    \Pi^{\ell}_{s,xyz}   +  \Pi^{\ell}_{m,xyz},
   \label{eq:Pi-xyz-s-ms}
\end{align}
where $ \phi = \sqrt{\ell^2 - \theta} $ and $\sqrt{\theta}$ correspond to all filter scales smaller than $\ell$,
and the subscripts $s$ and $m$ stand for single-scale and multiscale.
Equivalently, expressed in terms of the matrix trace operation, this is
\begin{align}
  \Pi^\ell_{xyz}
      & = 
    -\ell^2 \, \tr{(\ol{X}^\ell)^t \, \ol{Y}^\ell \,\, (\ol{Z}^\ell)^t} 
  \nonumber \\  
    & \quad - \; 
      \int_0^{\ell^2} \d\theta \, 
      \tr{ (\ol{X}^\ell)^t
         \left[
         \ol{ \ol{Y}^{\sqrt{\theta}} 
              (\ol{Z}^{\sqrt{\theta}})^t \,}^\phi 
           -
         \ol{\ol{Y}^{\sqrt{\theta}}\,}^\phi \;
         (\ol{\ol{Z}^{\sqrt{\theta}})^t\,}^\phi 
         \right]
         } 
.
  \label{eq:Pi-xyz-trace_based}
\end{align}
Thus, both the single-scale and the multi-scale terms may be expressed as 
 (integrals of)
the trace of the appropriate filterings and transposes of the product of the  three gradient tensors.

Splitting the gradient tensors
into their index symmetric and antisymmetric parts,
  e.g., $ \oll{\vX} = \oll{\vS}_X + \oll{\vct{\Omega}}_X $, 
produces a decomposition of
    eq.~\eqref{eq:Pi-xyz-trace_based}
that facilitates physical interpretation of the subterms.
For the single-scale terms one has, modulo the $-\ell^2$ factor, 
\begin{align}
\label{eq:XYZ_general}
    \mathrm{Tr}\left\{ \left(\oll{\vX} \right)^t \right.   &  \left.  \oll{\vY} \left(\oll{\vZ}\right)^t \right\} \notag \\ 
   = & \;
    \tr{ \left( \oll{\vS}_X - \oll{\bm \Omega}_X \right) \left( 
      \oll{\vS}_Y \oll{\vS}_Z      + \oll{\bm \Omega}_Y \oll{\vS}_Z  - \oll{\vS}_Y \oll{\bm \Omega}_Z 
      - \oll{\bm \Omega}_Y \oll{\bm \Omega}_Z  
    \right) } 
 \\
   = & \;
    \mathrm{Tr}\left\{   
          \oll{\vS}_X \oll{\vS}_Y \oll{\vS}_Z 
        + \left( \oll{\vS}_X - \oll{\bm \Omega}_X \right)
          \left( \oll{\bm \Omega}_Y \oll{\vS}_Z - \oll{\vS}_Y \oll{\bm \Omega}_Z \right)
        + \oll{\bm \Omega}_X \oll{\bm \Omega}_Y \oll{\bm \Omega}_Z 
    \right.
    \notag \\
     &
    \left.  \qquad
        - \oll{\vS}_X \oll{ \bm \Omega}_Y \oll{\bm \Omega}_Z 
        - \oll{ \bm \Omega}_X \oll{\vS}_Y \oll{\vS}_Z
    \right\}
 ,
\label{eq:XYZ_expanded}
\end{align}
which in general does not simplify further.

Simplifications do ensue, however, for special cases when one or more of $\vX$, $\vY$, $\vZ$ are equal.
One makes use of matrix properties like 
    $ {\bm \Omega}_Y \vS_Y + \vS_Y^t {\bm \Omega}_Y^t $
is a symmetric matrix and the square of any (square) matrix is a symmetric matrix.
For example, when $ \vY = \vZ $, 
as is relevant for the Inertial ($ \Pi^{I,\ell}_s $) 
and Maxwell ($ \Pi^{M,\ell}_s $) fluxes,
we obtain
\begin{equation}
   \tr{ \left(\oll{\vX} \right)^t  \oll{\vY}  \left(\oll{\vY} \right)^t } 
   =  
   \tr{ \oll{\vS}_X \oll{\vS}_Y \oll{\vS}_Y 
      - \oll{\vS}_X \oll{ \bm \Omega}_Y  \oll{ \bm \Omega}_Y
      + \oll{\vS}_X \left[  \oll{ \bm \Omega}_Y \oll{\vS}_Y 
      + \left(  \oll{ \bm \Omega}_Y \oll{\vS}_Y \right)^t \right]
    }.
 \label{eq:XYY-survives}
\end{equation}
Examples with $\vX = \vY$ and $ \vX = \vZ $ relate to the Advection and Dynamo magnetic energy SGS fluxes.  See sec.~\ref{sec:Pi-A}.

Turning to the multi-scale contributions in eq.~\eqref{eq:Pi-xyz-exact}, these
may of course be similarly decomposed.  
Since the filtering operation is linear the integrand can be split into the sum of four terms that each have the same structure as the original integrand, e.g.,
\begin{equation}
  \tau^{\phi} \left( 
        \ol{\vS}_{Y}^{\sqrt{\theta}},
        \ol{\bm \Omega}_{Z}^{\sqrt{\theta}}  \right)
   = 
  \ol{ \ol{\vS}_{Y}^{\sqrt{\theta}} \,\,
              \ol{\bm \Omega}_{Z}^{\sqrt{\theta}} \,}^\phi 
           -
         \ol{\ol{\vS}_{Y}^{\sqrt{\theta}}\,}^\phi \;
         \ol{\ol{\bm \Omega}_{Z}^{\sqrt{\theta}}\,}^\phi 
 .
  \label{Pi-multi-example}       
\end{equation}
After integration and contraction with 
  $ (\oll{\vX})^t = \oll{\vS}_X - \oll{\bm \Omega}_X $
this gives eight, in general distinct, multi-scale contributions.
Once again special cases such as $\vY=\vZ$ may mean some of these eight are zero, or equivalent, or cancel.  The needed particular instances are discussed in the subsections of sec.~\ref{sec:analysis}.

  \subsection{Special Cases}
  \label{app:general-specialcases}

Here, we list specific examples of 
    eq.~\eqref{eq:general_flux} 
that are relevant to the HD and/or MHD equations:
\begin{enumerate}   \itemsep=1ex
    \item  $\vx,\vy,\vz  \to  \vu$.  
    This yields the usual Navier--Stokes energy flux, 
        $ \Pi^\ell= -\ol{S}^\ell_{ij} \,\tau^\ell(u_i, u_j) $.
    Due to the index symmetry of the SGS stress tensor, only the symmetric part of the gradient tensor of $\vu$ plays a direct role, as noted previously. 
    \label{itm:ns}

    \item $ \vy,\vz  \to \vu$;
      \;
          $ \vx \to \vom = \nabla \times \vu$, the vorticity.
    This corresponds to the Navier--Stokes helicity flux, 
        $ \Pi^{H,\ell}= -2\,\ol{S}^\ell_{\! \omega \, ij} \,\tau^\ell(u_i, u_j) $. 
    As for the previous case, the symmetry of the SGS stress  means that the flux can be written in terms of just the symmetric part of the gradient tensor of vorticity, namely $ \ol{S}^\ell_{\! \omega \, ij} 
      = 
      ( \partial_j \ol{\omega}^\ell_{i} + \partial_i \ol{\omega}^\ell_{j} )/2$.
    \label{itm:ns_hlc}

    \item 
          $ \vy \to \vu, \vz \to \vb, $ 
        \;  
          $ \vx \to \va $ together with $ \vy \to \vb, \vz \to \vu = \nabla \times \va$; 
        \;  
          $ \vx \to \va $, where $\bm{a}$ s.t. $\vb = \nabla \times \va$ is the magnetic vector potential.
    Here the flux is that for the MHD magnetic helicity,
        $\Pi^{B,\ell}= -2\, \p_j \ol{a}^\ell_{i} \left( \tau^\ell(u_i, b_j) - \tau^\ell(b_i, u_j)  \right) $. 
    \label{itm:mhd_hlc}
    
    \item $\vx, \vy, \vz \in \{ \vu, \vb \}$: MHD kinetic energy, magnetic energy, and cross helicity fluxes.
    Regarding the energy fluxes, their
    exact decompositions and quantifications are discussed in detail in the main body of this work. 
    Decomposition of the MHD helicity fluxes will be examined in a future paper.
    \label{itm:mhd_gen}

    \item As an additional level of analysis, we may also consider various projections of the fields onto subspaces of particular interest. For instance,
    after decomposing the velocity field---using a basis constructed from eigenfunctions of the curl operator---into positively and negatively helical fields $\vu^\pm $, such that $\vu = \vu^+ + \vu^- $ \citep{Waleffe93, LessinnesEA11,LinkmannEA15, alexakis2017, AlexakisBiferale18, YangEA21-forcing},
    the following SGS stresses occur in the evolution equations for $\vu^\pm $
    \begin{align}
    \tau_{ij}^\ell(\vu^\pm, \vu^\pm) & = \ol{u_i^\pm u_j^\pm}^\ell - \ol{u_i^\pm}^\ell  \ol{u_j^\pm}^\ell \ , \\
    \tau_{ij}^\ell(\vu^\pm, \vu^\mp) & = \ol{u_i^\pm u_j^\mp}^\ell - \ol{u_i^\pm}^\ell  \ol{u_j^\mp}^\ell \ , \\ 
    \tau_{ij}^\ell(\vu^\mp, \vu^\pm) & = \ol{u_i^\mp u_j^\pm}^\ell - \ol{u_i^\mp}^\ell  \ol{u_j^\pm}^\ell \ .
    \end{align}

\end{enumerate} 

    \section{Two extended Betchov relations for MHD}
    \label{app:extended-betchov}


Here we derive two MHD analogs of the exact kinematic relation between components of the velocity-gradient tensor introduced by \cite{Betchov56}
and use them to obtain relations between several MHD energy subfluxes.

Recall that  \cite{Betchov56} showed that 
  $ \avg{ A_{ij} A_{jk} A_{ki} } = 0 $,
where $ A_{ij} = \partial_j u_i $, etc.
As a first step we wish to prove a similar relation between gradients of filtered fields, where the filtering scale on each field need not be the same.  Specifically we demonstrate that
\begin{equation} \label{eq:mhd_gen_Betchov}
    \avg{ 
      \Al_{ij}\, \Bm_{jk} \, \Bm_{ki} 
      }
      = 0 ,   
\end{equation}
where $\ell, m$ are two generic filtering scales, $B_{ij} = \p_j b_i$ is an (unfiltered) gradient tensor related to a solenoidal magnetic field. 
The above gradient tensors can be in principle calculated in different positions\footnote{For instance, $ \avg{ \Al_{ij}(\vx,t) \, \Bm_{jk}(\vx+ \vr',t) \, \Bm_{ki}(\vx+\vr'',t) } $ where $\vr', \vr''$ are displacement vectors.}. 
Using incompressibility and periodic boundary conditions one obtains
\begin{align}  \label{eq:procedure_multiscale}
 \avg{ \Al_{ij}  \Bm_{jk} \, \Bm_{ki} } 
=& 
  \avg{ \p_i  
    \left( \Al_{ij}  \Bm_{jk} \, \bfm_{k} \right)  }
 - 
  \avg{ \p_k 
    \left( \Al_{ij}  \Bm_{ji} \, \bfm_{k} \right)  }
 + 
  \avg{ \p_j
    \left( \Al_{ik}  \Bm_{ji} \, \bfm_{k} \right)  }\\
 &- 
   \avg{ 
      \Al_{ik}  \Bm_{ji} \, \Bm_{kj}   } .  
\end{align}
This yields eq.~\eqref{eq:mhd_gen_Betchov} since the averages of the gradients vanish when the boundary conditions are periodic (or the system is homogeneous),
and we are left with a quantity equal to its negative.

The next step is to decompose each gradient tensor of eq.~\eqref{eq:mhd_gen_Betchov} in terms of its symmetric and antisymmetric parts:
\begin{equation}
\avg{ 
 \left( \Sul_{ij} + \Ol_{ij} \right)  
 \left( \Sbm_{jk} + \Obm_{jk}  \right) 
 \left( \Sbm_{ki} + \Obm_{ki}  \right)  
 }
 = 0.
 \label{eq:mhd-Betchov-sym-anti}
\end{equation}
Exploiting the symmetries of the tensors involved yields the identity
\begin{equation}  \label{eq:gen_Betchov_decomp}
 -\avg{  \Sul_{ij} \, \Sbm_{jk} \, \Sbm_{ki} } 
= 
     \avg{ \Sul_{ij} \, \Obm_{jk} \, \Obm_{ki} }
 + 2 \avg{ \Ol_{ij}  \, \Sbm_{jk} \, \Obm_{ki} }
 .
\end{equation}
This can be considered as a generalized Betchov identity for MHD.
As a special case, we note that if the magnetic field becomes equal to the velocity field and we remove the filters, then eq.~\eqref{eq:gen_Betchov_decomp} collapses to the standard Betchov relation. 

Equation~\eqref{eq:gen_Betchov_decomp} 
is multi-scale but not in the form of energy fluxes.  To obtain such a relation we calculate its convolution with the Gaussian filter, with filter scale 
 $ \phi = \sqrt{\ell^2 - m^2} $,
and integrate over the filter scale $m$, 
following what was done in the RHS of eq.~\eqref{eq:tau-ell-bu}.
The result is
\begin{equation}
      \lan \Pi^{M,\ell}_{m,S\Sigma \Sigma} \ran  
=     \lan \Pi^{M,\ell}_{m,S J J} \ran 
  + 2 \lan \Pi^{M,\ell}_{m,\Omega \Sigma J} \ran
, 
\label{eq:multisc_betchov}
\end{equation}
which corresponds to eq.~\eqref{eq:extended_betchov}
in the main body of the paper. 
The single-scale version of this relation ($m$ subscripts replaced with $s$) also holds, as can be seen by setting $ \ell = m $ 
in  eq.~\eqref{eq:gen_Betchov_decomp}, i.e.,
\begin{equation}
      \lan \Pi^{M,\ell}_{s,S\Sigma \Sigma} \ran  
=     \lan \Pi^{M,\ell}_{s,S J J} \ran 
  - 2 \lan \Pi^{A,\ell}_{s,J \Sigma \Omega} \ran
 .
  \label{eq:multisc_betchov_local}
\end{equation}
Note that this mixes terms from the momentum equation with one from the induction equation.

In sec.~\ref{sec:Pi-M}, we infer from the simulation p.d.f.s that 
$\Pi^{M,\ell}_{m,S\Sigma \Sigma}  
  \approx 
   \Pi^{M,\ell}_{m,S J J } $ where both these terms appear as averaged quantities in eq.~\eqref{eq:multisc_betchov}.  We can recover the pointwise identity relative to the subfluxes appearing in eq.~\eqref{eq:multisc_betchov} taking into account the (not averaged) gradients from the RHS of eq.~\eqref{eq:procedure_multiscale}. As a consequence, $\Pi^{M,\ell}_{m,\Omega \Sigma J}$ should be cancelled by the contribution from the gradients.


In addition to
 eq.~\eqref{eq:mhd_gen_Betchov}, we can prove another exact identity that reads:
\begin{equation} \label{eq:mhd_gen_Betchov-2}
    \avg{ 
      \Bl_{ij}\, \left( \Bm_{jk} \, \Am_{ki} + \Am_{jk} \, \Bm_{ki} \right) 
      } = 0.
\end{equation}
This is obtained by employing incompressibility and periodic boundary conditions on
\begin{align}
\left\langle \Bl_{ij} \,  \Bm_{jk} \, \Am_{ki} \right\rangle 
=& 
\left\langle \p_k \left( \overline{b}^\ell_i \, \Bm_{kj} \, \Am_{ji} \right) \right\rangle 
- 
\left\langle \p_j \left( \overline{b}^\ell_{i} \, \overline{b}^m_k \, \overline{u}^m_{j,ik}  \right) \right\rangle 
 \notag \\
&+ 
\left\langle \p_i \left( \Bl_{ij} \, \overline{b}^m_k \, \Am_{jk}   \right) \right\rangle 
- 
\left\langle \Bl_{ij} \, \Am_{jk} \, \Bm_{ki}  \right\rangle 
 .
 \label{eq:mhd_gen_Betchov-2-partial}
\end{align}
Clearly the structure of this term may be of interest for Advection and Dynamo subfluxes. Decomposing each gradient tensor in terms of the symmetric and antisymmetric parts yields 
\begin{equation}
    \avg{ \Sbl_{ij} \, \overline{\Sigma}^m_{jk} \, \overline{S}^m_{ki} } + \avg{ \Sbl_{ij} \, \overline{J}^m_{jk} \, \overline{\Omega}^m_{ki} } + \avg{ \overline{J}^\ell_{ij} \, \overline{\Sigma}^m_{jk} \, \overline{\Omega}^m_{ki} } + \avg{ \overline{J}^\ell_{ij} \, \overline{J}^m_{jk} \, \overline{S}^m_{ki} } = 0 
 ,
\end{equation}
and, following manipulations similar to those yielding eq.~\eqref{eq:multisc_betchov}, this can be mapped into a relation between subfluxes
\begin{equation}
\lan \Pi^{A,\ell}_{m,\Sigma \Sigma S} \ran  + \lan \Pi^{A,\ell}_{m,J \Sigma \Omega} \ran  =  \lan \Pi^{A,\ell}_{m,\Sigma J \Omega} \ran + \lan \Pi^{A,\ell}_{m,J J S} \ran 
 ,
\label{eq:second_betchov}
\end{equation}
whose  single-scale  counterpart coincides with eq.~\eqref{eq:multisc_betchov_local}. 
These two relations may be used to write the decomposition of the total MHD energy flux more compactly and to assist with physical interpretations.

\subsection{Further observations}

From fig.~\ref{fig:fluxes_lorentz} it can be observed that the multiscale terms
 $ \langle \Pi^{M,\ell}_{m,S\Sigma \Sigma} \rangle $
and 
 $ \langle \Pi^{M,\ell}_{m,S J J} \rangle$ 
are approximately equal, albeit being very small compared to terms of type $SJ\Sigma$. 
This is reminiscent of the similar relation for two multiscale Inertial subfluxes discussed in 
 section~\ref{sec:inertial}.
However, in the present case the structure of the fields is different since the subfluxes of 
  $ \Pi^{M,\ell} $, 
are formed from one velocity gradient tensor and two magnetic gradient tensors. 

Since our numerical results indicate that
 $ \langle \Pi^{M,\ell}_{m,S\Sigma \Sigma} \rangle 
    \approx 
  \langle \Pi^{M,\ell}_{m,S J J }\rangle $, 
eq.~\eqref{eq:extended_betchov} implies that
 $\langle \Pi^{M,\ell}_{m,\Omega \Sigma J} \rangle \approx 0 $, 
as is also seen for the 
 $ \Pi^{I, \ell}_{\Omega S \Omega} $
Inertial term 
 (in both the HD and MHD cases)
that has the same symmetric/antisymmetric tensorial structure. Eq.~\eqref{eq:extended_betchov}
reveals that the difference between 
$ \langle \Pi^{M,\ell}_{m,S\Sigma \Sigma} \rangle$ 
and 
$ \langle \Pi^{M,\ell}_{m,S J J} \rangle$ 
is governed by
$ \langle \Pi^{M,\ell}_{m,\Omega \Sigma J} \rangle$,
a term
that does not contribute to the energy balance,
because in 
eq.~\eqref{eq:Pi-M} 
only the symmetric part of the gradient tensor survives after the contraction with the symmetric SGS stress 
  $ \tau^\ell(b_i,b_j) $. 
Thus we may posit a physical explanation for 
why 
 $ \langle \Pi^{M,\ell}_{m,\Omega \Sigma J} \rangle \approx 0$ 
by arguing that the values of 
    $ \langle \Pi^{M,\ell}_{m,S\Sigma \Sigma} \rangle$ 
and $ \langle \Pi^{M,\ell}_{m,S J J} \rangle$ 
are essentially determined by the energy balance of the system, and hence cannot be altered by a quantity that does not contribute to this.

Furthermore, we observe in fig.~\ref{fig:pdf_Lorentz}(b) that the p.d.f.s for 
    $ \Pi^{M,\ell}_{m,S\Sigma \Sigma} $
and $ \Pi^{M,\ell}_{m,S J J} $
are roughly coincident, especially along the tails. 
Recall that a similar feature was seen with the analogous Inertial  multi-scale subfluxes.
Further quantitative confirmation of this approximate congruence is given by the similarity of the relevant moments listed in table~\ref{tab:table_moments}.  
These types of approximate identity hold when there is an interplay of either three velocity gradient tensors 
  (Inertial term---HD and MHD) 
or one velocity and two magnetic gradient tensor 
 (Maxwell term). 
However, they do not occur when we study the same structure of subfluxes associated with one vorticity and two velocity gradient tensors in the context of helicity flux 
  (see \cite{CapocciEA23-Hk}). 
This suggests that the approximate identity is unlikely to be of kinematic origin, although the exact version, 
    eq.~\eqref{eq:extended_betchov},
is a kinematic result.

\section{Two-dimensional MHD}
  \label{app:mhd2d}

\subsection{Algebraic setting}

In the 2D case we can express the strain-rate and rotation-rate tensors associated with the incompressible field $\vX = (X_1, X_2)$ in the following way:
\begin{align}
&\vS_X =\begin{pmatrix}
& \p_1 X_1 &  \dfrac{\partial_2 X_1 + \partial_1 X_2}{2} \\
& \dfrac{\partial_2 X_1 + \partial_1 X_2}{2}  & -\p_1 X_1 \ ,
\end{pmatrix}  
\label{eq:2D_strain}\\
& \notag \\
&{\bm \Omega}_X = \dfrac{1}{2} \, \begin{pmatrix}
& 0 &  \partial_2 X_1 - \partial_1 X_2 \\
& -\partial_2 X_1 + \partial_1 X_2  & 0
\end{pmatrix} := \omega^{X} \begin{pmatrix}
0  & -1 \\
1  & 0 
\end{pmatrix} ,
\label{eq:2D_rotation}
\end{align}
where we have already enforced the incompressibility on the trace of eq.~\eqref{eq:2D_strain} and defined 
 $ \omega^X= \left( \partial_1 X_2 -  \partial_2 X_1 \right)/2 $ 
to make the notation more compact. Given an additional incompressible field $\vY$, it is straightforward to verify that $\vS_Y$ and $\bm \Omega_X$ satisfy the commutator algebra:
\begin{equation}
    [\vS_Y, \bm \Omega_X] = 2\,\left( \vS_Y \cdot \bm \Omega_X \right) ,
\label{eq:2D_comm}
\end{equation}
where, unlike the general 3D scenario, the product 
 $ \vS_Y \cdot \bm \Omega_X$ is a symmetric and traceless tensor. 
It is also useful to note that 
the product of two strain-rate tensors related to different gradient tensors can be decomposed as the sum of a symmetric tensor and an antisymmetric one:
\begin{align}
    \vS_X \cdot \vS_Y 
  &= 
    \left( \p_1 X_1 \p_1 Y_1  
          + \dfrac{1}{4} \pa \p_1 X_2  
          + \p_2 X_1 \pc\, \pa \p_1 Y_2 
          + \p_2 Y_1 \pc 
    \right) \cdot \, \mathbb{I}  
 \nonumber \\
 & \quad + 
   \dfrac{1}{2}
     \left( 
       - \p_1 X_1 \left(\p_1 Y_2 + \p_2 Y_1 \right) 
       + \p_1 Y_1\,\left( \p_1 X_2 + \p_2 X_1 \right) 
     \right) 
      \cdot 
    \begin{pmatrix}
       0  &  -1 \\
       1  &   0
    \end{pmatrix}  
  \label{eq:S_A_S_B_dec}
 \\
  & :=
  \sigma(\vX,\vY)\, \mathbb{I} 
   + \alpha(\vX,\vY)\, \hat{\Omega} 
,   
 \label{eq:S_A_times_S_B}
\end{align}
where $\mathbb{I}$ is the $ 2 \times 2$ identity matrix and $\hat{\bm \Omega}$ is the antisymmetric and traceless matrix that defines the 2D rotation-rate tensor of eq.~\eqref{eq:2D_rotation}. 
The (scalar) auxiliary functions $\sigma$ and $\alpha$ embody the functional part multiplying $\mathbb{I}$ and $\hat{\bm \Omega}$ respectively; moreover, they are respectively symmetric and antisymmetric under argument exchange symmetry, i.e.,
\begin{equation}
  \alpha(\vX,\vY)  = -\alpha(\vY,\vX), 
     \quad \quad \quad
  \sigma(\vX,\vY)  =  \sigma(\vY,\vX) .
\end{equation}
Thus, if we swap the fields $\vX$, $\vY$ in eq.~\eqref{eq:S_A_times_S_B}, we obtain
\begin{equation}
  \vS_Y \cdot \vS_X 
  = 
    \sigma(\vX,\vY)\, \mathbb{I} 
  - \alpha(\vX,\vY)\, \hat{\Omega} 
  .
\end{equation}

In terms of tensor traces, the decomposition of eq.~\eqref{eq:S_A_times_S_B} and eq.~\eqref{eq:2D_comm}
lead to three relevant identities, viz: 
\begin{align}
  &  \tr { \vS_{X} \vS_{Y} \vS_{Z} } = 0 
                       \label{eq:S_A_S_B_S_A} \\
  &  \tr { \vS_{X} \vS_{X} \vOmega_{Y} } = 0
                       \label{eq:S_A_S_A_O_A} \\
  &  \tr { \vS_{X} \vOmega_{Y} \vOmega_{Z} } = 0
                        \label{eq:S_A_O_B_O_C}
\end{align}
where~\eqref{eq:S_A_O_B_O_C} vanishes because 
  $ \vOmega_{Y} \vOmega_{Z} \propto \mathbb{I}$ and $\vS_X$ 
is traceless.

With these properties in mind we are ready to specialize
eqs.~\eqref{eq:Pi-I}--\eqref{eq:Pi-M}
to the 2D MHD situation.
This will involve appropriate simplifications of
the single/multi-scale decompositions like eq.~\eqref{eq:Pi-A-exact}.


    \subsection{SGS energy subfluxes}

As a consequence of 
 eqs.~\eqref{eq:S_A_S_B_S_A}--\eqref{eq:S_A_O_B_O_C}, 
in both single-scale and multi-scale cases, the terms involving the contraction of either 
three strain-rate tensors, 
three rotation rate tensors, 
or  
one strain-rate and two rotation rate tensors vanish. 
Hence the 2D Inertial SGS flux is solely due to
  $ \Pi_{m,S \Omega S}^{I,\ell} $
    \citep{Johnson21}:
\begin{equation}
\Pi^{I,\ell} 
  = 
- 2 \int_0^{\ell^2} \d \theta \, 
  \tr{  \vSul \left[ 
    \ol{\vOFu \vSFu}^\phi  -  \vOFFu \; \vSFFu  
             \right] 
  }
\label{eq:PI_2D}
\end{equation}
As mentioned in sec.~\ref{sec:inertial} this is the only Inertial term that ``survives'' in 3D MHD 
  (i.e., is not approximately zero; see figure~\ref{fig:fluxes_hydro_vs_mhd}), 
especially if we add a background magnetic field in the equations of motion (not shown).

The 2D Maxwell flux contains one single-scale and one multi-scale term,
\begin{align}
 & \Pi^{M,\ell} 
   =  - 2\,\ell^2 \tr{ \vSul \; \vSbl \; \vOb }
     + 2 \int_0^{\ell^2} \d \theta \, 
         \tr{ \vSul \left[ 
         \ol{\vOFb \vSFb}^\phi  -  \vOFFb \; \vSFFb  
                   \right] 
            } ,
 \label{eq:PM_2D}
\end{align}
and the Dynamo and the Advection fluxes formally contain two further multi-scale terms:
\begin{align}
\Pi^{A,\ell} 
    &= 
  -2 \, \ell^2 \tr{ \vSbl \; \vOb \; \vSul }
  - \int_0^{\ell^2} \d \theta \, 
    \mathrm{Tr} \Biggl\{  
      -\vOb 
      \left[ 
        \ol{\vSFb \vSFu}^\phi - \vSFFb \; \vSFFu  
      \right]  
 \label{eq:PD_2D}
 \\
   &\quad + 
      \vSbl \left[ 
            \ol{\vOFb \vSFu}^\phi - \vOFFb \; \vSFFu 
          \right] 
    - \vSbl \left[ 
            \ol{\vSFb \vOFu}^\phi - \vSFFb \; \vOFFu 
           \right] 
    \Biggr \}
,
\nonumber \\
\Pi^{D,\ell} 
    &= 
   \int_0^{\ell^2} \d \theta \, 
    \mathrm{Tr} \Biggl \{   
          -\vOb \left[ 
               \ol{\vSFu \vSFb}^\phi - \vSFFu \; \vSFFb
             \right] 
 \label{eq:PA_2A}
\\
    & \quad+ 
       \vSbl \left[ 
               \ol{\vOFu \vSFb}^\phi - \vOFFu \; \vSFFb 
             \right] 
     -
     \vSbl \left[ 
            \ol{\vSFu  \vOFb }^\phi - \vSFFu \; \vOFFb 
          \right] 
         \Biggr \}
.
\nonumber
\end{align}

After straightforward algebraic manipulations we obtain the expression for $\Pi^{\ell}$  that corresponds to the total energy flux for 2D MHD filtered at the scale $\ell$:
\begin{align}   
   \Pi^{\ell} 
  &=  
       \Pi^{I,\ell} + \Pi^{M,\ell} 
     + \Pi^{D,\ell} + \Pi^{A,\ell} 
\nonumber \\
  & = 
     - 4\ell^2 \tr{ \vSbl \; \vOb \; \vSul }  
     +  2 \int_0^{\ell^2} \d \theta \, 
     \mathrm{Tr} \Biggl\{   
        - \vSul \left[ 
               \ol{\vOFu \vSFu}^\phi - \vOFFu \; \vSFFu
              \right]  
\nonumber \\
  &  \quad +
        \vSul \left[ 
               \ol{\vOFb \vSFb}^\phi  -  \vOFFb \; \vSFFb  
          \right]  
     +  \vOb \left[ 
               \ol{\vSFb \vSFu}^\phi - \vSFFb \; \vSFFu  
            \right]   
        \Biggr\}
 . 
 \label{eq:sumP}
\end{align}
Clearly this has one single-scale and three distinct multi-scale contributions.

\section{Equipartition subrange p.d.f.s}
    \label{app:subrange-pdf}

Here we show some  p.d.f.s for the MHD energy subfluxes for a filter scale that lies in the region where there is approximate equipartition between the magnetic and kinetic energy fluxes. 
The p.d.f.s presented in the main body of the paper are calculated for a larger scale. 

\begin{figure}
    \begin{center} \includegraphics[width=.65\columnwidth]{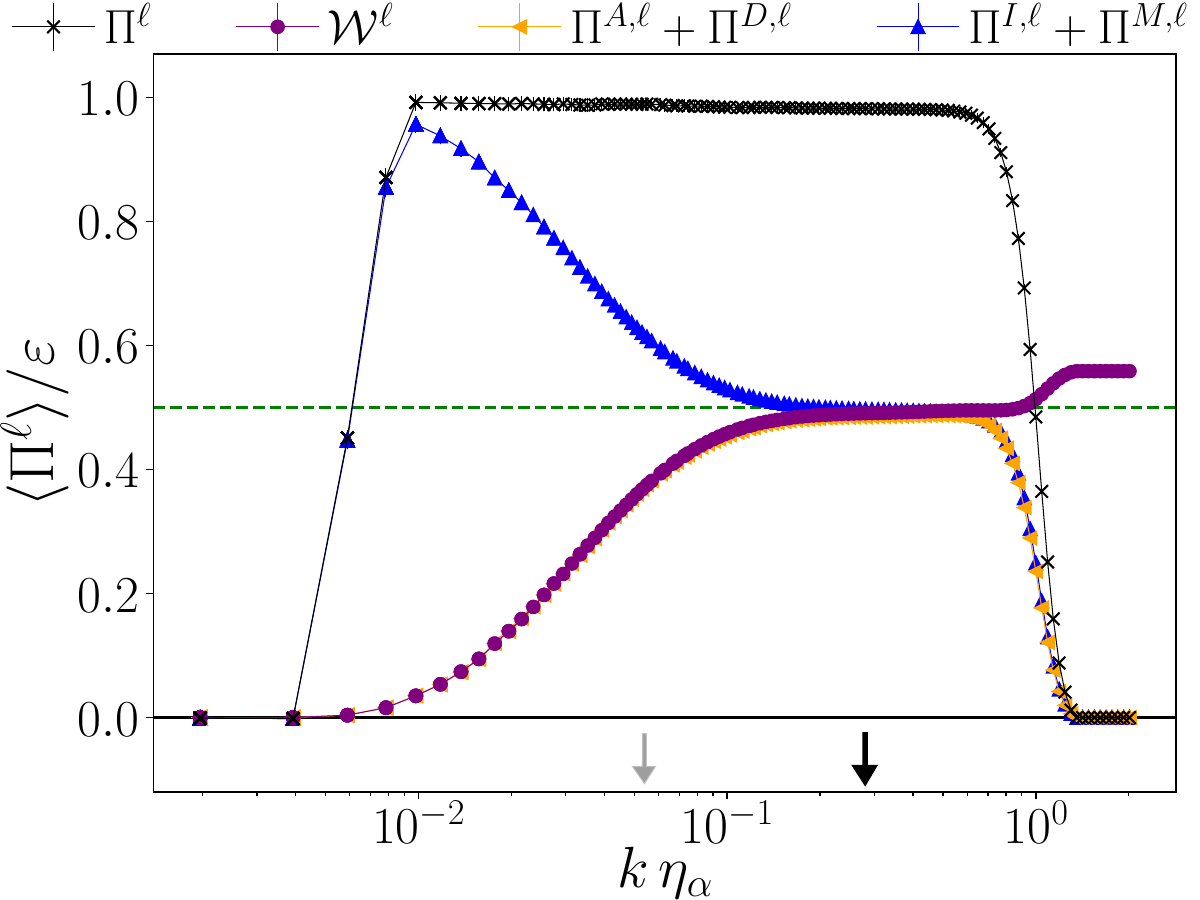}
    \end{center}
    \caption{Scale-filtered fluxes for the kinetic and magnetic energy,  normalized by the mean total energy dissipation rate 
       $ \varepsilon = \varepsilon_u + \varepsilon_b$,
    as a function of the adimensional parameter 
      $ k \eta_\alpha = \pi \eta_\alpha / \ell $. 
    The p.d.f.s shown in figures~\ref{fig:pdf_inertial_eqp}--\ref{fig:pdf_advection_eqp} are calculated for the value of $k \eta_\alpha $
    indicated by the thick (black) arrow, while 
    the thin (grey) arrow denotes the value of $k\eta_\alpha$ used to calculate the p.d.f.s shown in  sec.~\ref{sec:analysis}. 
    Also shown is the kinetic--magnetic energy conversion term $\cal{W}^\ell$. The green dashed horizontal line corresponds to the $y$-axis value of $0.5$. 
    The errorbars, although not fully visible, indicate one standard error.
    }
    \label{fig:sharp_filter}
\end{figure}
Figure~\ref{fig:sharp_filter} displays the net kinetic energy flux, 
  $ \Pi^{I,\ell} + \Pi^{M,\ell} $,
and the net magnetic energy flux,
  $ \Pi^{A,\ell} + \Pi^{D,\ell} $,
obtained using the Fourier filter and dataset A1.
  In essence it is a rearrangement of fig.~\ref{fig:sharp}. 
An equipartition region is evident for 
  $ 0.2 \lesssim k \eta_\alpha \lesssim 0.6 $,
where the magnetic and kinetic energy subfluxes reach approximately $50\%$ of the total energy flux, as indicated by the green dashed line.  
See 
   \cite{BianAluie19}
for discussion of this feature.
Moreover, this equipartition region is also the region where
the conversion term ${\cal W}^\ell$ of eq.~\eqref{eq:Eb-ls} saturates and becomes scale-independent.

\begin{figure}
 \begin{center}
     \includegraphics[width=.48\columnwidth]{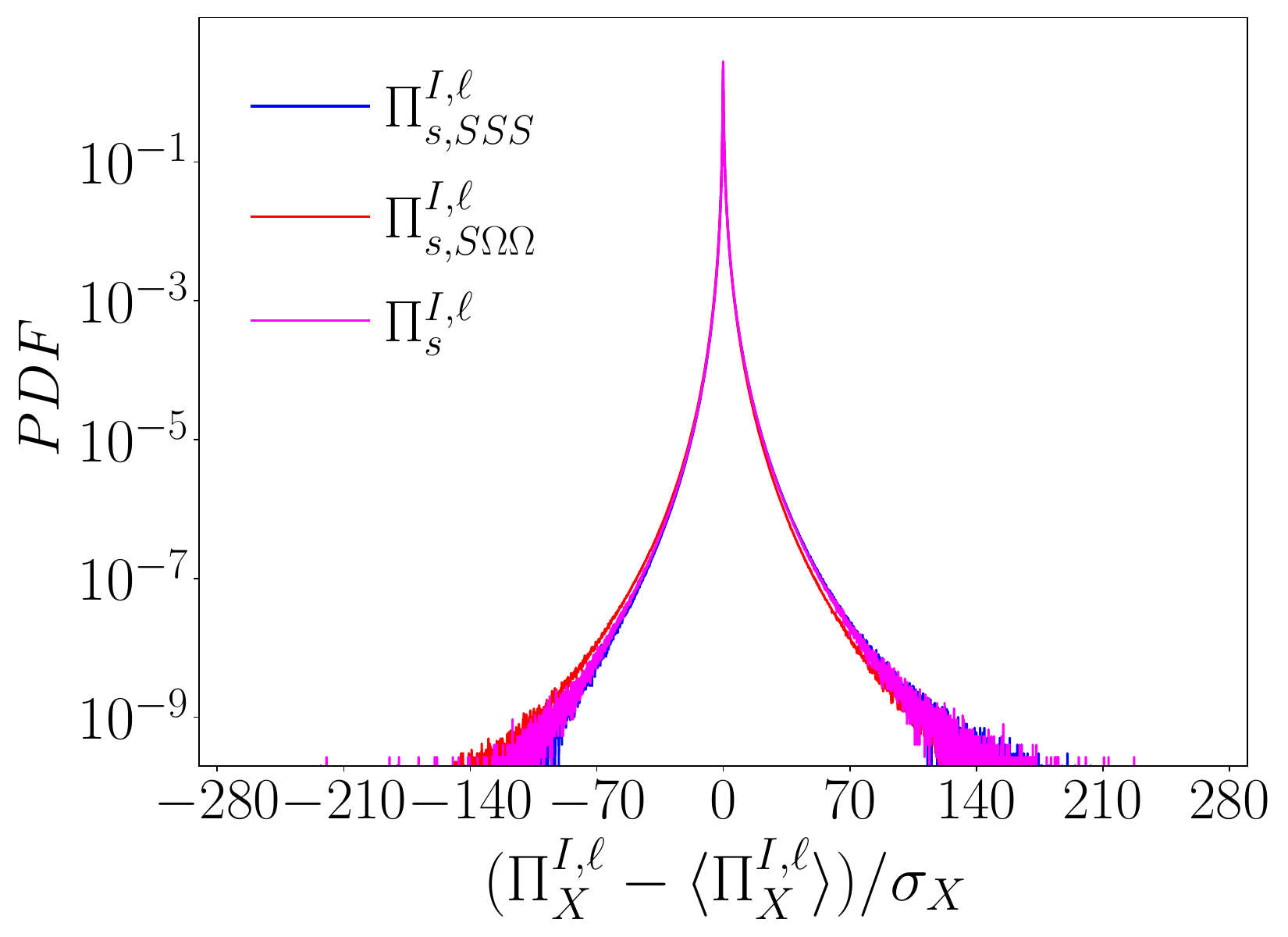} 
     \includegraphics[width=.48\columnwidth]{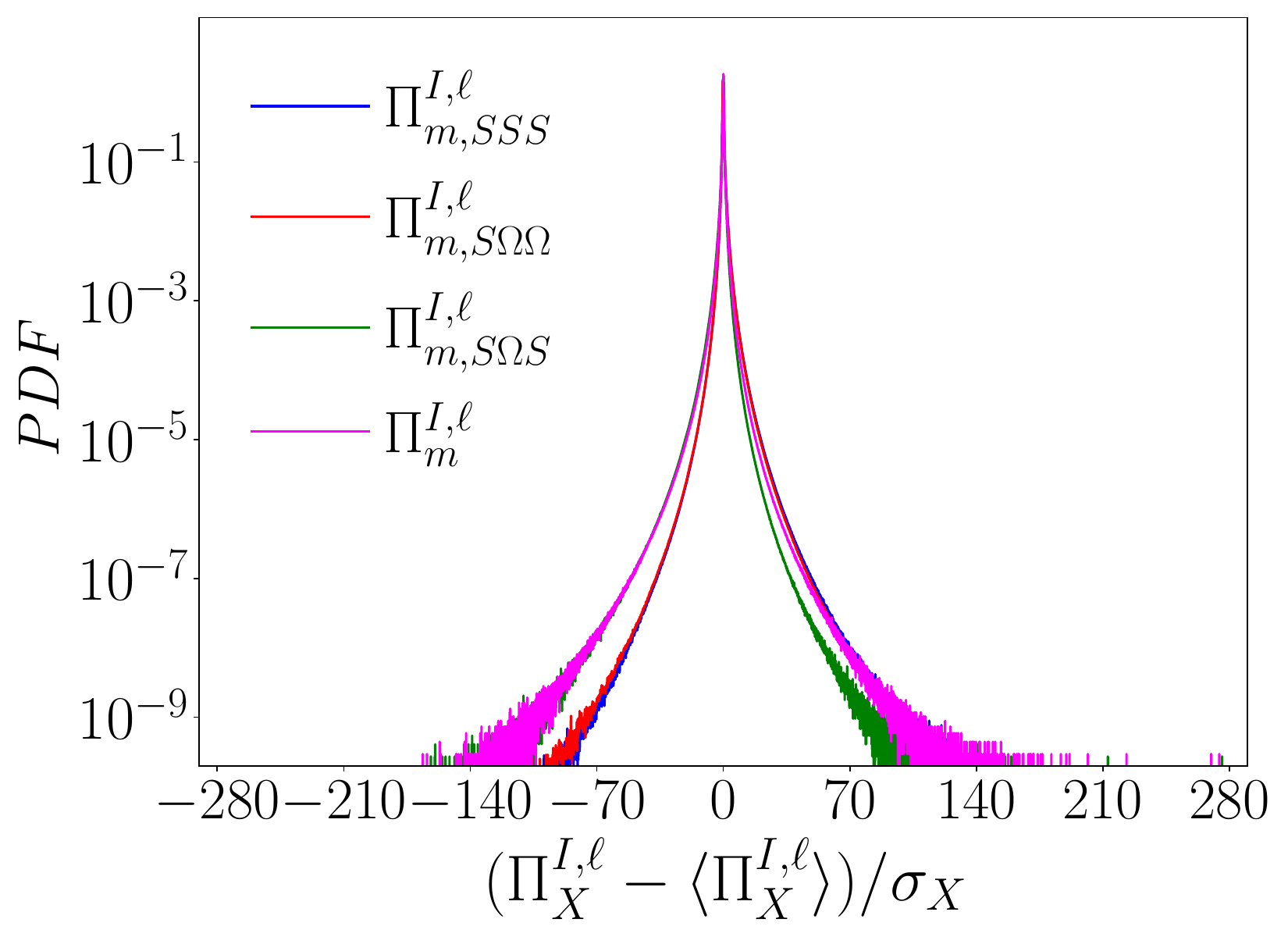} 
 \end{center}
	 \caption{Standardised p.d.f.s of Inertial subfluxes $\Pi^{I,\ell}_{X}$ at $k \eta = 0.27$, where $X$ represents the subflux identifier. Left: single-scale fluxes; 
     right: multi-scale fluxes. }
 \label{fig:pdf_inertial_eqp}
\end{figure}

Figure~\ref{fig:pdf_inertial_eqp} 
displays energy (sub)fluxes for MHD dataset A1, for the filter scale
    $ k \eta_\alpha = 0.27 $.
Comparing these figures to those presented in section~\ref{sec:analysis} 
it is apparent that the p.d.f.s in the equipartition region have fluctuations that are some three times larger.
Recall that the p.d.f.s in 
figures~\ref{fig:pdf_inertial}, 
  \ref{fig:pdf_Lorentz}, 
and
  \ref{fig:pdf_advection}, 
were calculated for the larger scale
    $ k \eta_\alpha = 5.4 \times 10^{-2} $. 
As we progress further into the inertial range the p.d.f.s 
develop even broader tails (not shown).

\begin{figure}
 \begin{center}
     \includegraphics[width=.48\columnwidth]{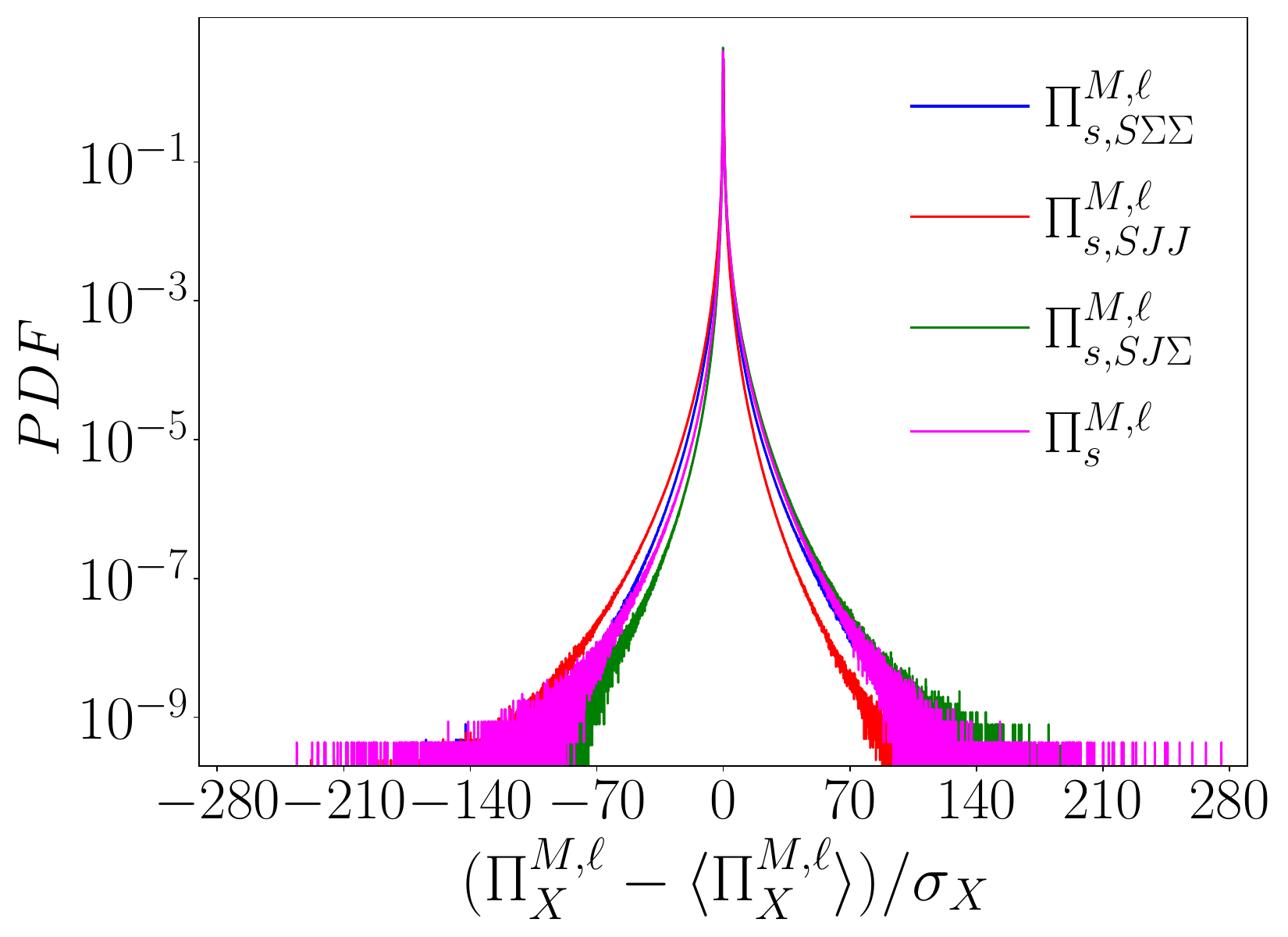} 
     \includegraphics[width=.48\columnwidth]{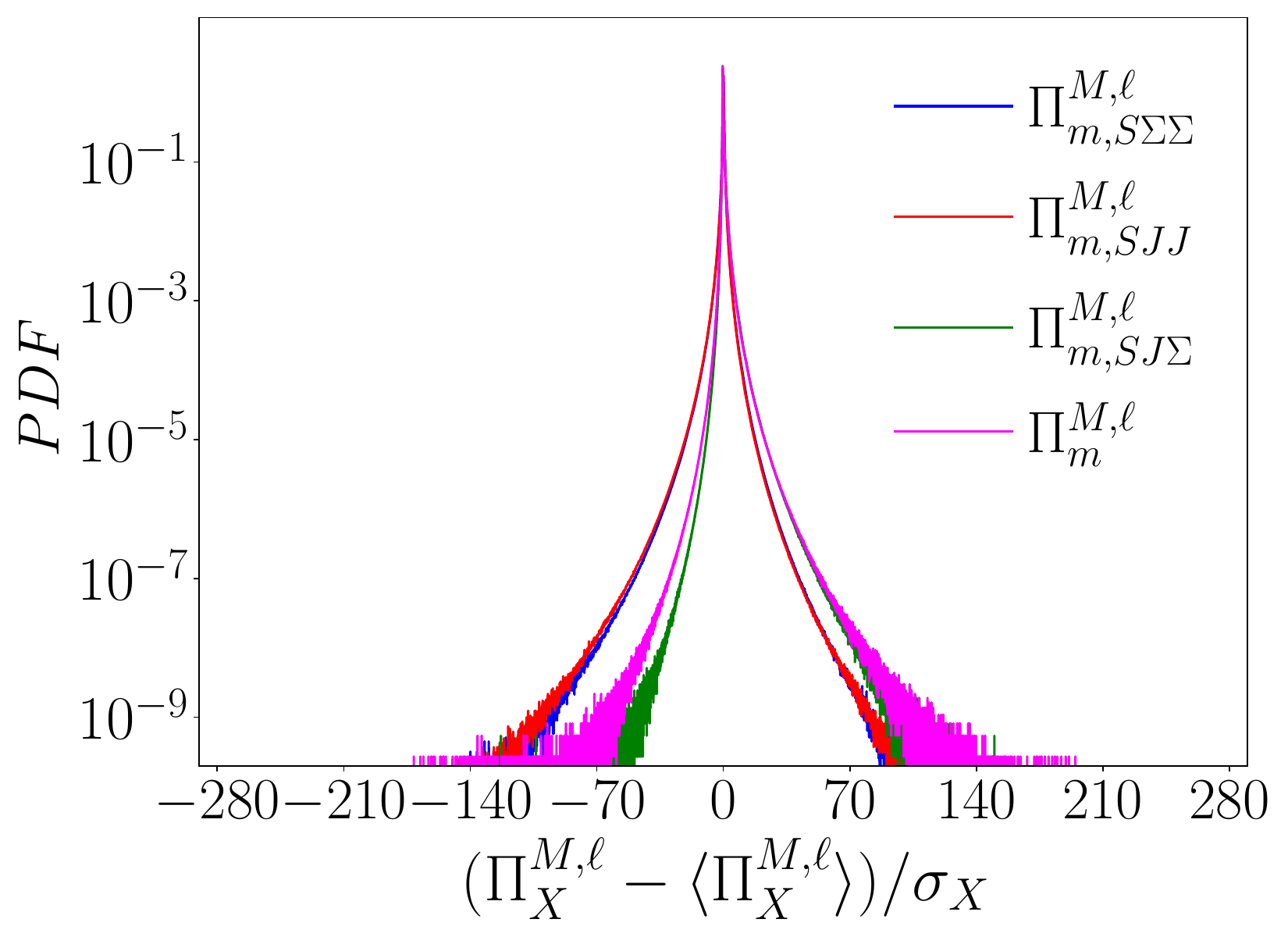} 
 \end{center}
	 \caption{As for figure~\ref{fig:pdf_inertial_eqp} for the Maxwell energy fluxes,  $\Pi^{M,\ell}_{X}$. }
\label{fig:pdf_Lorentz_eqp}
\end{figure}

\begin{figure}
 \begin{center}
     \includegraphics[width=.48\columnwidth]{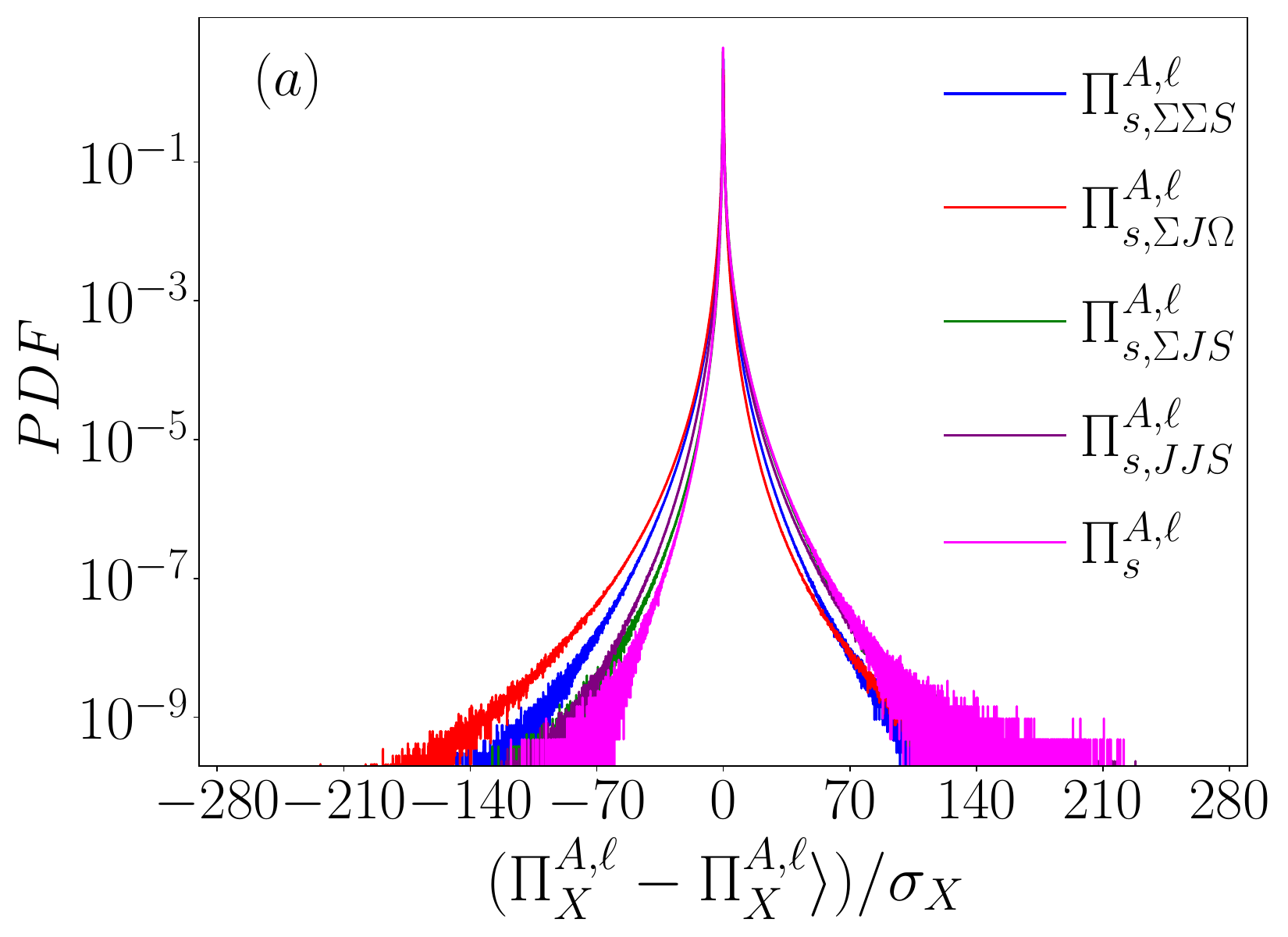} 
     \includegraphics[width=.48\columnwidth]{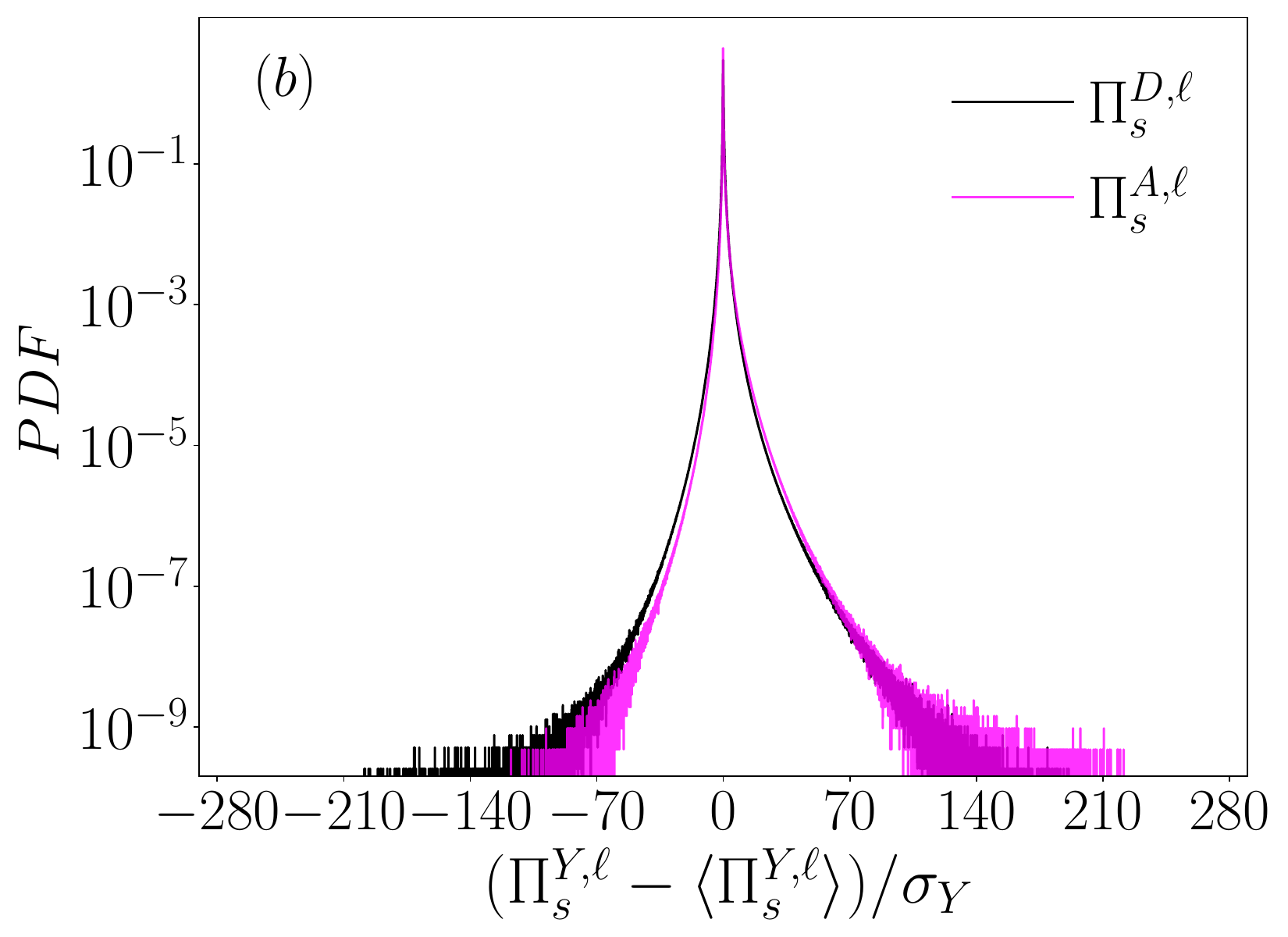} 
      \includegraphics[width=.48\columnwidth]{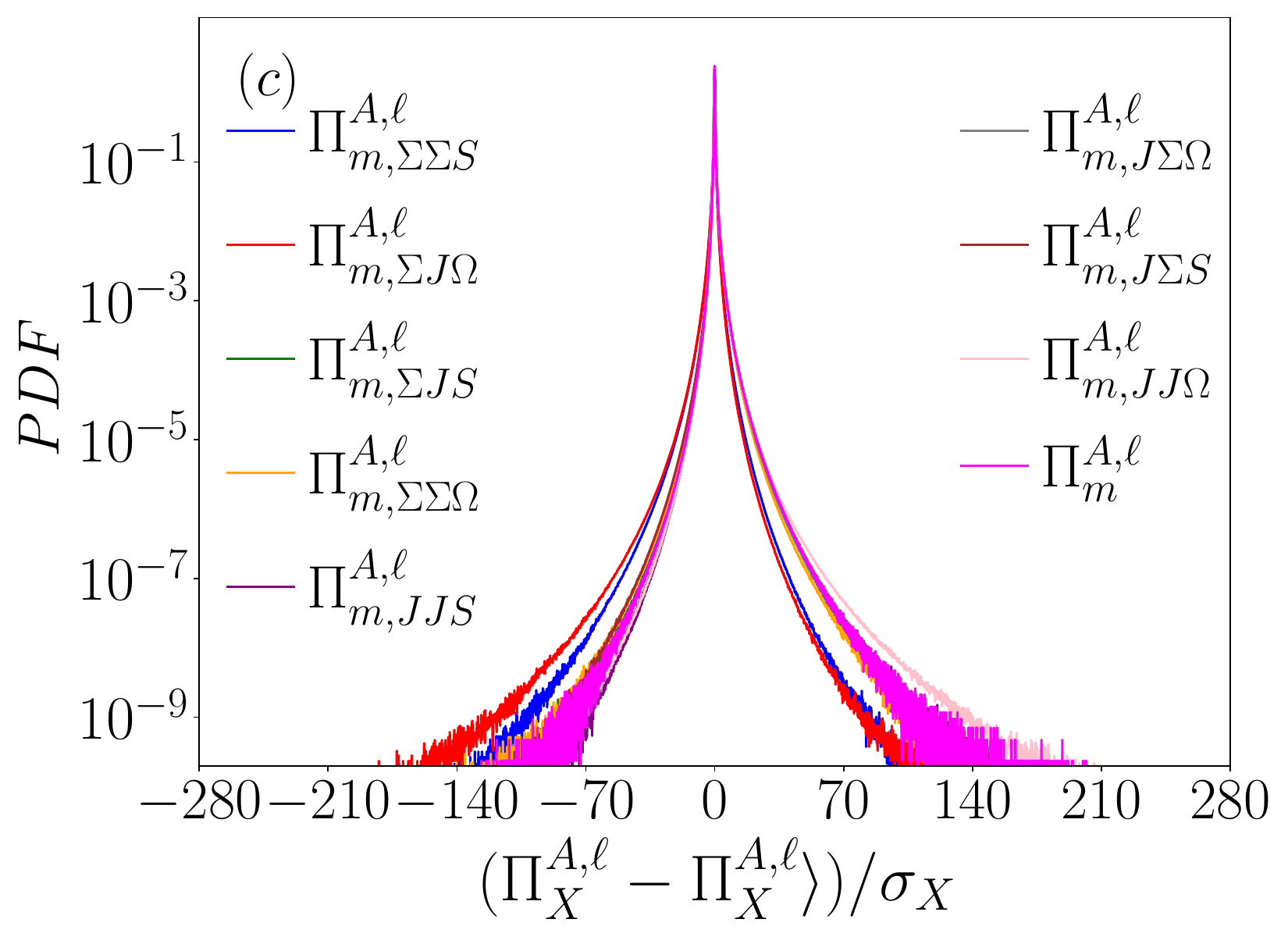} 
    \includegraphics[width=.48\columnwidth]{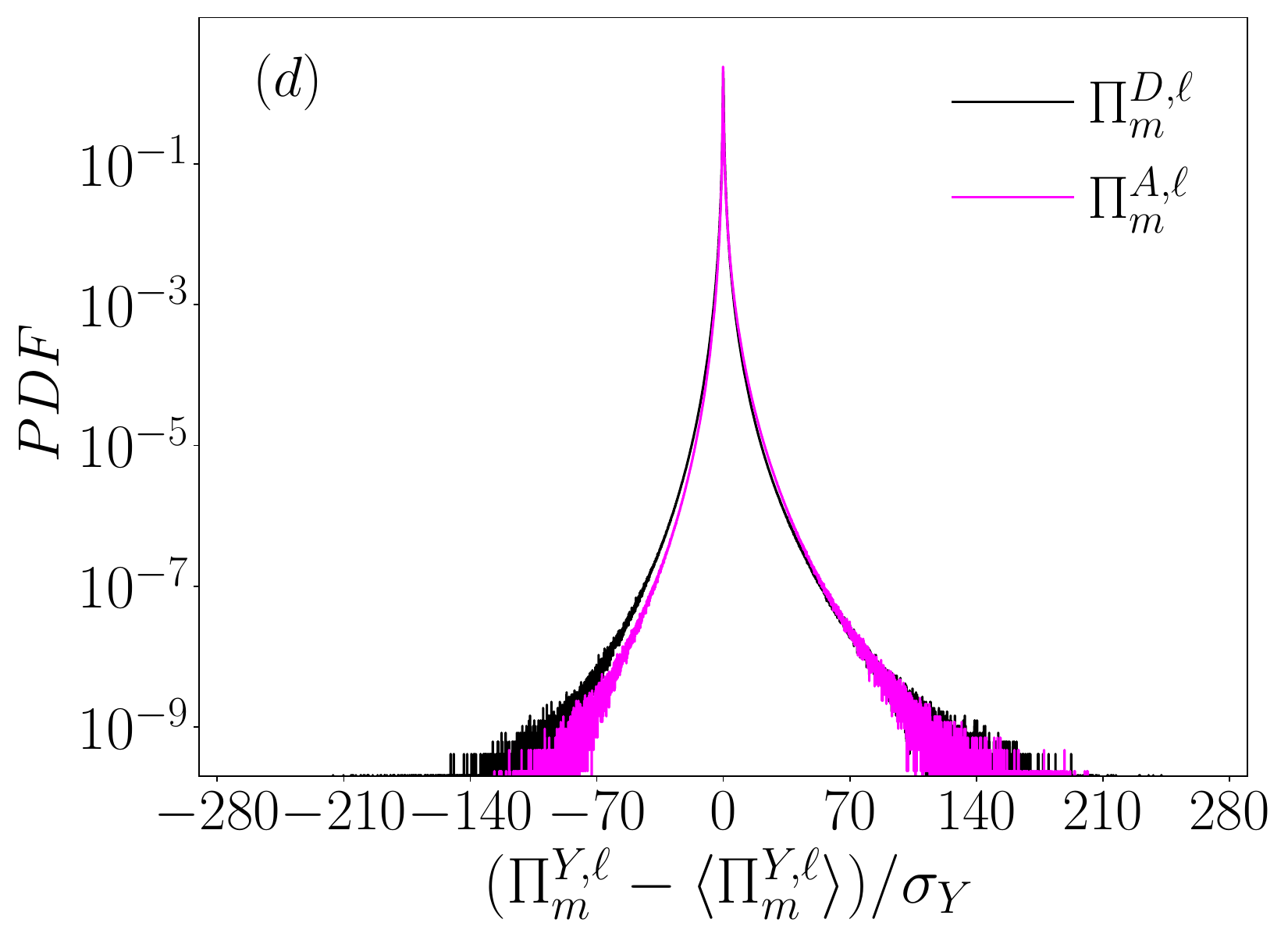} 
 \end{center}
	 \caption{Standardised p.d.f.s. at $ k \eta_\alpha = 0.27 $.
(a) single-scale Advection subfluxes $\Pi^{A,\ell}_{s,X}$, where $X$ represents the subflux identifier. 
(b) single-scale net fluxes for the Dynamo and Advection terms, respectively $ \Pi^{D,\ell}_{s} $ and $ \Pi^{A,\ell}_{s} $, where $Y=D$ or $A$ identifies the term. 
The subflux $\Pi_s^{A,\ell}$ is shown in both panels. 
(c),(d) p.d.f.s of multi-scale Advection and Dynamo subfluxes, respectively. 
}
\label{fig:pdf_advection_eqp}
\end{figure}



   \clearpage

\section{Subfluxes definitions}
\label{app:defintions}

In this section we provide the definitions of all the subfluxes appearing in the decomposition of the MHD energy fluxes. 
As highlighted in sec.~\ref{sec:methods}, for the  subfluxes 
$\Pi^I_{s, S \Omega S}$, 
$\Pi^I_{m, S \Omega S}$ 
and 
$\Pi^M_{s, SJ \Sigma}$, 
$\Pi^M_{m, SJ \Sigma}$ 
there is an extra factor of two that arises from the symmetry of the corresponding SGS stress tensors. Because in eqs.\eqref{eq:Eu-ls} and \eqref{eq:Eb-ls} 
the fluxes appear with the same leading signs,
both the Maxwell and the Dynamo subfluxes in the definition below acquires an additional minus sign.

\subsection{Inertial}
The following subfluxes are identical to the hydrodynamic counterpart from \cite{Johnson20,Johnson21}.
\begin{align}
   &\Pi^{I,\ell}_{s,S S S}    
     = 
    -  \ell^2 \, \tr{\big( \overline{\vS}^\ell \big)^t \,
       \overline{\vS}^\ell \big(\overline{\vS}^\ell\big)^t } \\
   & \Pi^{I,\ell}_{m,S S S } 
     =
    - \int_0^{\ell^2} \! \! \d\theta \,
       \tr{\big( \overline{\vS}^\ell \big)^t \Big(
       \ol{ \overline{\vS}^{\sqrt{\theta}} 
              \big(\overline{\vS}^{\sqrt{\theta}} \big)^{t} \,}^\phi 
           -
            \ol{\overline{\vS}^{\sqrt{\theta}}\,}^\phi\;
             \ol{ \big( \overline{\vS}^{\sqrt{\theta}} \big)^{\! t} \,}^\phi  
       \Big)} \\
       & \notag  \\
     &\Pi^{I,\ell}_{s,S \Omega \Omega}    
     = 
    -  \ell^2 \, \tr{\big( \overline{\vS}^\ell \big)^t \,
       \overline{\vOmega}^\ell \big(\overline{\vOmega}^\ell\big)^t } \\
   & \Pi^{I,\ell}_{m,S \Omega \Omega } 
     =
    - \int_0^{\ell^2} \! \! \d\theta \,
       \tr{\big( \overline{\vS}^\ell \big)^t \Big(
       \ol{ \overline{\vOmega}^{\sqrt{\theta}} 
              \big(\overline{\vOmega}^{\sqrt{\theta}} \big)^{t} \,}^\phi 
           -
            \ol{\overline{\vOmega}^{\sqrt{\theta}}\,}^\phi\;
             \ol{ \big( \overline{\vOmega}^{\sqrt{\theta}} \big)^{\! t} \,}^\phi  
       \Big)} \\
       &   \notag \\
   &\Pi^{I,\ell}_{s,S \Omega S}    
     = 
    - 2 \ell^2 \, \tr{\big( \overline{\vS}^\ell \big)^t \,
       \overline{\vOmega}^\ell \big(\overline{\vS}^\ell\big)^t } \equiv 0 \\
   & \Pi^{I,\ell}_{m,S \Omega S} 
     =
    - 2\int_0^{\ell^2} \! \! \d\theta \,
       \tr{\big( \overline{\vS}^\ell \big)^t \Big(
       \ol{ \overline{\vOmega}^{\sqrt{\theta}} 
              \big(\overline{\vS}^{\sqrt{\theta}} \big)^{t} \,}^\phi 
           -
            \ol{\overline{\vOmega}^{\sqrt{\theta}}\,}^\phi\;
             \ol{ \big( \overline{\vS}^{\sqrt{\theta}} \big)^{\! t} \,}^\phi  
       \Big)} 
\end{align}

\subsection{Maxwell}
\begin{align}
   &\Pi^{M,\ell}_{s,S \Sigma \Sigma}    
     = 
      \ell^2 \, \tr{\big( \overline{\vS}^\ell \big)^t \,
       \overline{\vSigma}^\ell \big(\overline{\vSigma}^\ell\big)^t } \\
   & \Pi^{M,\ell}_{m,S \Sigma \Sigma} 
     =
     \int_0^{\ell^2} \! \! \d\theta \,
       \tr{\big( \overline{\vS}^\ell \big)^t \Big(
       \ol{ \overline{\vSigma}^{\sqrt{\theta}} 
              \big(\overline{\vSigma}^{\sqrt{\theta}} \big)^{t} \,}^\phi 
           -
            \ol{\overline{\vSigma}^{\sqrt{\theta}}\,}^\phi\;
             \ol{ \big( \overline{\vSigma}^{\sqrt{\theta}} \big)^{\! t} \,}^\phi  
       \Big)} \\
       & \notag  \\
   &\Pi^{M,\ell}_{s,S J J}    
     = 
      \ell^2 \, \tr{\big( \overline{\vS}^\ell \big)^t \,
       \overline{\vJ}^\ell \big(\overline{\vJ}^\ell\big)^t } \\
   & \Pi^{M,\ell}_{m,S J J} 
     =
     \int_0^{\ell^2} \! \! \d\theta \,
       \tr{\big( \overline{\vS}^\ell \big)^t \Big(
       \ol{ \overline{\vJ}^{\sqrt{\theta}} 
              \big(\overline{\vJ}^{\sqrt{\theta}} \big)^{t} \,}^\phi 
           -
            \ol{\overline{\vJ}^{\sqrt{\theta}}\,}^\phi\;
             \ol{ \big( \overline{\vJ}^{\sqrt{\theta}} \big)^{\! t} \,}^\phi  
       \Big)}\\
   &   \notag \\
   &\Pi^{M,\ell}_{s,S J \Sigma}    
     = 
     2 \ell^2 \, \tr{\big( \overline{\vS}^\ell \big)^t \,
       \overline{\vJ}^\ell \big(\overline{\vSigma}^\ell\big)^t } \\
   & \Pi^{M,\ell}_{m,S J \Sigma} 
     =
     2\int_0^{\ell^2} \! \! \d\theta \,
       \tr{\big( \overline{\vS}^\ell \big)^t \Big(
       \ol{ \overline{\vJ}^{\sqrt{\theta}} 
              \big(\overline{\vSigma}^{\sqrt{\theta}} \big)^{t} \,}^\phi 
           -
            \ol{\overline{\vJ}^{\sqrt{\theta}}\,}^\phi\;
             \ol{ \big( \overline{\vSigma}^{\sqrt{\theta}} \big)^{\! t} \,}^\phi  
       \Big)}
\end{align}

\subsection{Advection}
\begin{align}
  &\Pi^{A,\ell}_{s,\Sigma \Sigma S}
     = 
    - \ell^2 \, \tr{\big( \overline{\vSigma}^\ell \big)^t \,
       \overline{\vSigma}^\ell \big(\overline{\vS}^\ell\big)^t } \\
  &\Pi^{A,\ell}_{m,\Sigma \Sigma S}
     =
       \; - \;
       \int_0^{\ell^2} \d\theta \,
       \tr{\big( \overline{\vSigma}^\ell \big)^t \Big(
       \ol{ \overline{\vSigma}^{\sqrt{\theta}} 
              \big(\overline{\vS}^{\sqrt{\theta}} \big)^{t} \,}^\phi 
           -
            \ol{\overline{\vSigma}^{\sqrt{\theta}}\,}^\phi\;
             \ol{ \big( \overline{\vSigma}^{\sqrt{\theta}} \big)^{\! t} \,}^\phi  
       \Big)}\\
  &  \notag \\
  &\Pi^{A,\ell}_{s,\Sigma J S}
     = 
    - \ell^2 \, \tr{\big( \overline{\vSigma}^\ell \big)^t \,
       \overline{\vJ}^\ell \big(\overline{\vS}^\ell\big)^t }\\
  &\Pi^{A,\ell}_{m,\Sigma J S}
  =
       \; - \;
       \int_0^{\ell^2} \d\theta \,
       \tr{\big( \overline{\vSigma}^\ell \big)^t \Big(
       \ol{ \overline{\vJ}^{\sqrt{\theta}} 
              \big(\overline{\vS}^{\sqrt{\theta}} \big)^{t} \,}^\phi 
           -
            \ol{\overline{\vJ}^{\sqrt{\theta}}\,}^\phi\;
             \ol{ \big( \overline{\vS}^{\sqrt{\theta}} \big)^{\! t} \,}^\phi  
       \Big)} \\
  &  \notag \\
  &\Pi^{A,\ell}_{s,\Sigma \Sigma \Omega}
     = - \ell^2 \, \tr{\big( \overline{\vSigma}^\ell \big)^t \,
       \overline{\vSigma}^\ell \big(\overline{\vOmega}^\ell\big)^t } \equiv 0 \\ 
  &\Pi^{A,\ell}_{m,\Sigma \Sigma \Omega} =     
  \; - \;
       \int_0^{\ell^2} \d\theta \,
       \tr{\big( \overline{\vSigma}^\ell \big)^t \Big(
       \ol{ \overline{\vSigma}^{\sqrt{\theta}} 
              \big(\overline{\vOmega}^{\sqrt{\theta}} \big)^{t} \,}^\phi 
           -
            \ol{\overline{\vSigma}^{\sqrt{\theta}}\,}^\phi\;
             \ol{ \big( \overline{\vOmega}^{\sqrt{\theta}} \big)^{\! t} \,}^\phi  
       \Big)}\\
  &  \notag \\
  &\Pi^{A,\ell}_{s,\Sigma J \Omega}   
     = 
    - \ell^2 \, \tr{\big( \overline{\vSigma}^\ell \big)^t \,
       \overline{\vJ}^\ell \big(\overline{\vOmega}^\ell\big)^t } \\
  &\Pi^{A,\ell}_{m,\Sigma J \Omega}
       =
       \; - \;
       \int_0^{\ell^2} \d\theta \,
       \tr{\big( \overline{\vSigma}^\ell \big)^t \Big(
       \ol{ \overline{\vJ}^{\sqrt{\theta}} 
              \big(\overline{\vOmega}^{\sqrt{\theta}} \big)^{t} \,}^\phi 
           -
            \ol{\overline{\vJ}^{\sqrt{\theta}}\,}^\phi\;
             \ol{ \big( \overline{\vOmega}^{\sqrt{\theta}} \big)^{\! t} \,}^\phi  
       \Big)}\\
  & \notag  \\
  &\Pi^{A,\ell}_{s,J \Sigma S}
      = 
    - \ell^2 \, \tr{\big( \overline{\vJ}^\ell \big)^t \,
       \overline{\vSigma}^\ell \big(\overline{\vS}^\ell\big)^t } \\
  &\Pi^{A,\ell}_{m,J \Sigma S}
      =
       \; - \;
       \int_0^{\ell^2} \d\theta \,
       \tr{\big( \overline{\vJ}^\ell \big)^t \Big(
       \ol{ \overline{\vSigma}^{\sqrt{\theta}} 
              \big(\overline{\vS}^{\sqrt{\theta}} \big)^{t} \,}^\phi 
           -
            \ol{\overline{\vSigma}^{\sqrt{\theta}}\,}^\phi\;
             \ol{ \big( \overline{\vS}^{\sqrt{\theta}} \big)^{\! t} \,}^\phi  
       \Big)}\\
       & \notag  \\
  &\Pi^{A,\ell}_{s,J \Sigma \Omega}
      = 
    - \ell^2 \, \tr{\big( \overline{\vJ}^\ell \big)^t \,
       \overline{\vSigma}^\ell \big(\overline{\vOmega}^\ell\big)^t } \\
  &\Pi^{A,\ell}_{m,J \Sigma \Omega}
      =
       \; - \;
       \int_0^{\ell^2} \d\theta \,
       \tr{\big( \overline{\vJ}^\ell \big)^t \Big(
       \ol{ \overline{\vSigma}^{\sqrt{\theta}} 
              \big(\overline{\vOmega}^{\sqrt{\theta}} \big)^{t} \,}^\phi 
           -
            \ol{\overline{\vSigma}^{\sqrt{\theta}}\,}^\phi\;
             \ol{ \big( \overline{\vOmega}^{\sqrt{\theta}} \big)^{\! t} \,}^\phi  
       \Big)}\\
       & \notag   \\
  &\Pi^{A,\ell}_{s,J J S}
     =  
      - \ell^2 \, \tr{\big( \overline{\vJ}^\ell \big)^t \,
       \overline{\vJ}^\ell \big(\overline{\vS}^\ell\big)^t }  \\
  &\Pi^{A,\ell}_{m,J J S}   
    =
    \; - \;
       \int_0^{\ell^2} \d\theta \,
       \tr{\big( \overline{\vJ}^\ell \big)^t \Big(
       \ol{ \overline{\vJ}^{\sqrt{\theta}} 
              \big(\overline{\vS}^{\sqrt{\theta}} \big)^{t} \,}^\phi 
           -
            \ol{\overline{\vJ}^{\sqrt{\theta}}\,}^\phi\;
             \ol{ \big( \overline{\vS}^{\sqrt{\theta}} \big)^{\! t} \,}^\phi  
       \Big)}\\
  & \notag   \\
  &\Pi^{A,\ell}_{s,J J \Omega}
     =  
      - \ell^2 \, \tr{\big( \overline{\vJ}^\ell \big)^t \,
       \overline{\vJ}^\ell \big(\overline{\vOmega}^\ell\big)^t } \equiv 0 \\
  &\Pi^{A,\ell}_{m,J J \Omega}   
    =
    \; - \;
       \int_0^{\ell^2} \d\theta \,
       \tr{\big( \overline{\vJ}^\ell \big)^t \Big(
       \ol{ \overline{\vJ}^{\sqrt{\theta}} 
              \big(\overline{\vOmega}^{\sqrt{\theta}} \big)^{t} \,}^\phi 
           -
            \ol{\overline{\vJ}^{\sqrt{\theta}}\,}^\phi\;
             \ol{ \big( \overline{\vOmega}^{\sqrt{\theta}} \big)^{\! t} \,}^\phi  
       \Big)}
\end{align}

\subsection{Dynamo}
\begin{align}
  &\Pi^{D,\ell}_{s,\Sigma S \Sigma}
     = 
     \ell^2 \, \tr{\big( \overline{\vSigma}^\ell \big)^t \,
       \overline{\vS}^\ell \big(\overline{\vSigma}^\ell\big)^t } \\
  &\Pi^{D,\ell}_{m,\Sigma S \Sigma}
     =
       \;  \;
       \int_0^{\ell^2} \d\theta \,
       \tr{\big( \overline{\vSigma}^\ell \big)^t \Big(
       \ol{ \overline{\vS}^{\sqrt{\theta}} 
              \big(\overline{\vSigma}^{\sqrt{\theta}} \big)^{t} \,}^\phi 
           -
            \ol{\overline{\vS}^{\sqrt{\theta}}\,}^\phi\;
             \ol{ \big( \overline{\vSigma}^{\sqrt{\theta}} \big)^{\! t} \,}^\phi  
       \Big)}\\
  &  \notag \\
  &\Pi^{D,\ell}_{s,\Sigma \Omega \Sigma}
     = 
     \ell^2 \, \tr{\big( \overline{\vSigma}^\ell \big)^t \,
       \overline{\vOmega}^\ell \big(\overline{\vSigma}^\ell\big)^t } \equiv 0 \\
  &\Pi^{D,\ell}_{m,\Sigma \Omega \Sigma}
     =
       \;  \;
       \int_0^{\ell^2} \d\theta \,
       \tr{\big( \overline{\vSigma}^\ell \big)^t \Big(
       \ol{ \overline{\vOmega}^{\sqrt{\theta}} 
              \big(\overline{\vSigma}^{\sqrt{\theta}} \big)^{t} \,}^\phi 
           -
            \ol{\overline{\vOmega}^{\sqrt{\theta}}\,}^\phi\;
             \ol{ \big( \overline{\vSigma}^{\sqrt{\theta}} \big)^{\! t} \,}^\phi  
       \Big)}\\
  &  \notag \\
    &\Pi^{D,\ell}_{s,\Sigma S J}
     = 
     \ell^2 \, \tr{\big( \overline{\vSigma}^\ell \big)^t \,
       \overline{\vS}^\ell \big(\overline{\vJ}^\ell\big)^t } \\
  &\Pi^{D,\ell}_{m,\Sigma S J}
     =
       \;  \;
       \int_0^{\ell^2} \d\theta \,
       \tr{\big( \overline{\vSigma}^\ell \big)^t \Big(
       \ol{ \overline{\vS}^{\sqrt{\theta}} 
              \big(\overline{\vJ}^{\sqrt{\theta}} \big)^{t} \,}^\phi 
           -
            \ol{\overline{\vS}^{\sqrt{\theta}}\,}^\phi\;
             \ol{ \big( \overline{\vJ}^{\sqrt{\theta}} \big)^{\! t} \,}^\phi  
       \Big)}\\
  &  \notag \\
    &\Pi^{D,\ell}_{s,\Sigma \Omega J}
     = 
     \ell^2 \, \tr{\big( \overline{\vSigma}^\ell \big)^t \,
       \overline{\vOmega}^\ell \big(\overline{\vJ}^\ell\big)^t } \\
  &\Pi^{D,\ell}_{m,\Sigma \Omega J}
     =
       \;  \;
       \int_0^{\ell^2} \d\theta \,
       \tr{\big( \overline{\vSigma}^\ell \big)^t \Big(
       \ol{ \overline{\vOmega}^{\sqrt{\theta}} 
              \big(\overline{\vJ}^{\sqrt{\theta}} \big)^{t} \,}^\phi 
           -
            \ol{\overline{\vOmega}^{\sqrt{\theta}}\,}^\phi\;
             \ol{ \big( \overline{\vJ}^{\sqrt{\theta}} \big)^{\! t} \,}^\phi  
       \Big)}\\
  &  \notag \\
      &\Pi^{D,\ell}_{s,J S \Sigma}
     = 
     \ell^2 \, \tr{\big( \overline{\vJ}^\ell \big)^t \,
       \overline{\vS}^\ell \big(\overline{\vSigma}^\ell\big)^t } \\
  &\Pi^{D,\ell}_{m,J S \Sigma}
     =
       \;  \;
       \int_0^{\ell^2} \d\theta \,
       \tr{\big( \overline{\vJ}^\ell \big)^t \Big(
       \ol{ \overline{\vS}^{\sqrt{\theta}} 
              \big(\overline{\vSigma}^{\sqrt{\theta}} \big)^{t} \,}^\phi 
           -
            \ol{\overline{\vS}^{\sqrt{\theta}}\,}^\phi\;
             \ol{ \big( \overline{\vSigma}^{\sqrt{\theta}} \big)^{\! t} \,}^\phi  
       \Big)}\\
  &  \notag \\
      &\Pi^{D,\ell}_{s,J \Omega \Sigma}
     = 
     \ell^2 \, \tr{\big( \overline{\vJ}^\ell \big)^t \,
       \overline{\vOmega}^\ell \big(\overline{\vSigma}^\ell\big)^t } \\
  &\Pi^{D,\ell}_{m,J \Omega \Sigma}
     =
       \;  \;
       \int_0^{\ell^2} \d\theta \,
       \tr{\big( \overline{\vJ}^\ell \big)^t \Big(
       \ol{ \overline{\vOmega}^{\sqrt{\theta}} 
              \big(\overline{\vSigma}^{\sqrt{\theta}} \big)^{t} \,}^\phi 
           -
            \ol{\overline{\vOmega}^{\sqrt{\theta}}\,}^\phi\;
             \ol{ \big( \overline{\vSigma}^{\sqrt{\theta}} \big)^{\! t} \,}^\phi  
       \Big)}\\
  &  \notag \\
      &\Pi^{D,\ell}_{s,J S J}
     = 
     \ell^2 \, \tr{\big( \overline{\vJ}^\ell \big)^t \,
       \overline{\vS}^\ell \big(\overline{\vJ}^\ell\big)^t } \\
  &\Pi^{D,\ell}_{m,J S J}
     =
       \;  \;
       \int_0^{\ell^2} \d\theta \,
       \tr{\big( \overline{\vJ}^\ell \big)^t \Big(
       \ol{ \overline{\vS}^{\sqrt{\theta}} 
              \big(\overline{\vJ}^{\sqrt{\theta}} \big)^{t} \,}^\phi 
           -
            \ol{\overline{\vS}^{\sqrt{\theta}}\,}^\phi\;
             \ol{ \big( \overline{\vJ}^{\sqrt{\theta}} \big)^{\! t} \,}^\phi  
       \Big)}\\
       &  \notag \\
      &\Pi^{D,\ell}_{s,J \Omega J}
     = 
     \ell^2 \, \tr{\big( \overline{\vJ}^\ell \big)^t \,
       \overline{\vOmega}^\ell \big(\overline{\vJ}^\ell\big)^t } \equiv 0  \\
  &\Pi^{D,\ell}_{m,J \Omega J}
     =
       \;  \;
       \int_0^{\ell^2} \d\theta \,
       \tr{\big( \overline{\vJ}^\ell \big)^t \Big(
       \ol{ \overline{\vOmega}^{\sqrt{\theta}} 
              \big(\overline{\vJ}^{\sqrt{\theta}} \big)^{t} \,}^\phi 
           -
            \ol{\overline{\vOmega}^{\sqrt{\theta}}\,}^\phi\;
             \ol{ \big( \overline{\vJ}^{\sqrt{\theta}} \big)^{\! t} \,}^\phi  
       \Big)}
\end{align}

  \bibliographystyle{jpp}
  \bibliography{ag,hl,mp,qz,extras}
\end{document}